\documentclass[final,3p]{elsarticle}
\usepackage[utf8]{inputenc}
\usepackage{float}
\usepackage{comment}
\usepackage{graphicx}
\usepackage{amsmath}
\usepackage[version=4]{mhchem}
\usepackage{siunitx}
\usepackage{longtable,tabularx}
\usepackage{array} 
\usepackage{multirow}

\usepackage{amssymb}
\usepackage[T1]{fontenc}
\usepackage{amsthm}

\usepackage{float}
\usepackage{amsmath}
\usepackage{bm}
\usepackage{rotating}
\usepackage{hyperref}
\usepackage{tabularx}
\usepackage{subfigure}
\usepackage{psfrag}
\usepackage{amssymb}
\usepackage{wasysym}
\usepackage[normalem]{ulem}
\usepackage{psfrag}
\usepackage{stmaryrd}
\usepackage{stackrel}
\usepackage{multirow}
\usepackage{mathdots}
\usepackage{pdflscape}

\usepackage{xcolor}
\usepackage{algorithm,algpseudocode}

\begin{document}

\begin{frontmatter}



\title{Curvilinear Moving Overset Method for High-order Non-dissipative Schemes}

 \author[NS]{Minhazul Islam}

 \author[NS]{and Nek Sharan}
 \ead{nsharan@auburn.edu}

\address[NS]{Department of Aerospace Engineering, Auburn University, Auburn, AL 36849, USA} 


\begin{abstract}
This paper presents a non-dissipative, high-order, moving overset method for curvilinear grids to simulate unsteady compressible flows in complex geometries with moving components. The presented method is based on the weak overset interface treatment for static grids introduced in \cite{sharan2016time,sharan2018time}. This interface treatment differs from the conventional overset interface treatment (where conservative or primitive variables at the interface (or fringe) points are overwritten with the interpolated values) by employing a characteristic decomposition and imposing only the incoming characteristic variables for inviscid flows (consistent with the hyperbolic character of the Euler equations). For viscous flows, the viscous fluxes are imposed in addition to the incoming characteristic variables. This study extends and evaluates the robustness of weak interface treatment for moving overset grids. Moreover, instead of using multiple layers of fringe points at the interface, as is typical in conventional overset methods, high-order accuracy and stability are demonstrated with a single layer, thus reducing the parallel communication costs at each timestep. Non-dissipative schemes that are up to sixth-order accurate in the interior are used without introducing artificial dissipation explicitly (using numerical filters) or implicitly (using upwind schemes). Stability is assessed through long-time simulations without solution filtering and, for a one-way-coupled scalar model problem under periodic grid motion, through the fully discrete propagator and its Floquet multipliers. The proposed approach is validated over a range of canonical and practical unsteady flow problems involving moving grids. Grid-convergence studies for inviscid problems (1-D scalar advection and 2-D vortex convection through a rotating overset grid) confirm the expected $p+1$ global order of accuracy, and the viscous cases show agreement with reference force histories and long-time stability. 

\end{abstract}


\end{frontmatter}

\section{Introduction\label{sec:Introduction}}
Overset grids offer a simple and cost-effective approach to simulate fluid flows in complex configurations involving moving components. They enable independently generated body-fitted component grids to move through a static background grid, providing opportunities for selective mesh refinement around solid (moving) bodies without grid distortion. 
The first application of moving overset grids was reported in \cite{DoughertyBenekSteger1985} for an inviscid, two-dimensional store separation simulation, where a store grid moved relative to a background grid and the overset connectivity was updated 
via hole cutting, donor identification, and inter-grid interpolation \cite{doi:10.2514/6.1989-637}. 
A more general moving overset framework for viscous, three-dimensional simulations of multiple aerodynamic bodies in relative motion was presented by Meakin and Suhs \cite{doi:10.2514/6.1989-1996}, who 
applied it to perform tiltrotor simulations \cite{Meakin1993}, among other flow problems \cite{meakin1996chimera,meakin1999unsteady}. These applications demonstrated the feasibility of dynamically updating overset connectivity during grid motion. 

Subsequent efforts broadened the application of moving overset methods to flow-feature tracking with moving (overset) tracker grids. For example, \cite{Chawla1993} used locally refined tracker grids to follow shocks and vortical structures, 
Good accuracy can be maintained if overset connectivity remains robust under grid motion. As simulation scales became larger and more complex, studies increasingly focused on developing efficient and scalable methods to handle overset connectivity in parallel computations \cite{Prewitt2000}. These developments have led to solver-independent overset assemblers, such as PEGASUS \cite{rogers2003pegasus} and SUGGAR \cite{Noack2005Suggar}, which formalized moving overset connectivity as a reusable, automated capability, with later works improving parallel performance and data management strategies \cite{Chan2009,roget2014robust}.

Addressing overset connectivity and parallel scalability represents only part of the challenge of deploying overset methods; robust moving overset grid simulations of practical turbulent flows also require high-order accurate and time-stable discretizations. The application of high-order methods \cite{Canuto1987,Lele1992,Tam1993,Hirsch1975,Kim1987} has long been recognized to alleviate resolution and computational cost limitations associated with low-order schemes in applications involving turbulence, aeroacoustics, and unsteady aerodynamics. However, the use of non-dissipative high-order methods in non-canonical/practical domains 
is often limited by the lack of accurate and stable boundary and interface procedures suitable for general curvilinear and moving/deforming meshes \cite{carpenter1993stability,sharan2018time}. To be effective, overset interface treatments must preserve the global order of accuracy and stability properties of the underlying scheme; otherwise, interface-related errors can become the primary limitation on solution quality. Existing high-order approaches span finite-difference/volume, discontinuous Galerkin, and flux-reconstruction schemes. These approaches differ in the way the interface conditions are enforced, \textit{i.e.}, strongly or weakly, as further discussed in Sections \ref{subsec:strong_treatment} and \ref{subsec:weak_treatment}. They also vary in the strategy used to ensure numerical stability, including artificial dissipation, upwind fluxes, dynamic filtering, or slope/flux limiting. In addition, different strategies (e.g., multiple fringe layers \cite{chan2002overgrid}, supermesh projection \cite{crabill2016high}, etc.) are often adopted to preserve high-order accuracy near overset interfaces.


Alabi and Ladeinde \cite{Alabi2007} formulated a high-order dynamic overset procedure using a strong (Dirichlet) interface treatment in a finite-difference discretization, where the 4th and 6th order compact differencing was used in the interior for low subsonic flows and WENO/MUSCL schemes at high Mach numbers when strong flow gradients were present. To preserve the order of accuracy of the interior scheme, multiple layers of overset interface (fringe) points were required, with high-order discretizations necessitating two or more layers. With careful choice of boundary closure and moving overset interface treatment, we show in this study that high-order accuracy can be ensured with a single layer of fringe points. Moreover, as discussed in Section \ref{subsec:2D_Isentropic_Vortex}, the strong imposition/overwriting of solutions at the interface points can lead to numerical instabilities at the moving overset interface with non-dissipative interior schemes. To address the issue, this study investigates a weak interface treatment with characteristic decomposition.


Brazell et al. \cite{Brazell2016JCP} developed a moving overset mesh approach to solve problems with bodies in relative motion using high-order Discontinuous Galerkin (DG) methods on hybrid unstructured meshes. 
They evaluated the advective terms of the compressible flow equations using a Riemann solver with Lax-Friedrichs fluxes and 
the overset interface condition was enforced strongly by inserting the interpolated data into the boundary flux integrals. 
Duan and Wang \cite{DuanWang2020} developed a high-order moving-overset method using a flux reconstruction (FR) method, where elementwise polynomial solutions were advanced using common Riemann fluxes evaluated at flux points. 
They utilized a hybrid approach employing structured Cartesian background grids and unstructured near-body grids, which allows the solver to handle complex moving geometries while efficiently propagating wake features.
The interpolated data at the overset interface were applied weakly by imposing a Riemann flux at the boundary flux points of the receiver elements. 
The common Riemann fluxes (e.g., Roe, HLLC, Rusanov) used in the FR method introduce upwinding, as they rely on directional wave propagation and apply numerical dissipation based on characteristic speeds. The numerical dissipation is essential for stability in this framework, providing the necessary upwind bias to prevent oscillations at overset interfaces.

Applications such as bio-inspired flapping flight \cite{engels2015numerical}, wind turbine aerodynamics \cite{zahle2008overset}, and rotorcraft flows \cite{strawn1999rotorcraft} involve complex unsteady flows dominated by vortical structures whose accurate resolution requires computational methods that introduce minimal numerical dissipation over long integration times. Unsteady flow computations employing overset grids, as discussed above, often require numerical dissipation or spatial filtering to ensure long-time stability; these ad hoc treatments can nonphysically alter critical flow features and dampen vortical structures/shear turbulence in the wake. It is a well-recognized limitation of overset grid methods that they are prone to numerical instabilities when combined with low-dissipative, high-order methods, since it cannot be guaranteed that interpolation at overlapping interfaces will not increase the overall energy of the solution. Efforts to develop stable high-order overset methods for large-eddy simulation (LES) of turbomachinery (e.g., \cite{marty2015and,de2018numerical}) have similarly found that minimizing artificial dissipation and dispersion of turbulent structures at moving interfaces is a central challenge, with accuracy at overset interfaces limited by the interpolation scheme employed. High-order centered finite-difference methods are particularly well-suited for turbulence and aeroacoustic computations due to their low dissipation and dispersion errors, but their combination with moving overset interfaces, without any stabilizing upwinding or filtering, remains an open problem. 


\paragraph{Relation to previous work and contributions}
The present method builds on the static overset method of \cite{sharan2016time,sharan2018time}). The following
elements are taken from that work and are not claimed as new here:
diagonal-norm summation-by-parts finite-difference operators up to the 3--6--3 operator;
weak imposition of Lagrange-interpolated donor data at the receiver
boundary points; the characteristic decomposition of the inviscid flux
Jacobian, so that only incoming characteristics are penalized at the
interface; and a comparison of the weak treatment with injection on static
grids. Since the weak imposition act only on the receiver boundary points, the
use of a single receiver (fringe) layer also follows from that framework.
For static grids, \cite{sharan2018time} proved time stability for
one-dimensional overlaps of arbitrary size by the energy method, with a
stability-derived dissipative operator added at the interpolation donor
points, and used that operator in its multidimensional computations.

The contributions of the present work concern the moving-grid setting:
\begin{enumerate}
\item The weak characteristic interface treatment is formulated for moving
curvilinear grids. The grid velocity enters the contravariant velocities
and the energy flux through the time metrics, and the characteristic split
is applied to the moving-grid flux Jacobians. Which characteristics are
incoming at a receiver point is therefore decided by wave speeds relative to
the moving interface. For viscous flows, interpolated Cartesian viscous
fluxes are transformed with the receiver-grid metrics at the current grid
position.
\item Donor cells and interpolation weights are recomputed as the grids
move, through a curvilinear inverse mapping (Section~\ref{sec:Interface
treatment in overset methods}). The resulting semi-discrete system is
therefore non-autonomous.
\item In contrast to \cite{sharan2018time}, no dissipative operator is added
at the donor points. The only interface terms are receiver-side weak interface terms,
and the interior scheme on the donor grid is left unmodified.
\item Because the system matrix varies in time, the static stability
arguments of \cite{sharan2018time} do not apply. Stability is instead
assessed with the fully discrete propagator and its Floquet multipliers for
a one-dimensional model problem with one-way coupling under periodic grid
motion. This analysis does not establish stability of the compressible-flow
treatment, which is assessed through long-time simulations.
\item The comparison between injection and the weak treatment is extended
to moving grids, including a viscous three-dimensional plunging finite wing
with curved, non-aligned overset interfaces, and the parallel cost of one
versus multiple fringe layers is quantified.
\end{enumerate}
No new stabilizing term is introduced for the moving case. Instead, the
receiver-side characteristic treatment, applied relative to the moving
grid, gave long-time stable solutions in all cases considered, with
centered interior schemes up to sixth order and without donor-point
dissipation, filtering or upwinding.

The practical advantages of the weak characteristic
treatment are assessed through controlled comparisons
with the strong treatment under otherwise identical
numerical settings. These comparisons demonstrate
improved robustness across problems of differing
physical complexity, from two-dimensional inviscid
vortex convection to highly unsteady three-dimensional
viscous flow with pronounced spanwise variations,
without numerical filtering. Furthermore, the
fringe-layer study quantifies the reductions in
overset data volume, Message Passing Interface (MPI) exchange time, and total
solver time achieved by the single-layer configuration
relative to the tested multilayer configurations. Together, these
comparisons establish the robustness and computational
benefits of the proposed treatment for the moving-grid
problems considered.

The paper is organized as follows. Section~\ref{sec:Interface treatment in overset methods} describes the curvilinear mapping, the interpolation procedure, and the strong/weak interface treatments. Section~\ref{sec:Numerical discretization using weak interface treatment} presents the detailed derivation of the numerical discretization based on the weak overset interface treatment with characteristic decomposition for moving curvilinear meshes. The stability of the proposed discretization is assessed numerically for a one-dimensional scalar advection problem under periodic grid motion, using the norm of the fully discrete propagator and its Floquet multipliers, supported by frozen-coefficient eigenvalue spectra of the semi-discrete system matrix. 
Section~\ref{sec:Numerical-results} discusses the results for a wide range of 2-D and 3-D inviscid and viscous flow problems involving moving bodies/grids. Section~\ref{sec:fringe_communication} examines the effect of fringe-layer count on communication volume and runtime. Section~\ref{sec:Conclusions} provides the conclusions of this study.
\vspace{0.0cm}

\section{Interface Treatment in Overset Methods \label{sec:Interface treatment in overset methods}}

\begin{figure}
\begin{centering}
\includegraphics[width=0.48\textwidth]{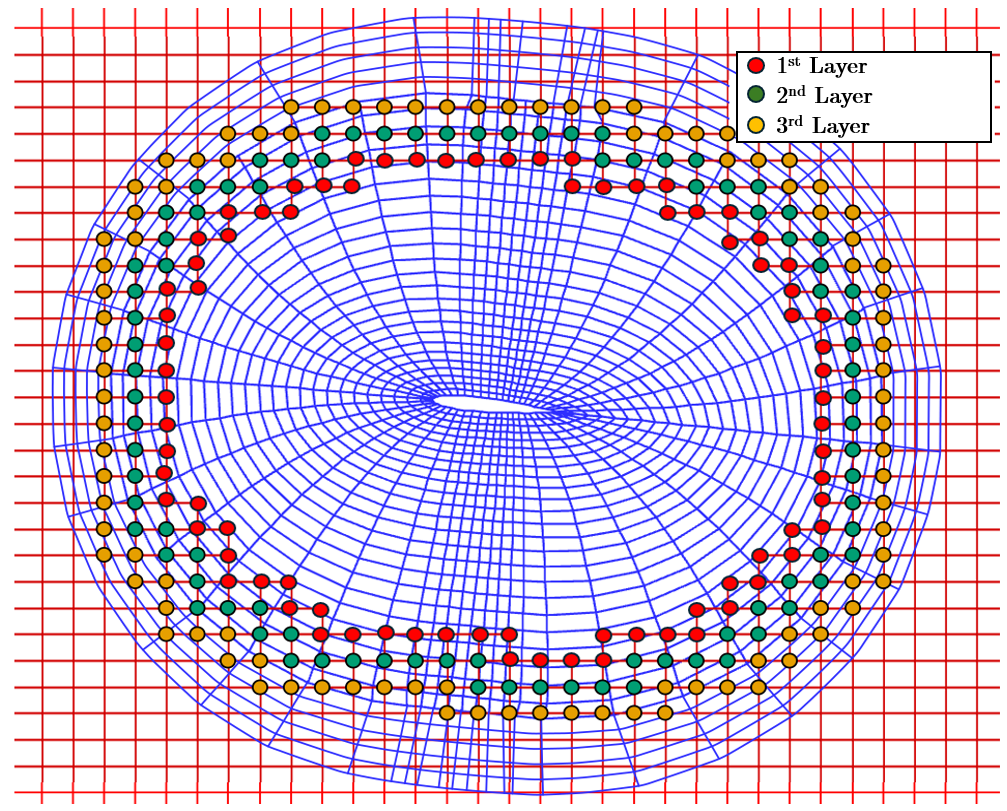}\includegraphics[width=0.52\textwidth]{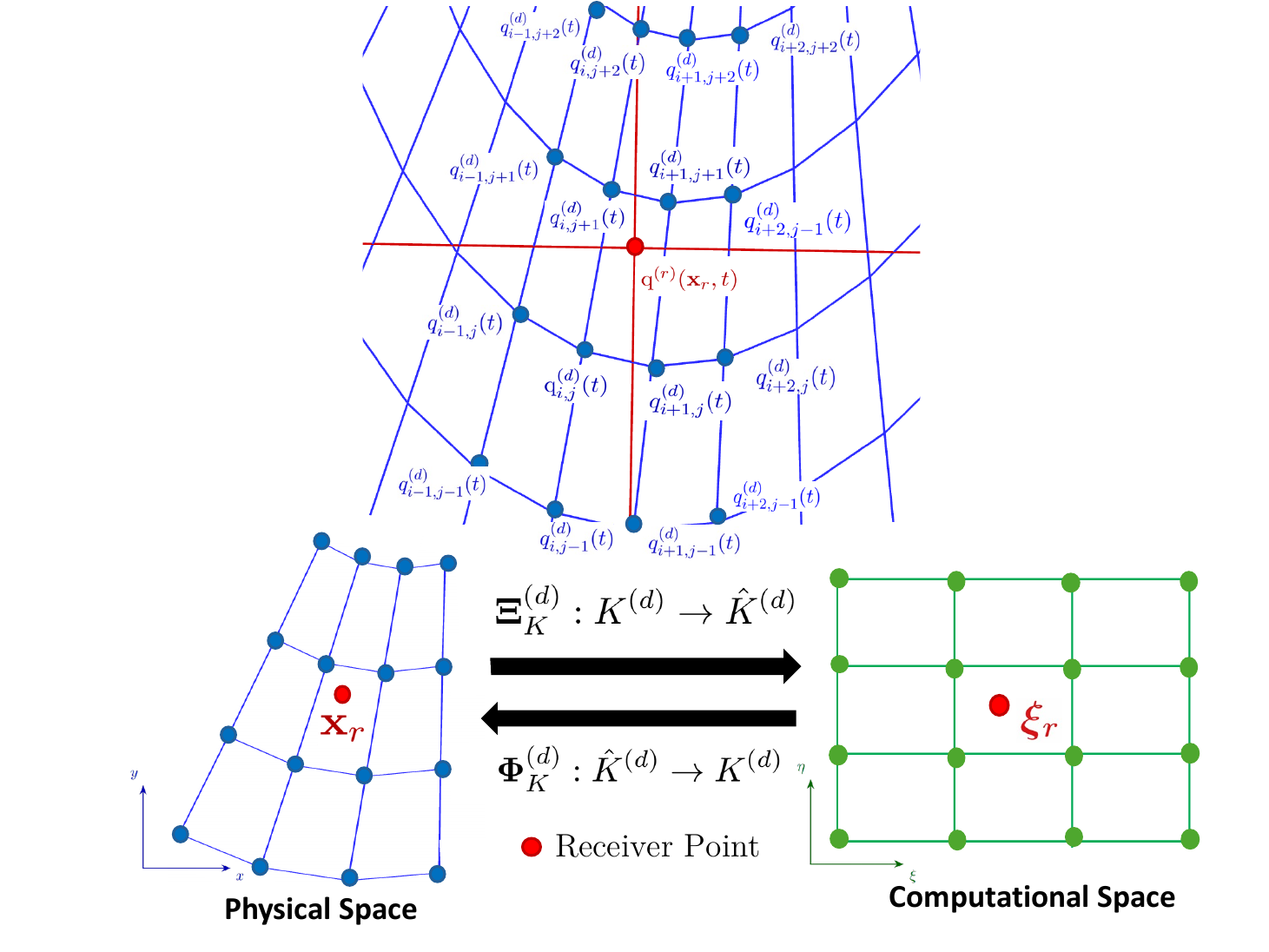}
\par\end{centering}
\begin{centering}
(a)\qquad{}\qquad{}\qquad{}\qquad{}\qquad{}\qquad{}\qquad{}\qquad{}\qquad{}\qquad{}\qquad{}\qquad{}(b)
\par\end{centering}
    
    
    \caption{Overset-grid interpolation framework. (a) Global view of the curvilinear overset grids for flow simulations over an airfoil; the body-conforming (O-grid) is shown in blue, whereas the background Cartesian grid in red. The fringe (receiver) points on the background grid are color-coded by layer: the first layer is marked in red, the second in green, and the third in orange, with each additional layer retaining all preceding layers. (b) An overset interface showing the donor (blue) and receiver (red) grids and the curvilinear mapping used for overset interpolation.}
    \label{fig:interpolation_1}
\end{figure}

To discuss the interface treatment for an overset grid configuration, consider a receiver (also called the fringe or interface) point located at the physical coordinate $\mathbf x_r \in \mathbb{R}^{\mathcal{N}}$, where $\mathcal{N}=1,2,3$ denotes the number of spatial dimensions, as shown for a $\mathcal{N}=2$ (two-dimensional) overset interface in Fig.~\ref{fig:interpolation_1}. Let $\mathbf q^{(r)}(\mathbf x_r,t)$ denote the 
numerical solution at the receiver point and $\hat{\mathbf q}(\mathbf x_r,t)$ be the interpolated solution evaluated at $\mathbf x_r$ from the donor grid. 
The evaluation of the interpolated solution at $\mathbf x_r$ is carried out in two steps. First, the donor grid interpolation stencil is mapped from the physical space to a reference (computational) space, as shown in Fig.~\ref{fig:interpolation_1}(b), to obtain the computational coordinates, $\bm{\xi}_r$, of the receiver point in the donor stencil. Then, the interpolated solution, $\hat{\mathbf q}(\mathbf x_r,t)$, is evaluated using Lagrange interpolation on the donor stencil in the  computational coordinate system. The physical to computational space mapping with details of the interpolation procedure is discussed below.

Let $K^{(d)}$ denote the donor grid interpolation stencil (in the physical space) 
for the receiver point $\mathbf x_r$, shown as blue nodes in Fig.~\ref{fig:interpolation_1}(b). The interpolation stencil is mapped to the computational space through a smooth curvilinear mapping
\begin{equation}
\bm{\Xi}^{(d)}_{K} : K^{(d)} \rightarrow \hat{K}^{(d)} ,
\qquad
\bm{\xi} = \bm{\Xi}^{(d)}_{K}(\mathbf x),
\label{eq:cell_mapping-1}
\end{equation}
with its inverse
\begin{equation}
\bm{\Phi}^{(d)}_{K} : \hat{K}^{(d)} \rightarrow K^{(d)} ,
\qquad
\mathbf x = \bm{\Phi}^{(d)}_{K}(\bm{\xi}).
\label{eq:cell_mapping}
\end{equation}
For general curvilinear grids, the mapping \(\bm{\Phi}^{(d)}_{K}\) is constructed from the coordinates of the donor points using a Lagrange interpolant over donor cells in computational space, given by
\begin{equation}
\bm{\Phi}^{(d)}_{K}(\bm{\xi})
=
\sum_{a=1}^{N_s} \mathbf x_a \, L_a(\bm{\xi}),
\label{eq:lagrange_mapping}
\end{equation}
where \(\mathbf x_a=(x_a,y_a,z_a)\in K^{(d)}\) are the physical coordinates of the donor points (\textit{i.e.}, the nodes in the interpolation stencil), \(L_a(\bm{\xi})\) are the Lagrange basis functions, and \(N_s\) is the number of donor points. Here, \(\bm{\xi}=(\xi,\eta,\zeta)\in\hat{K}^{(d)}\) denotes the computational coordinates local to the mapped donor stencil, shown in green in Fig.~\ref{fig:interpolation_1}(b) for a two-dimensional configuration, and \(\bm{\Xi}^{(d)}_{K}=\left[\bm{\Phi}^{(d)}_{K}\right]^{-1}\). The Lagrange basis functions, \(L_a(\bm{\xi})\), are given by
\begin{equation}
L_a(\bm{\xi})
\equiv
L_{\alpha\beta\gamma}(\xi,\eta,\zeta)
=
\ell_\alpha(\xi)\,
\ell_\beta(\eta)\,
\ell_\gamma(\zeta),
\label{eq:geom_basis_tensor}
\end{equation}
where \(a=(\alpha,\beta,\gamma)\) denotes the donor point indices. The one-dimensional Lagrange polynomials in (\ref{eq:geom_basis_tensor}) are
\begin{equation}
\ell_\alpha(\xi)=\prod_{\substack{m\in\mathcal I\\ m\neq \alpha}}
\frac{\xi-\xi_m}{\xi_\alpha-\xi_m},\qquad
\ell_\beta(\eta)=\prod_{\substack{n\in\mathcal J\\ n\neq \beta}}
\frac{\eta-\eta_n}{\eta_\beta-\eta_n},\qquad
\ell_\gamma(\zeta)=\prod_{\substack{p\in\mathcal K\\ p\neq \gamma}}
\frac{\zeta-\zeta_p}{\zeta_\gamma-\zeta_p},
\label{eq:geom_basis_1d}
\end{equation}
with $\{\xi_m\}_{m\in\mathcal I}$, $\{\eta_n\}_{n\in\mathcal J}$, and $\{\zeta_p\}_{p\in\mathcal K}$ denoting the donor point locations in $\hat{K}^{(d)}$ along the $\xi$, $\eta$, and $\zeta$ directions, respectively. The stencil index sets $\mathcal I$, $\mathcal J$, and $\mathcal K$ are chosen according to the interpolation order. For Lagrange linear interpolation in each direction,
\begin{equation}
\mathcal I=\{0,1\},\qquad
\mathcal J=\{0,1\},\qquad
\mathcal K=\{0,1\},
\label{eq:linear_interp_indices}
\end{equation}
whereas for cubic interpolation in each direction,
\begin{equation}
\mathcal I=\{0,1,2,3\},\qquad
\mathcal J=\{0,1,2,3\},\qquad
\mathcal K=\{0,1,2,3\}.
\label{eq:cubic_interp_indices}
\end{equation}

\noindent We assume that the mapping \(\bm{\Phi}^{(d)}_{K}\) is smooth, one-to-one, and orientation-preserving, with a nonsingular Jacobian matrix given by
\begin{equation}
\mathcal{J} (\bm{\xi})
=
\frac{\partial \bm{\Phi}^{(d)}_{K}}{\partial \bm{\xi}}
=
\begin{bmatrix}
\partial_\xi x & \partial_\eta x & \partial_\zeta x \\
\partial_\xi y & \partial_\eta y & \partial_\zeta y \\
\partial_\xi z & \partial_\eta z & \partial_\zeta z
\end{bmatrix},
\label{eq:mapping_jacobian}
\end{equation}
such that \(\det(\mathcal J(\boldsymbol\xi))>0\). Accordingly, the entries of the Jacobian matrix are evaluated by differentiating (\ref{eq:lagrange_mapping}), given by
\begin{equation}
\partial_\xi \bm{\Phi}^{(d)}_{K}(\bm{\xi})
=
\sum_{a=1}^{N_s} \mathbf x_a \, \partial_\xi L_a(\bm{\xi}),
\label{eq:Jacobian_derv_calculation}
\end{equation}
and similarly for $\partial_\eta \bm{\Phi}^{(d)}_{K}$ and $\partial_\zeta \bm{\Phi}^{(d)}_{K}$. 

The computational coordinates corresponding to the receiver location, $\mathbf x_r$, is then obtained by solving the nonlinear inverse problem
\begin{equation}
\bm{\Phi}^{(d)}_{K}(\bm{\xi}_r) = \mathbf x_r ,
\qquad
\bm{\xi}_r = (\xi_r,\eta_r,\zeta_r).
\label{eq:inverse_mapping}
\end{equation}

\noindent For general curvilinear grids, \eqref{eq:inverse_mapping} does not have a 
closed-form solution and is solved iteratively using the Newton's method. 
Given an initial guess $\bm{\xi}^{(0)}$, successive iterations are computed using
\begin{equation}
\bm{\xi}^{(h+1)}
=
\bm{\xi}^{(h)}
-
\mathcal{J}^{-1}\!\left(\bm{\xi}^{(h)}\right)
\left(
\bm{\Phi}^{(d)}_{K}\!\left(\bm{\xi}^{(h)}\right)
-
\mathbf x_r
\right),
\label{eq:newton_iteration}
\end{equation}
where the inverse of the Jacobian is calculated by inverting (\ref{eq:mapping_jacobian}) and using (\ref{eq:Jacobian_derv_calculation}) for derivative calculations. The iteration is terminated once the physical residual satisfies
\begin{equation}
\left\|
\bm{\Phi}^{(d)}_{K}(\bm{\xi}^{(h)}) - \mathbf x_r
\right\|_2 < \varepsilon,
\end{equation}
and $\bm{\xi}^{(h)} \in \hat{K}$, where $\varepsilon = 10^{-10}$ is chosen. The converged solution 
$\bm{\xi}_r = (\xi_r,\eta_r,\zeta_r)$ uniquely identifies the location of the receiver point in the computational space, $\hat{K}^{(d)}$.

With the computational coordinates $\bm{\xi}_r=(\xi_r,\eta_r,\zeta_r)$ determined, the donor solution is evaluated using Lagrange interpolation, similar to (\ref{eq:lagrange_mapping}), given by
\begin{equation}
\hat{\mathbf q}(\mathbf x_r,t)
=
\sum_{a=1}^{N_s} \mathbf q^{(d)}_{a} \, L_a(\bm{\xi}_r),
\label{eq:lagrange_interp_tensor}
\end{equation}
where $\mathbf q^{(d)}_{a} \equiv \mathbf q^{(d)}(\mathbf x_a,t)$ is the vector of conserved variables at the donor node indexed by \(a=(\alpha,\beta,\gamma)\) and the Lagrange basis functions \(L_a(\bm{\xi}_r)\) are given by (\ref{eq:geom_basis_tensor}) and (\ref{eq:geom_basis_1d}). The interpolated solution, $\hat{\mathbf q}(\mathbf x_r,t)$, obtained from (\ref{eq:lagrange_interp_tensor}) is then applied at the interface (fringe) points strongly or weakly, as discussed below.

\subsection{Strong Overset Interface Treatment}\label{subsec:strong_treatment}
As illustrated in Fig.~\ref{fig:interpolation_1}, let $\mathbf{x}_r$ denote a receiver (fringe/interface) point on an overset interface. The receiver-grid numerical solution at this point is $\mathbf q^{(r)}(\mathbf{x}_r,t)$, and the overlapping donor grid provides an interpolated solution $\hat{\mathbf q}(\mathbf{x}_r,t)$. 
In the strong interface treatment (\textit{e.g.}, \cite{Alabi2007,Brazell2016JCP,KOOMULLIL2008618}, the interpolated solution is injected at the receiver point, 
\begin{equation}
\mathbf q^{(r)}(\mathbf{x}_r,t) \;\leftarrow\; \hat{\mathbf q}(\mathbf{x}_r,t),
\label{eq:strong_injection}
\end{equation}
overwriting the solution at each time step (or substep). Although simple to implement, overwriting all conservative variables at each fringe/interface point ignores the hyperbolic nature of the inviscid terms in the flow equations that specify the direction of propagation of the characteristic waves. Hence, overwriting all conservative variables often leads to numerical instabilities that originate at the overset interface and reflect back and forth between the domain boundaries and overset interfaces, amplifying over time and eventually destabilizing the entire simulation \cite{sharan2018time}. This requires the introduction of numerical filtering or artificial dissipation for long-time stability. Moreover, the (ad hoc) overwriting at the end of each time step also complicates the theoretical stability analysis because this interface treatment cannot be easily incorporated in the system of ordinary differential equations resulting from the numerical semi-discretization. 

\vspace{0.1cm}
\subsection{Weak Overset Interface Treatment}\label{subsec:weak_treatment}
In the weak interface treatment, the interpolated solution is imposed using a penalty term added to the discretization of the governing equations. Unlike strong overset injection, which replaces the receiver grid point solution with interpolated data from the donor grid, the weak approach computes the solution at the interface/fringe points. Let $\mathbf Q$ denote the vector of numerical solution at all points on the receiver grid  and $\hat{\mathbf Q}$ denote a vector containing donor interpolated data at the (overset interface) receiver points. A general form of semi-discretization for governing equations on the receiver grid can then be written as
\begin{equation}
\frac{d\mathbf Q}{dt}
=
\mathcal{G}(\mathbf Q)
-
\mathcal{P}
\big(
\mathbf Q - \hat{\mathbf Q}
\big),
\end{equation}
where $\mathcal{G}(\mathbf Q)$ denotes a discretization of the spatial derivative terms in the governing equations and the operator $\mathcal{P}$ ensures that the penalty term is applied only to the receiver grid points at the overset interface. $\mathcal{P}$ can also incorporate the eigendecomposition to impose only the incoming characteristic variables. The explicit forms of $\mathcal{G}$ and
$\mathcal{P}$ for the 1-D scalar advection equation
and the 3-D compressible Navier-Stokes equations
are detailed in Sections \ref{subsec:1-D Scalar Advection}
and \ref{subsec:Compressible Navier-Stokes equations}, respectively.

\section{Moving Overset Grid Discretization Using the Weak Interface Treatment\label{sec:Numerical discretization using weak interface treatment}}
The weak interface treatment is employed in this study because of its flexibility in accommodating a characteristic decomposition and enabling numerical stability analysis, which helps avoid the use of ad hoc numerical dissipation for stabilization. In the following, we discuss the numerical discretization and a numerical stability assessment, based on the fully discrete propagator, for a linear scalar advection problem (Section \ref{subsec:1-D Scalar Advection}) and the extension of the approach to solving the compressible Navier-Stokes equations (Section \ref{subsec:Compressible Navier-Stokes equations}).
\subsection{Moving Overset Interface Treatment for the 1-D Scalar Advection Equation\label{subsec:1-D Scalar Advection}}

\begin{figure}
    \centering
    \includegraphics[width=1.0\textwidth]{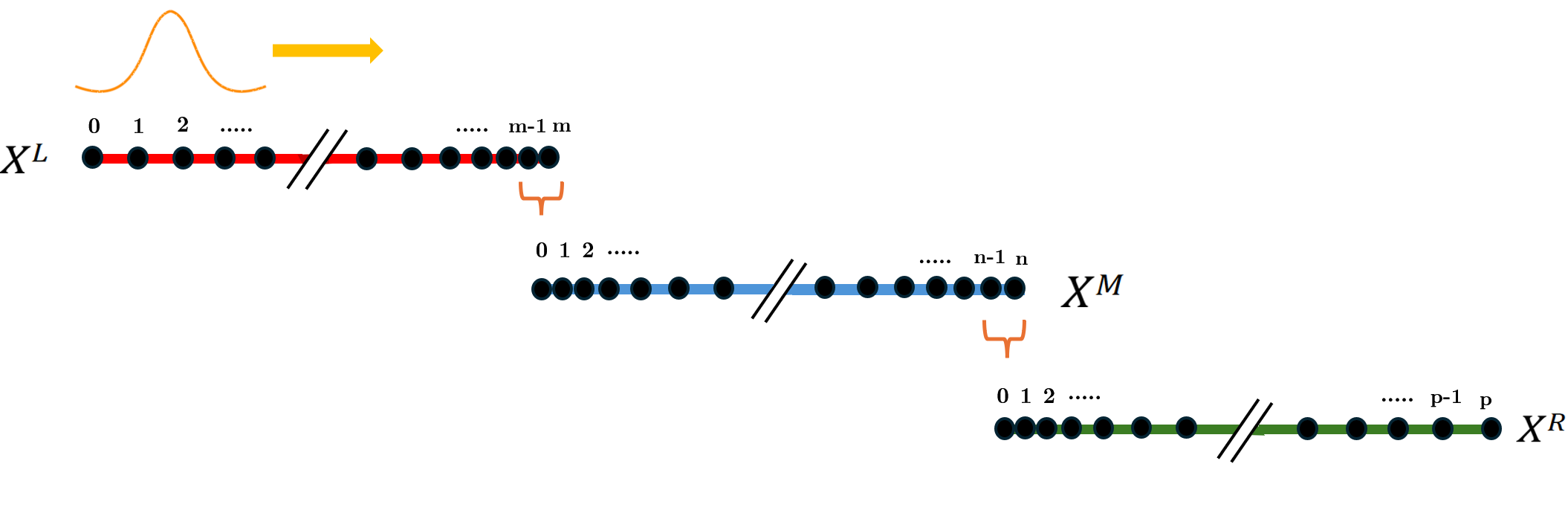}
    \caption{Schematic of the one-dimensional overlapping grids on which Eq.~(\ref{eq:1d_advection}) is solved.}
    \label{fig:1D-overlapping-grid}
\end{figure}

Consider the one-dimensional linear advection equation
\begin{equation}
\frac{\partial u}{\partial t}
+
c \frac{\partial u}{\partial x}
=
0,
\qquad -1 \le x \le 1,\qquad t \ge 0,
\label{eq:1d_advection}
\end{equation}
on three overlapping (overset) grids: a left grid
\(
-1=a_L \le x \le b_L
\),
a middle grid
\(
a_M(t)\le x \le b_M(t)
\),
and a right grid
\(
a_R \le x \le b_R=1
\), as shown in Fig.~\ref{fig:1D-overlapping-grid}.
The left and right grids are static, whereas the middle grid moves according to
\begin{equation}
\frac{d x_M}{dt}=\beta(t),
\label{eq:1d_advection_motion_eq}
\end{equation}
such that its left and right boundaries remain within the neighboring grids. 
\noindent Assuming \(c>0\), the initial and boundary conditions are, respectively,
\begin{equation}
u(x,0)=f(x),\qquad u(-1,t)=g(t).
\label{eq:1d_adv_icbc}
\end{equation}

\noindent To accommodate moving and spatially stretched grids, we introduce the
general curvilinear coordinate transformation
\begin{equation}
\tau=t,
\qquad
\xi=\xi(x,t).
\label{eq:1d_curv_transform}
\end{equation}
Applying the chain rule, the Cartesian derivatives can be expressed in
terms of the curvilinear derivatives using
\begin{equation}
\frac{\partial}{\partial t}
=
\frac{\partial}{\partial \tau}
+
\xi_t\frac{\partial}{\partial \xi},
\qquad
\frac{\partial}{\partial x}
=
\xi_x\frac{\partial}{\partial \xi},
\label{eq:1d_chain_rule}
\end{equation}
where the shorthand notation \(\xi_t=\partial \xi/\partial t\) and \(\xi_x=\partial \xi/\partial x\) is used. Similarly, the curvilinear derivatives can be expressed in terms of the Cartesian derivatives using
\begin{equation}
\frac{\partial}{\partial \tau}
=
\frac{\partial}{\partial t}
+
x_{\tau}\frac{\partial}{\partial x},
\qquad
\frac{\partial}{\partial \xi}
=
x_{\xi}\frac{\partial}{\partial x}.
\label{eq:1d_chain_rule-1}
\end{equation}
Equations (\ref{eq:1d_chain_rule}) and (\ref{eq:1d_chain_rule-1}) can be written as, respectively,
\begin{equation}
\begin{bmatrix}\partial/\partial t\\
\partial/\partial x
\end{bmatrix}=\begin{bmatrix}1 & \xi_{t}\\
0 & \xi_{x}
\end{bmatrix}\begin{bmatrix}\partial/\partial\tau\\
\partial/\partial\xi
\end{bmatrix} \qquad \mathrm{and} \qquad 
\begin{bmatrix}\partial/\partial\tau\\
\partial/\partial\xi
\end{bmatrix}=\begin{bmatrix}1 & x_{\tau}\\
0 & x_{\xi}
\end{bmatrix}\begin{bmatrix}\partial/\partial t\\
\partial/\partial x
\end{bmatrix}.\label{eq:derv_transform_mat}
\end{equation}
Inverting the second equation in (\ref{eq:derv_transform_mat}) and comparing it with the first equation provides the metric relations
\begin{equation}
\xi_t=-x_\tau \xi_x,
\qquad
\xi_x=\frac{1}{x_\xi}.
\label{eq:1d_metric_relations}
\end{equation}
Substituting \eqref{eq:1d_chain_rule} into \eqref{eq:1d_advection} provides
\begin{equation}
\frac{\partial u}{\partial \tau}
+
\left(\xi_t+c\xi_x\right)\frac{\partial u}{\partial \xi}
=
0,\label{eq:1d_curv_form-1}
\end{equation}
and using (\ref{eq:1d_metric_relations}) in (\ref{eq:1d_curv_form-1}) yields
\begin{equation}
\frac{\partial u}{\partial \tau}
+
\left(-x_\tau+c\right)\xi_x \frac{\partial u}{\partial \xi}
=
0.
\label{eq:1d_curv_form}
\end{equation}
Here, \(x_\tau\) incorporates the grid motion. Assuming $c>0$ and \(
-x_\tau+c>0
\), a semi-discretization of \eqref{eq:1d_curv_form} with the boundary condition (\ref{eq:1d_adv_icbc}) using a weak boundary/interface treatment on the three overset grids shown in Fig. \ref{fig:1D-overlapping-grid} can be written as
\begin{equation}
\frac{d\mathbf{u}}{d\tau}
=
-\mathcal{X}_L D_L \mathbf{u}
-\tau_L \mathcal{X}_L H_L^{-1}\mathbf{e}_0^L (u_0-g),
\label{eq:left_adv_match}
\end{equation}
\begin{equation}
\frac{d\mathbf{v}}{d\tau}
=
-\mathcal{X}_M(\tau) D_M \mathbf{v}
-\tau_M \mathcal{X}_M(\tau) H_M^{-1}\mathbf{e}_0^M
\left(v_0-\boldsymbol{I}_L(\tau)^T\mathbf{u}\right),
\label{eq:middle_adv_match}
\end{equation}
\begin{equation}
\frac{d\mathbf{w}}{d\tau}
=
-\mathcal{X}_R D_R \mathbf{w}
-\tau_R \mathcal{X}_R H_R^{-1}\mathbf{e}_0^R
\left(w_0-\boldsymbol{I}_M(\tau)^T\mathbf{v}\right),
\label{eq:right_adv_match}
\end{equation}
where
\[
\mathbf{u}=
\begin{bmatrix}
u_0 & \cdots & u_m
\end{bmatrix}^T,\qquad
\mathbf{v}=
\begin{bmatrix}
v_0 & \cdots & v_n
\end{bmatrix}^T,\qquad
\mathbf{w}=
\begin{bmatrix}
w_0 & \cdots & w_p
\end{bmatrix}^T
\]
are the solution vectors on the left, middle, and right grids, respectively. In \eqref{eq:left_adv_match}--\eqref{eq:right_adv_match}, $\mathcal{X}_{\kappa}$ ($\kappa=L,M,R$) are diagonal matrices whose
$i^{\textrm{}}$-th diagonal entry is an approximation to $\left|-x_{\tau}+c\right|\xi_{x}$
at the $i^{\textrm{}}$-th grid point of the respective grids. The
left and right grids are static, therefore, $x_{\tau}$ and $\xi_{t}$
is non-zero only for the middle grid. $D_{\kappa}$ denote summation-by-parts
first-derivative approximations of the form $D_{\kappa}=H_{\kappa}^{-1}Q_{\kappa}$ \cite[Appendix B.1]{gustafsson2007high},
where $H_{\kappa}$ is a diagonal, symmetric positive definite (s.p.d.)
norm matrix and 
\begin{equation}
Q_\kappa + Q_\kappa^T = \mathrm{diag}(-1,0,\cdots,0,1).
\label{eq:Q_property}
\end{equation}
The penalty parameter $\tau_{\kappa} \geq 1/2$, the vector $\mathbf{e}_{0}^{\kappa}=\begin{bmatrix}1 & 0 & \cdots & 0\end{bmatrix}^{T}$ selects the left boundary node,
and $\boldsymbol{I}_{L}(\tau)$ contains the interpolation weights for the solution
interpolation from the left grid to the first node of the middle
grid. Similarly, $\boldsymbol{I}_{M}(\tau)$ enables interpolation from
the middle grid to the first node of the right grid.
The discretization \eqref{eq:middle_adv_match} for the middle grid assumes
$c-x_\tau>0$; therefore, the interpolated data is
weakly imposed at the left boundary. For $c<0$ and
$c-x_\tau<0$, the boundary data would instead be
imposed at the right boundary. Thus, the location
of the imposed data follows the direction of
characteristic propagation relative to the grid.

The scheme \eqref{eq:left_adv_match}--\eqref{eq:right_adv_match} can be collectively written as
\begin{equation}
\frac{d\mathbf{\boldsymbol{\Phi}}}{d\tau}=M(\tau)\mathbf{\boldsymbol{\Phi}}+\mathbf{b}(\tau),\qquad\mathbf{\Phi}=\begin{bmatrix}\mathbf{u}\\
\mathbf{v}\\
\mathbf{w}
\end{bmatrix}.\label{eq:semi_disc}
\end{equation}

\noindent where the matrix $M$, called the system or discretization matrix, is given by
\begin{equation}
M(\tau)=\begin{bmatrix}-\mathcal{X}_{L}H_{L}^{-1}Q_{L}-\tau_{L}\mathcal{X}_{L}H_{L}^{-1}E_{0}^{L} & 0 & 0\\
\tau_{M}\mathcal{X}_{M}(\tau)H_{M}^{-1}\mathbf{e}_{0}^{M}\boldsymbol{I}_{L}(\tau)^{T} & -\mathcal{X}_{M}(\tau)H_{M}^{-1}Q_{M}-\tau_{M}\mathcal{X}_{M}(\tau)H_{M}^{-1}E_{0}^{M} & 0\\
0 & \tau_{R}\mathcal{X}_{R}H_{R}^{-1}\mathbf{e}_{0}^{R}\boldsymbol{I}_{M}(\tau)^{T} & -\mathcal{X}_{R}H_{R}^{-1}Q_{R}-\tau_{R}\mathcal{X}_{R}H_{R}^{-1}E_{0}^{R}
\end{bmatrix},\label{eq:M_matrix}
\end{equation}

\noindent and

\begin{equation}
E_{0}^{\kappa}=\mathbf{e}_{0}^{\kappa}(\mathbf{e}_{0}^{\kappa})^{T},\qquad\textrm{}\qquad\mathbf{b}(\tau)=\begin{bmatrix}\tau_{L}\mathcal{X}_{L}H_{L}^{-1}\mathbf{e}_{0}^{L}g(\tau)\\
\mathbf{0}\\
\mathbf{0}
\end{bmatrix}.\label{eq:E0_mat}
\end{equation}



\noindent \textit{Relation to the static scheme of \cite{sharan2018time}.} For static grids, the scheme \eqref{eq:left_adv_match}--\eqref{eq:right_adv_match}
reduces to Eqs.~(13)--(14) of \cite{sharan2018time} without the dissipative
operator $\Upsilon_L$ that \cite{sharan2018time} adds at the donor points.
In that case, $M$ is constant and block lower triangular, and its
eigenvalues are those of the diagonal blocks, each of which is stable for
$\tau_\kappa\ge 1/2$ \cite[Section~3, Remark]{sharan2018time}. Two features of
the moving-grid problem prevent this argument, and the energy argument of
\cite{sharan2018time}, from being applied directly. First, the coefficients
$\mathcal X_M(\tau)$ and the interpolation vectors $\bm I_L(\tau)$,
$\bm I_M(\tau)$ change in time. Second, the boundary at which data are
imposed is set by the sign of the grid-relative wave speed $c-x_\tau$
rather than of $c$.

Under grid motion, the grid metrics, interpolation weights, and overset connectivity vary in time, so that \eqref{eq:semi_disc} is a non-autonomous system. Negative real parts of the eigenvalues of $M(\tau)$ at each frozen time are not sufficient for stability of such a system, since time-varying eigenvectors and non-normal coupling can produce growth even when every instantaneous spectrum lies in the left half-plane. We therefore assess stability numerically using the fully discrete propagator. With $\mathbf{b}=\mathbf{0}$, one time step of the time integrator can be written as $\mathbf{\Phi}^{n+1}=G_n\mathbf{\Phi}^n$, where $G_n$ depends on $M$ at the stage times, and the solution after $n$ steps is $\mathbf{\Phi}^{n}=P_n\mathbf{\Phi}^0$ with $P_n=G_{n-1}\cdots G_0$. The Euclidean norm $|P_n|_2$, which is not a frozen-time quantity, bounds the amplification of all possible initial perturbations, including transient growth due to non-normality and the time dependence of $M$. For periodic grid motion with period $T=N\Delta t$, the discrete system is periodic, and the eigenvalues of the one-period propagator $P_T=P_N$  are the Floquet multipliers, and $\rho(P_T)<1$ implies that perturbations decay as $t \rightarrow \infty$. Frozen-time eigenvalue spectra of $M(\tau)$ are reported as a supporting diagnostic only. The results are presented in Section~\ref{subsec:1-D Advection Result}.

In the next section, the treatment is extended to the three-dimensional compressible flow equations on moving overset grids.

\subsection{Moving Overset Interface Treatment for the 3-D Compressible Navier-Stokes Equations\label{subsec:Compressible Navier-Stokes equations}}

The compressible Navier-Stokes equations in general curvilinear coordinates are given by \cite{pulliam1981diagonal}
\begin{equation}
\frac{\partial \mathbf{Q}}{\partial \tau}
+ \frac{\partial \mathbf{F}}{\partial \xi}
+ \frac{\partial \mathbf{G}}{\partial \eta}
+ \frac{\partial \mathbf{H}}{\partial \zeta}
= 0,
\label{eq:ns}
\end{equation}
where 
\begin{equation}
\mathbf{Q}
= \frac{\mathbf{q}}{J} = 
\frac{1}{J}
\begin{bmatrix}
\rho \\
\rho u \\
\rho v \\
\rho w \\
E
\end{bmatrix},\qquad
\mathbf{F} = \mathbf{F}_c - \mathbf{F}_v, \qquad
\mathbf{G} = \mathbf{G}_c - \mathbf{G}_v, \qquad
\mathbf{H} = \mathbf{H}_c - \mathbf{H}_v.
\label{eq:q_and_fluxes}
\end{equation}
In \eqref{eq:q_and_fluxes}, $\mathbf{q}$ denotes the physical conservative variables and $\mathbf{Q}$ the transformed variables. $\rho$ denotes the density, $\mathbf{u}=(u_{1},u_{2},u_{3})=(u,v,w)$
are the Cartesian velocity components, 
$E = p/(\gamma-1)+\frac{1}{2}\rho(u^{2}+v^{2}+w^{2})$ is the total energy per unit volume, $p$ denotes the pressure, and $J=\det\left(\partial \bm{\xi}/\partial \mathbf{x}\right)$ is the Jacobian of the coordinate transformation. We assume the time to be invariant, therefore, $\tau=t$. The flow variables are non-dimensionalized by a reference
length scale $L^{*}$, velocity scale $a_{\infty}^{*}$ (ambient speed
of sound), density scale $\rho_{\infty}^{*}$, pressure scale $\rho_{\infty}^{*}a_{\infty}^{*^{2}}$,
temperature scale $a_{\infty}^{*^{2}}/C_{p,\infty}^{*}=(\gamma-1)T_{\infty}^{*}$,
and viscosity $\mu_{\infty}^{*}$. $*$ denotes a dimensional variable,
while $\infty$ denotes an ambient quantity. $\gamma=C_{p}^{*}/C_{v}^{*}$
is the ratio of the specific heat at a constant pressure to the specific
heat at a constant volume. The Reynolds number and the Prandtl number
are defined as $Re=\rho_{\infty}^{*}a_{\infty}^{*}L^{*}/\mu_{\infty}^{*}$
and $Pr=\mu^{*}C_{p}^{*}/k^{*}$, respectively, where $k^{*}$ denotes
the thermal conductivity. $\mathbf{F}_c$, $\mathbf{G}_c$, and $\mathbf{H}_c$ in \eqref{eq:q_and_fluxes} are the inviscid fluxes, given by
\begin{equation}
\mathbf{F}_c=\frac{1}{J}\begin{bmatrix}\rho U\\
\rho uU+p\xi_{x}\\
\rho vU+p\xi_{y}\\
\rho wU+p\xi_{z}\\
(E+p)U-\xi_{t}p
\end{bmatrix},\quad \mathbf{G}_c=\frac{1}{J}\begin{bmatrix}\rho V\\
\rho uV+p\eta_{x}\\
\rho vV+p\eta_{y}\\
\rho wV+p\eta_{z}\\
(E+p)V-\eta_{t}p
\end{bmatrix},\quad \mathbf{H}_c=\frac{1}{J}\begin{bmatrix}\rho W\\
\rho uW+p\zeta_{x}\\
\rho vW+p\zeta_{y}\\
\rho wW+p\zeta_{z}\\
(E+p)W-\zeta_{t}p
\end{bmatrix},
\end{equation}
where the contravariant velocities are
\begin{equation}
U=\xi_t+\xi_x u+\xi_y v+\xi_z w,\quad
V=\eta_t+\eta_x u+\eta_y v+\eta_z w,\quad
W=\zeta_t+\zeta_x u+\zeta_y v+\zeta_z w.
\label{eq:contravariant_vel}
\end{equation}

\noindent The metric time-derivative terms appearing in the contravariant velocities account for grid motion, and are further discussed below. $\mathbf{F}_v$, $\mathbf{G}_v$, and $\mathbf{H}_v$ are the viscous fluxes, given by
\begin{equation}
\mathbf{F}_v=\frac{1}{J}\left(\xi_x \tilde{\mathbf{F}}_v+\xi_y \tilde{\mathbf{G}}_v+\xi_z \tilde{\mathbf{H}}_v\right),\quad
\mathbf{G}_v=\frac{1}{J}\left(\eta_x \tilde{\mathbf{F}}_v+\eta_y \tilde{\mathbf{G}}_v+\eta_z \tilde{\mathbf{H}}_v\right),\quad
\mathbf{H}_v=\frac{1}{J}\left(\zeta_x \tilde{\mathbf{F}}_v+\zeta_y \tilde{\mathbf{G}}_v+\zeta_z \tilde{\mathbf{H}}_v\right),
\label{eq:viscous_fluxes}
\end{equation}
where
\begin{equation}
\tilde{\mathbf{F}}_v=\begin{bmatrix}0\\
\tau_{11}\\
\tau_{12}\\
\tau_{13}\\
u_{j}\tau_{1j}-q_{1}
\end{bmatrix},\quad\tilde{\mathbf{G}}_v=\begin{bmatrix}0\\
\tau_{21}\\
\tau_{22}\\
\tau_{23}\\
u_{j}\tau_{2j}-q_{2}
\end{bmatrix},\quad\tilde{\mathbf{H}}_v=\begin{bmatrix}0\\
\tau_{31}\\
\tau_{32}\\
\tau_{33}\\
u_{j}\tau_{3j}-q_{3}
\end{bmatrix}.\label{eq:cartesian_visc_flux}
\end{equation}
are the Cartesian viscous fluxes, and the stress tensor and the heat flux are given by
\begin{equation}
\tau_{ij}=\frac{\mu}{Re}\left(\frac{\partial u_{i}}{\partial x_{j}}+\frac{\partial u_{j}}{\partial x_{i}}\right)+\frac{\lambda}{Re}\frac{\partial u_{k}}{\partial x_{k}}\delta_{ij},\qquad q_{i}=-\frac{\mu}{Re\,Pr}\frac{\partial T}{\partial x_{i}}.
\label{eq:stress_heatTransport}
\end{equation}
The viscosity is modeled by the Sutherland law and $\lambda=\mu_{B}-\frac{2}{3}\mu$, where $\mu_{B}=0.6\mu$ is chosen as a model for the bulk viscosity of air. The non-dimensional ideal gas law is
\begin{equation}
p=\frac{\gamma-1}{\gamma}\rho T.
\label{eq:ideal_EOS}
\end{equation}

\noindent For more details on the flow governing equations in curvilinear coordinates, see \cite{kim2012adjoint,sharan2016time}.
\subsubsection{Metric Terms to Account for Grid Motion}
\begin{figure}
    \centering
    \includegraphics[width=0.9\textwidth]{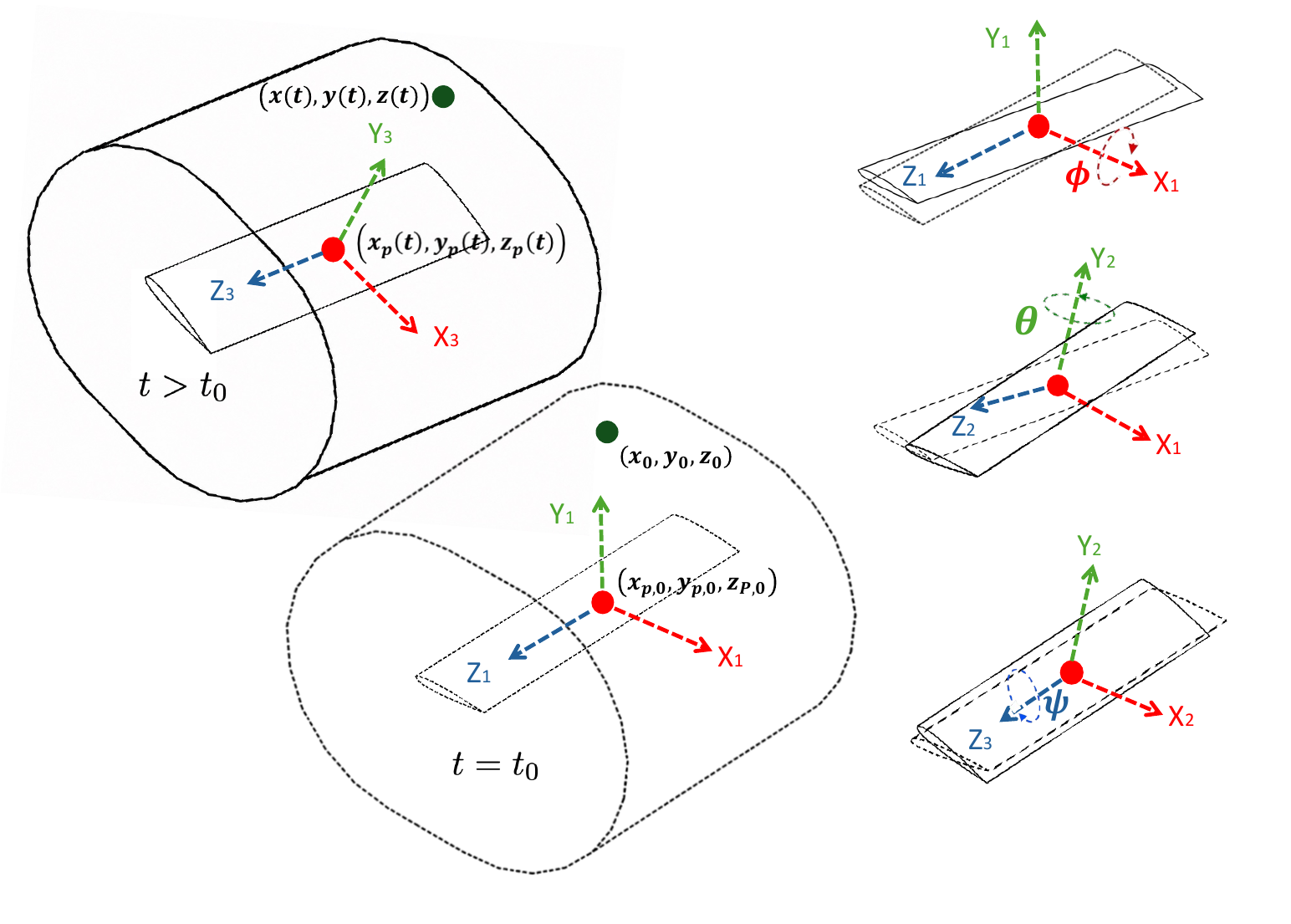}
\caption{Schematic of a wing undergoing rigid-body motion. The right side shows the rotational sequence ending at $(X_3,Y_3,Z_3)$, while the left side shows the complete translation and rotation. The dashed and solid wings on the left denote the initial (at time $t=t_0$) and translated/rotated configuration (at time $t>t_0$), respectively.}
    \label{fig:overlapping-grid}
\end{figure}

For moving grids, time derivative of the curvilinear coordinates \((\xi,\eta,\zeta)\) in \eqref{eq:contravariant_vel} are non-zero, and are given by \cite{pulliam1981diagonal}
\begin{equation}
\begin{aligned}
\xi_t   &= -\left( x_t\,\xi_x + y_t\,\xi_y + z_t\,\xi_z \right), \\
\eta_t  &= -\left( x_t\,\eta_x + y_t\,\eta_y + z_t\,\eta_z \right), \\
\zeta_t &= -\left( x_t\,\zeta_x + y_t\,\zeta_y + z_t\,\zeta_z \right),
\end{aligned}
\label{eq:metric_time}
\end{equation}
where $x_t$, $y_t$, and $z_t$ depend on the grid movement (translation/rotation) in physical coordinates. We consider rigid body motion in this study, where, as illustrated in Fig. \ref{fig:overlapping-grid}, the location of a grid point, initially (at time $t_0$) located at \(\mathbf{x}_0=(x(t_0),y(t_0),z(t_0))^T\equiv(x_0,y_0,z_0)^T\), on the moving grid can be obtained from
\begin{equation}
\begin{bmatrix}
x(t)\\ y(t)\\ z(t)
\end{bmatrix}
=
\begin{bmatrix}
x_p(t)\\ y_p(t)\\ z_p(t)
\end{bmatrix}
+
\mathbf{R}(\phi(t),\theta(t),\psi(t))
\left(
\begin{bmatrix}
x_0\\ y_0\\ z_0
\end{bmatrix}
-
\begin{bmatrix}
x_{p,0}\\ y_{p,0}\\ z_{p,0}
\end{bmatrix}
\right),
\label{eq:rigid_map}
\end{equation}
where the grid's translational motion 
is captured by the displacement of the pivot point \(\mathbf{x}_p(t)=\left(x_p(t),y_p(t),z_p(t)\right)^T\) 
and rotational motion by rotation about the pivot point using the (time-dependent) rotation matrix \(\mathbf{R}(\phi,\theta,\psi)\). The (Euler) angles \(\phi(t)\), \(\theta(t)\), and \(\psi(t)\) denote rotation about the axes \(x\), \(y\), and \(z\), respectively; positive angles turn points counter-clockwise by the right-hand rule, looking down each positive axis toward the origin. $\mathbf{x}_{p,0}=(x_p(t_0),y_p(t_0),z_p(t_0))^T\equiv(x_{p,0},y_{p,0},z_{p,0})^T$ is the initial location of the pivot point. The rotation matrix is given by
\begin{equation}
\mathbf{R}(\phi,\theta,\psi)=
\begin{bmatrix}
\cos\psi \cos\theta &
\cos\psi \sin\theta \sin\phi - \sin\psi \cos\phi &
\cos\psi \sin\theta \cos\phi + \sin\psi \sin\phi \\

\sin\psi \cos\theta &
\sin\psi \sin\theta \sin\phi + \cos\psi \cos\phi &
\sin\psi \sin\theta \cos\phi - \cos\psi \sin\phi \\

-\sin\theta &
\cos\theta \sin\phi &
\cos\theta \cos\phi
\end{bmatrix}.
\label{eq:R_full}
\end{equation}

\noindent Equation~\eqref{eq:rigid_map} assumes that the grid is generated in its initial orientation, \textit{i.e.}, $\phi(0)=\theta(0)=\psi(0))=0$, so that $\mathbf{R}(\phi(0),\theta(0),\psi(0))=\mathbf{I}$ and $\mathbf{x}(t_0)=\mathbf{x}_0$; any initial pitch or offset is therefore built into the reference coordinates $\mathbf{x}_0$ and not into the rotation. The time derivative of~\eqref{eq:rigid_map} provides the grid velocity components for the RHS of \eqref{eq:metric_time}, given by
\begin{equation}
\begin{aligned}
x_t \equiv \dot x &= \dot x_p
      + \dot{\mathbf{R}}_{11} (x_0-x_{p,0})
      + \dot{\mathbf{R}}_{12} (y_0-y_{p,0})
      + \dot{\mathbf{R}}_{13} (z_0-z_{p,0}), \\
y_t \equiv \dot y &= \dot y_p
      + \dot{\mathbf{R}}_{21} (x_0-x_{p,0})
      + \dot{\mathbf{R}}_{22} (y_0-y_{p,0})
      + \dot{\mathbf{R}}_{23} (z_0-z_{p,0}), \\
z_t \equiv \dot z &= \dot z_p
      + \dot{\mathbf{R}}_{31} (x_0-x_{p,0})
      + \dot{\mathbf{R}}_{32} (y_0-y_{p,0})
      + \dot{\mathbf{R}}_{33} (z_0-z_{p,0}),
\end{aligned}
\label{eq:grid_vel}
\end{equation}
where the overdot denotes time derivative and \(\dot{\mathbf{R}}_{ij}\) represents the time derivative of the rotation-matrix components.

\subsubsection{Weak Interface Treatment with Characteristic Decomposition}
The weak overset interface treatment that respects the direction of characteristic wave propagation, as discussed for the scalar advection problem in Section \ref{subsec:1-D Scalar Advection}, is adapted in this section for the compressible Navier-Stokes equations to impose the interpolated data at the fringe (or receiver) points; see Fig.~\ref{fig:interpolation_1}(a). 
To discuss the discretization at an overset interface, consider a fringe point on, say, the $\kappa^{\pm}$ boundary where $\kappa=\xi,\eta$ or $\zeta$. $\kappa$ is the direction normal to the face on which the interface point lies and $\pm$ indicates a lower ($+$) or upper ($-$) computational-coordinate boundary. Let the solution at the fringe point with index $\left(i,j,k\right)$ be denoted by $\mathbf{Q}_{ijk}=\mathbf{q}_{ijk}/J_{ijk}$ and the interpolated physical conservative variables, from the donor grid, be given by $\hat{\mathbf{q}}_{ijk}$. To 
impose only the incoming characteristics at the fringe point, the inviscid flux Jacobians are obtained from
\begin{equation}
A_{\xi}(\mathbf Q) = \frac{\partial \mathbf{F}_c}{\partial \mathbf Q}, \qquad
A_{\eta}(\mathbf Q) = \frac{\partial \mathbf{G}_c}{\partial \mathbf Q}, \qquad
A_{\zeta}(\mathbf Q) = \frac{\partial \mathbf{H}_c}{\partial \mathbf Q},
\label{eq:Ak_def}
\end{equation}
and decomposed using
\begin{equation}
A_{\kappa}
=
S_{\kappa}\,\Lambda_{\kappa}\,S_{\kappa}^{-1}\qquad \text{for} \qquad \kappa=\xi,\eta ~\text{or} ~\zeta,
\label{eq:eig_decomp}
\end{equation}
where $S_{\kappa}$ is the matrix of the right eigenvectors and $\Lambda_{\kappa}$ is the diagonal
matrix of eigenvalues \cite{pulliam1981diagonal}.
The eigenvalues are split to identify incoming and outgoing characteristic waves using
\begin{equation}
\Lambda_{\kappa}^{\pm}
=
\frac{1}{2}\left(\lvert \Lambda_{\kappa} \rvert \pm \Lambda_{\kappa} \right),
\qquad
K_{\kappa^{\pm}} = S_{\kappa}\,\Lambda_{\kappa}^{\pm}\,S_{\kappa}^{-1}.
\label{eq:K_split}
\end{equation}
For viscous flows, in addition to the physical conservative variables, the Cartesian viscous fluxes $\mathbf{\tilde{F}}_v$, $\mathbf{\tilde{G}}_v$, and $\mathbf{\tilde{H}}_v$, defined in \eqref{eq:cartesian_visc_flux}, are interpolated  from the donor grid using the interpolation procedure in \eqref{eq:lagrange_interp_tensor}. The interpolated Cartesian viscous fluxes are then transformed using \eqref{eq:viscous_fluxes} with the metric terms evaluated at the receiver point to obtain $\mathbf{\hat{F}}_v$, $\mathbf{\hat{G}}_v$, and $\mathbf{\hat{H}}_v$. The interpolated values are imposed weakly at the fringe point using a semi-discretization given by
\begin{equation}
\frac{d\mathbf Q_{ijk}}{dt}
=
-\left(D_{\xi_m}\mathbf F_m\right)_{ijk}
-\frac{J_{ijk}^{-1}}{h_0}
\Bigl(
\sigma_I\,K_{\kappa^{\pm}} + \sigma_1^V I_5
\Bigr)
\left(\mathbf q_{ijk}-\hat{\mathbf q}_{ijk}\right)
+\frac{\sigma_2^V}{h_0}
\Bigl[
\left(\mathbf F_v^{\kappa}\right)_{ijk}
-
\left(\hat{\mathbf F}_v^{\kappa}\right)_{ijk}
\Bigr],
\label{eq:SAT_overset_NS}
\end{equation}
where $D_{\xi_m}\mathbf F_m$ is the discrete approximation to the spatial (flux) derivatives of \eqref{eq:ns} and summation over $m=1,2,3$ is implied. $h_0$ is the appropriate endpoint weight of the norm matrix $H$~\cite[Appendix B.1]{gustafsson2007high} in the \(\kappa\) direction
and $I_5$ is the $5\times5$ identity matrix. 
In the first and last terms of the RHS of \eqref{eq:SAT_overset_NS}, the notation $(\mathbf{F}_1,\mathbf{F}_2,\mathbf{F}_3)\equiv(\mathbf{F},\mathbf{G},\mathbf{H})$ and $(\mathbf{F}_v^1,\mathbf{F}_v^2,\mathbf{F}_v^3)\equiv(\mathbf{F}_v,\mathbf{G}_v,\mathbf{H}_v)$, respectively, is used. 
Consistent with the static overset grid discretization \cite{sharan2016time}, the penalty parameters are chosen as
\begin{equation}
\sigma_I \ge \frac{1}{2},\qquad
\sigma_1^V = \frac{2\mu}{Re}\,\left(\kappa_x^2+\kappa_y^2+\kappa_z^2\right),\qquad
\sigma_2^V =
\begin{cases}
+\dfrac{1}{2}, & \kappa^{+}\ \text{(lower boundary)},\\[4pt]
-\dfrac{1}{2}, & \kappa^{-}\ \text{(upper boundary)}.
\end{cases}
\label{eq:sigmas_NS}
\end{equation}

\noindent If the interface point lies on an edge or a corner, then each normal direction ($2$ for an edge and $3$ for a corner) is treated separately and the interface terms for each direction is added to the RHS of \eqref{eq:SAT_overset_NS}.

\section{Numerical Results\label{sec:Numerical-results}}
The above moving overset interface treatment is implemented in the in-house compressible flow code \cite{sharan2018time,sharan2021investigation,sharan2022high} to solve various inviscid/viscous flow problems on moving grids. This section details the stability and accuracy results for those problems. 
 Time integration in all simulations is performed using the classical fourth-order Runge-Kutta (RK4) method. Spatial derivatives are approximated using the diagonal-norm summation-by-parts operators \cite{gustafsson2007high}, denoted here by the notation $p$--$2p$--$p$, where $p$ is the order-of-accuracy of the boundary scheme and $2p$ is that of the interior scheme.

\subsection{1-D Scalar Advection \label{subsec:1-D Advection Result}}
The scalar advection equation \eqref{eq:1d_curv_form-1} is solved on overlapping grids as shown in Fig. \ref{fig:1D-overlapping-grid}, where the left and right grids are static with \((a_L,b_L)=(-1,-0.1)\) and \((a_R,b_R)=(0.1,1)\). The middle grid undergoes a prescribed sinusoidal translation, given by
\begin{equation}
x_M(t)=A\sin(\omega t),
\label{eq:adv_motion_used}
\end{equation}
with amplitude $A=0.1$ and angular frequency $\omega=2\pi/T$, where a time period $T=1$ is chosen. The extent of the moving middle grid is then given by 
\begin{equation}
   \left(a_M(t),b_M(t)\right)
   =
   \left(-0.35+x_M(t),\,0.35+x_M(t)\right).
   \label{eq:middle_grid_interval}
\end{equation}

\noindent The initial/boundary condition \eqref{eq:1d_adv_icbc} is specified using the exact solution
\begin{equation}
   u(x,t) = f(x-ct)=e^{-\beta(x-x_0-ct)^2},
\end{equation}
where the constant advection speed \(c=1\), the scaling parameter $\beta = 80$, and the initial location of the Gaussian $x_0 = -0.4$. 

For this one-dimensional problem, the grid motion reduces to pure translation in the $x$-direction. 
Consequently,
$y_t=z_t=0$ and only a single curvilinear coordinate $\xi$ is required. The metric time derivatives~\eqref{eq:metric_time} therefore simplify to
\begin{equation}
\xi_t = - x_t \,\xi_x,
\label{eq:time_metric_1Dadv}
\end{equation}
where $x_t$ denotes the grid velocity in physical coordinates. The sinusoidal translation~\eqref{eq:adv_motion_used} implies
\begin{equation}
x_t \;=\; \frac{dx_{M}}{dt} \;=\; A\omega\cos(\omega t)
\end{equation}
for the middle grid and $x_t=0$ for the left and right grids. 

To preserve the spatial accuracy of the underlying scheme, Lagrange linear interpolation is used with the $1$--$2$--$1$
scheme, while cubic interpolation is used with the $2$--$4$--$2$
and $3$--$6$--$3$ schemes. The interpolation and the donor--receiver
mapping procedure were described in Section~\ref{sec:Interface treatment in overset methods}.

\begin{figure}
\begin{centering}
\includegraphics[width=0.5\textwidth]{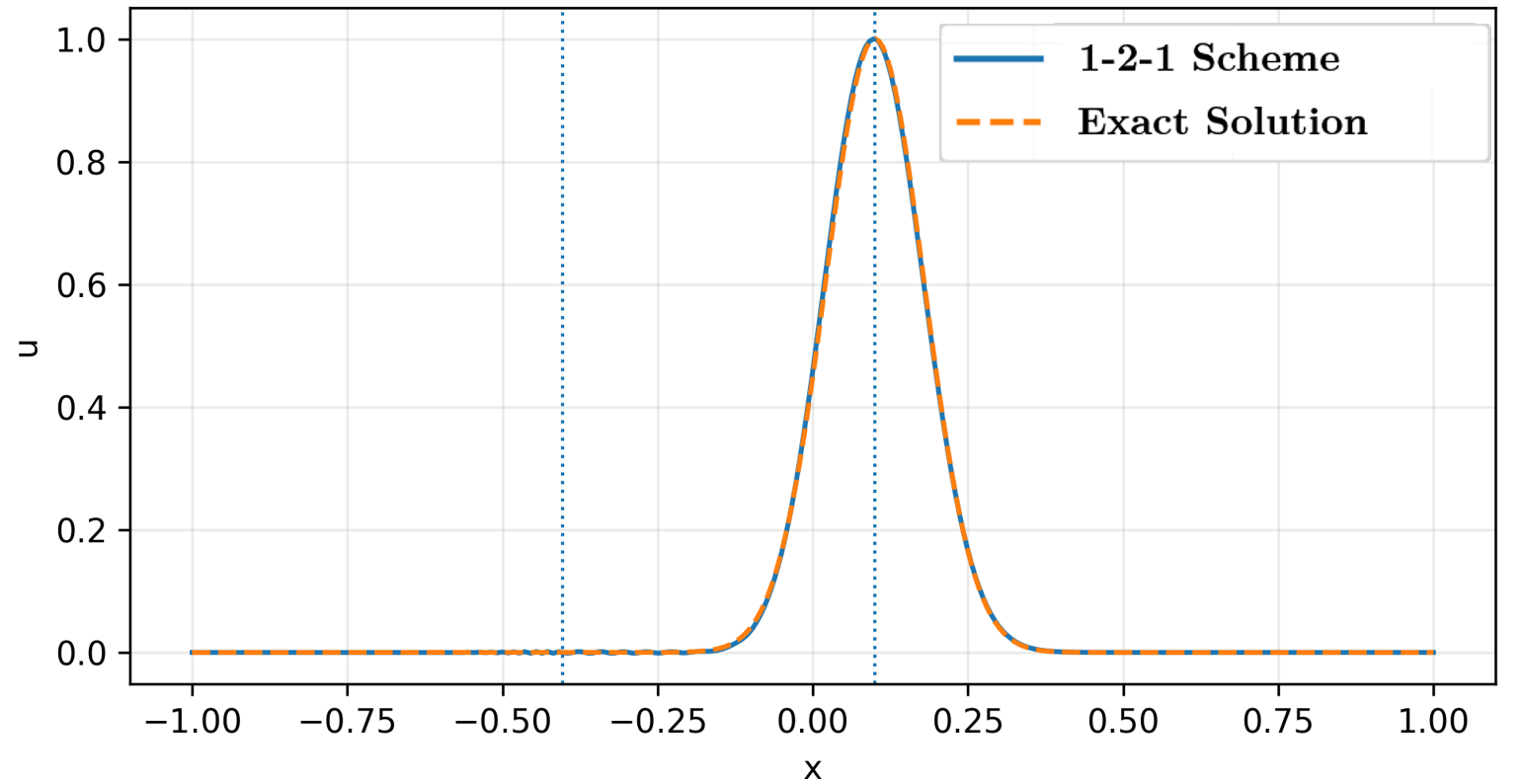}\includegraphics[width=0.49\textwidth]{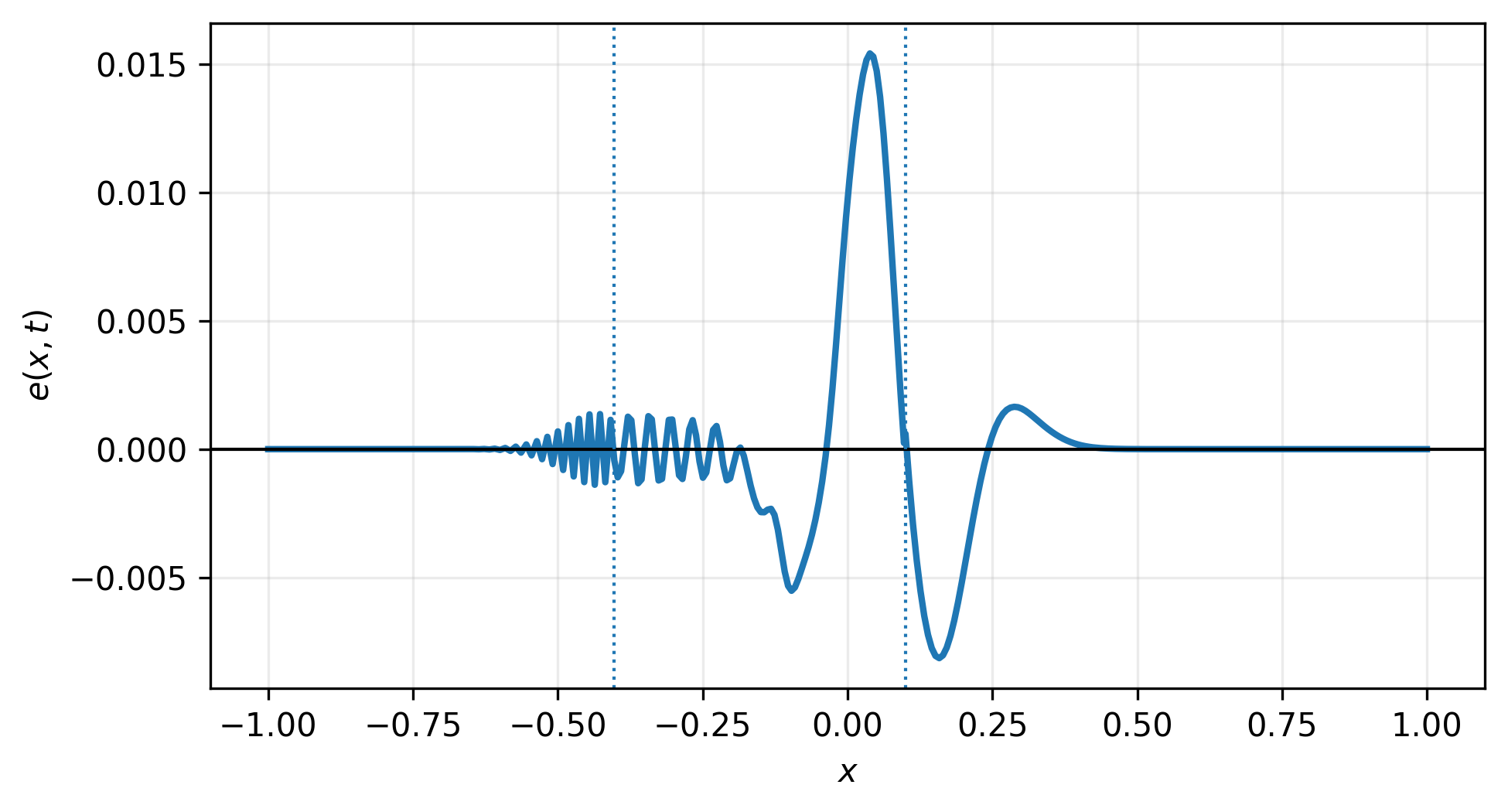}
\par\end{centering}
\begin{centering}
\qquad{}(a)\qquad{}\qquad{}\qquad{}\qquad{}\qquad{}\qquad{}\qquad{}\qquad{}\qquad{}\qquad{}\qquad{}(b)
\par\end{centering}
\begin{centering}
\includegraphics[width=0.5\textwidth]{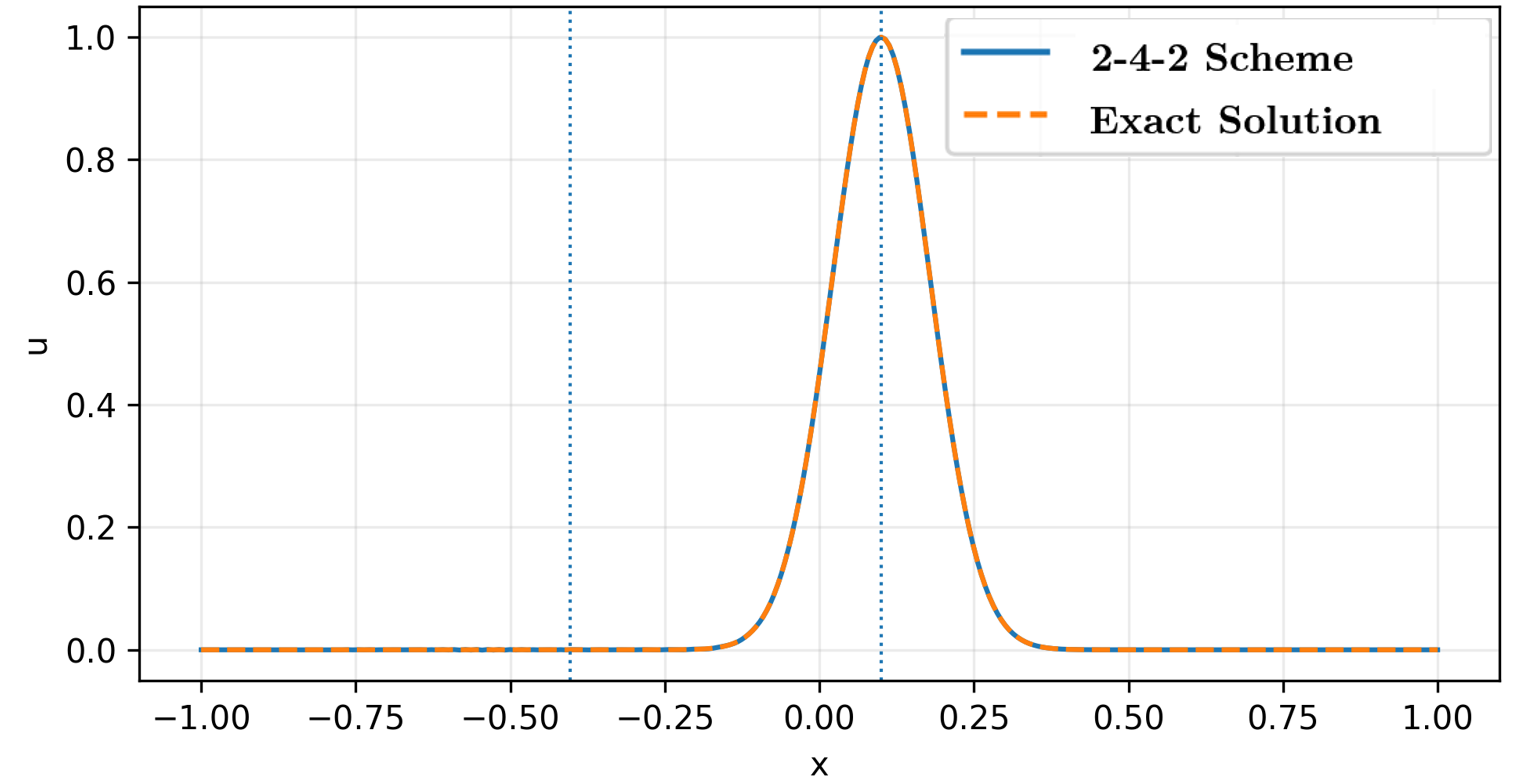}\includegraphics[width=0.49\textwidth]{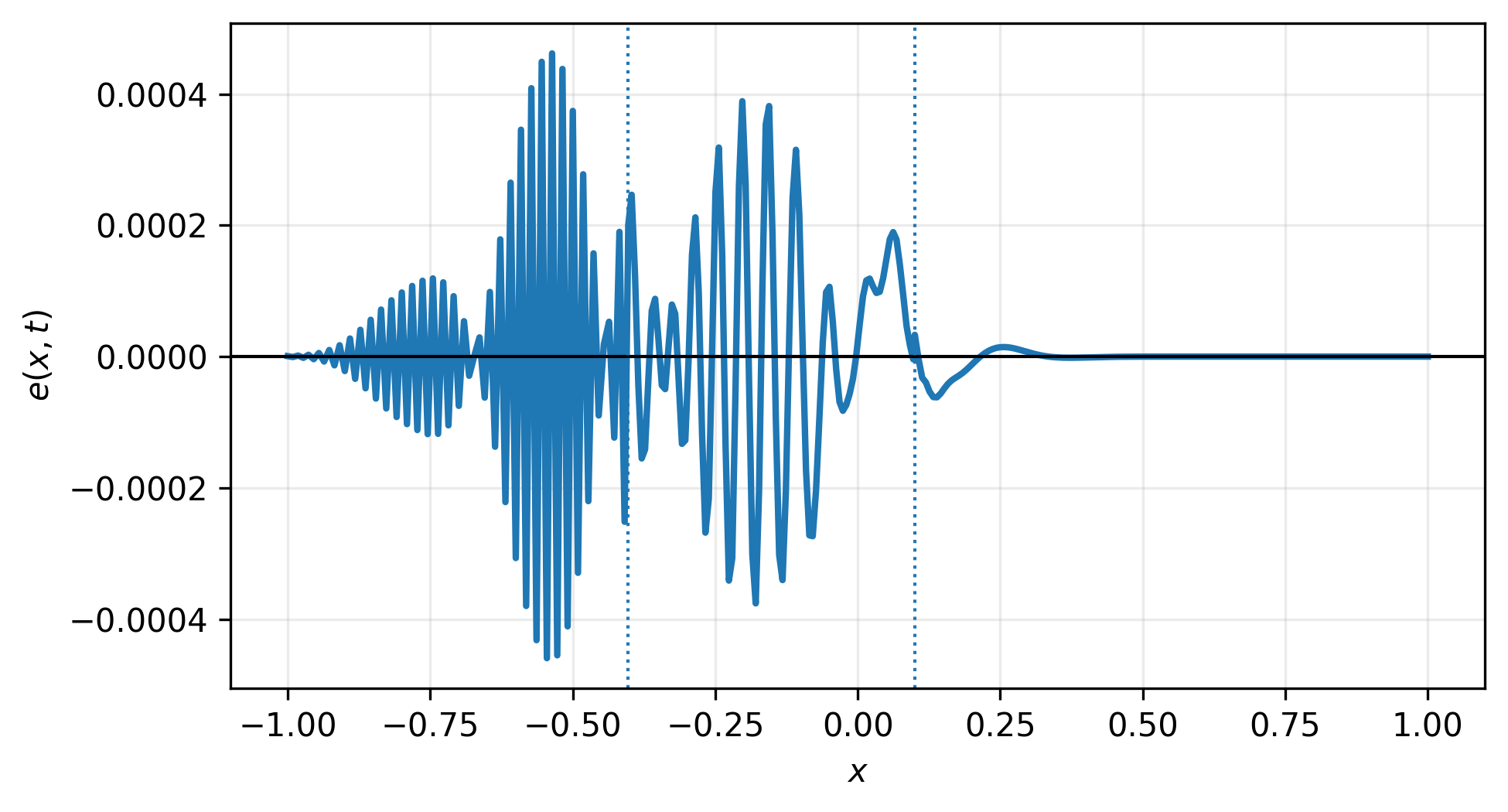}
\par\end{centering}
\begin{centering}
\qquad{}(c)\qquad{}\qquad{}\qquad{}\qquad{}\qquad{}\qquad{}\qquad{}\qquad{}\qquad{}\qquad{}\qquad{}(d)
\par\end{centering}
\begin{centering}
\includegraphics[width=0.5\textwidth]{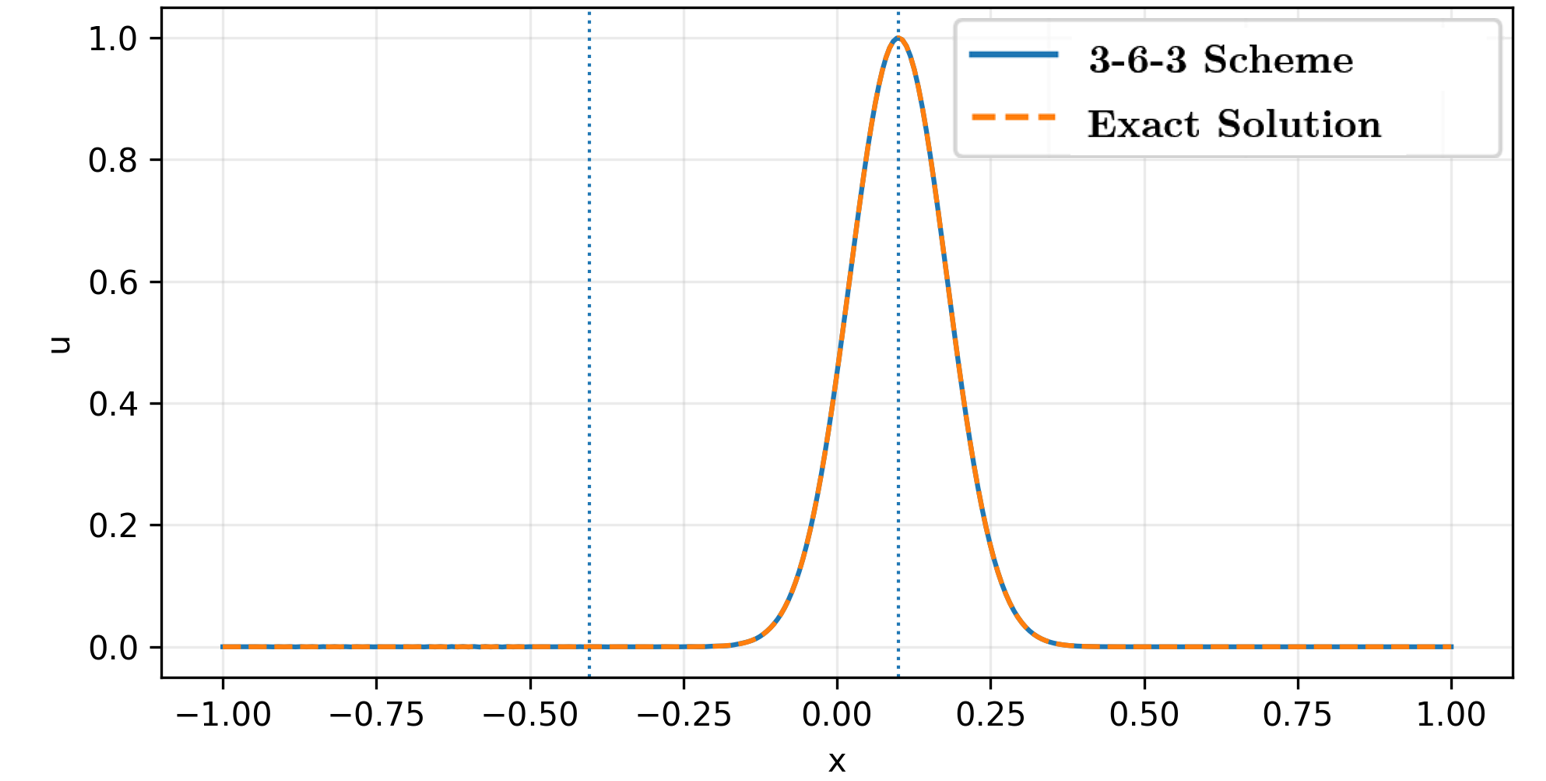}\includegraphics[width=0.49\textwidth]{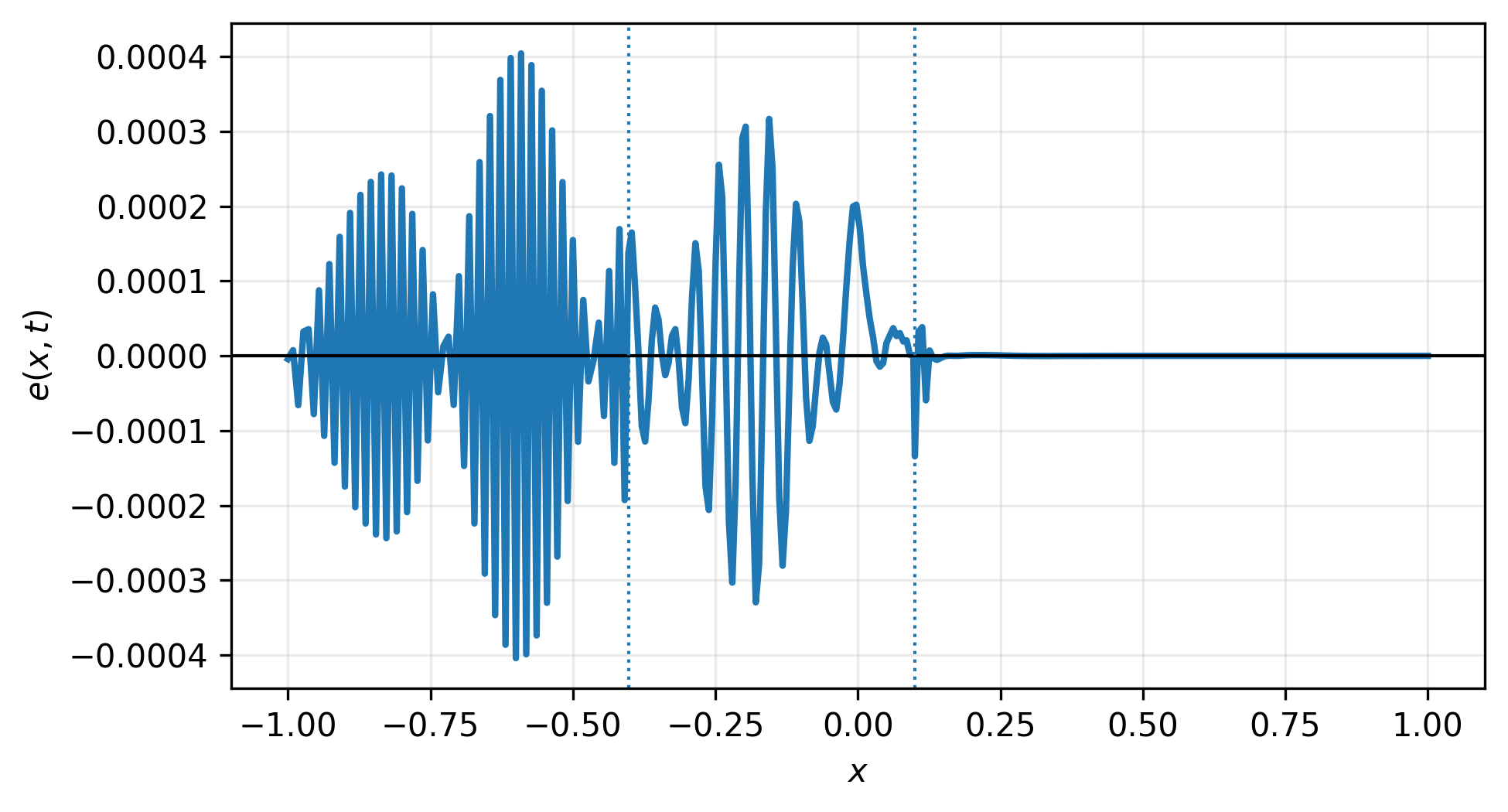}
\par\end{centering}
\begin{centering}
\qquad{}(e)\qquad{}\qquad{}\qquad{}\qquad{}\qquad{}\qquad{}\qquad{}\qquad{}\qquad{}\qquad{}\qquad{}(f)
\par\end{centering}
    \centering

    





    \caption{1-D scalar advection results at $t=0.5$ for various schemes: (a) \& (b) $1$--$2$--$1$ scheme, (c) \& (d) $2$--$4$--$2$ scheme, and (e) \& (f) $3$--$6$--$3$ scheme. The left column shows the solutions, while the right column shows the spatial errors. The vertical dashed line indicates the overset interface of the moving middle grid.}
    \label{fig:advection_interface_error_grid}
\end{figure}

Figure~\ref{fig:advection_interface_error_grid} shows the numerical solution and the corresponding spatial error for the 
moving overset grid configuration described above. 
$80$ grid points are used on the left grid, $100$ on the middle grid, and $120$ on the right grid. The results are reported at $t=0.5$, chosen so that the advected solution has traversed the moving overset interface and is located at the right boundary/interface of the middle grid. The dashed vertical lines mark the instantaneous location of the moving middle-grid overset interfaces. The rows correspond to various $p-2p-p$ schemes. As evident, the error is not concentrated around the overset interfaces, and hence the proposed treatment does not compromise the accuracy of the underlying scheme.

\begin{figure}
\begin{centering}
\includegraphics[width=0.49\textwidth]{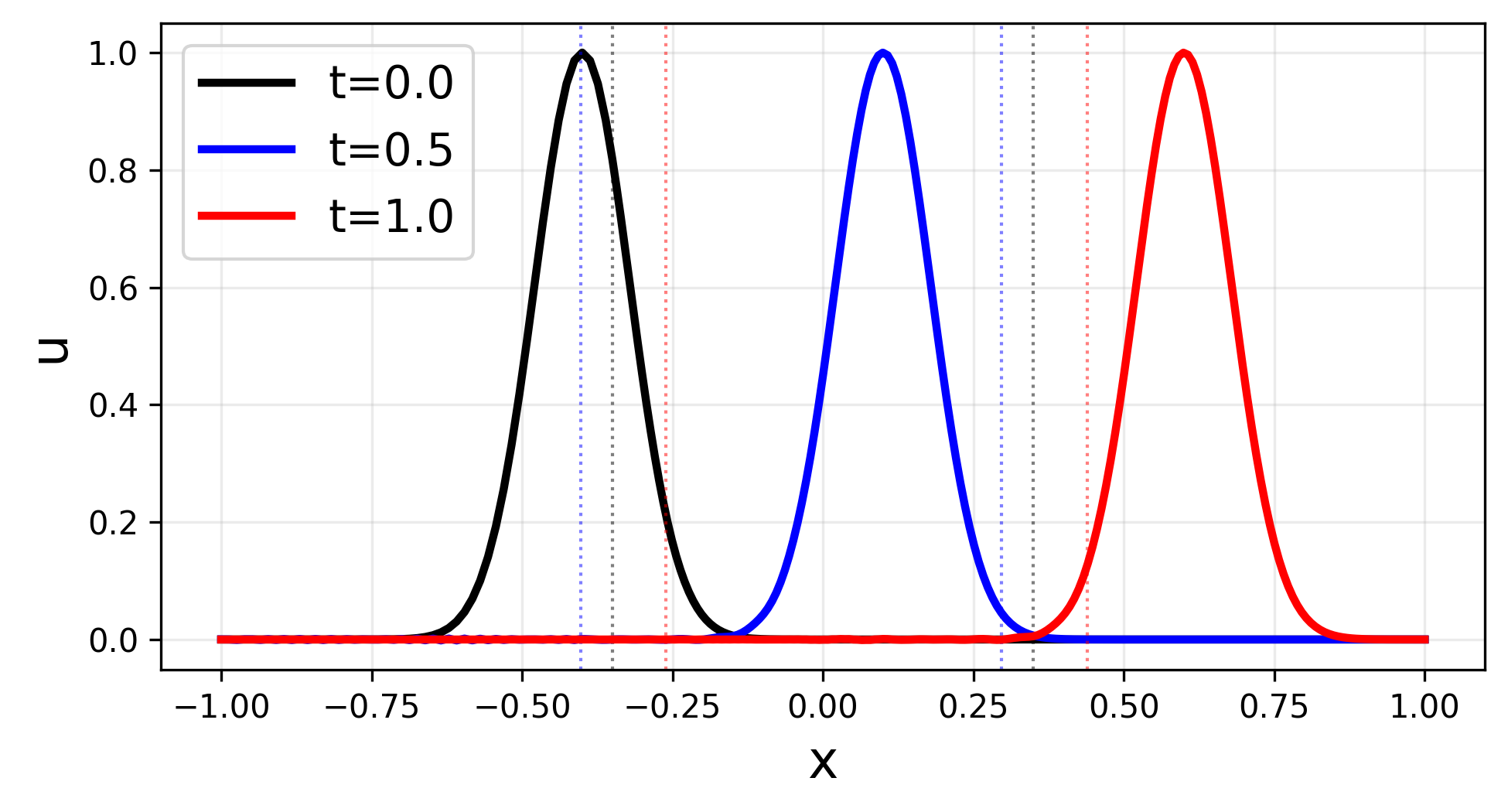}\includegraphics[width=0.5\textwidth]{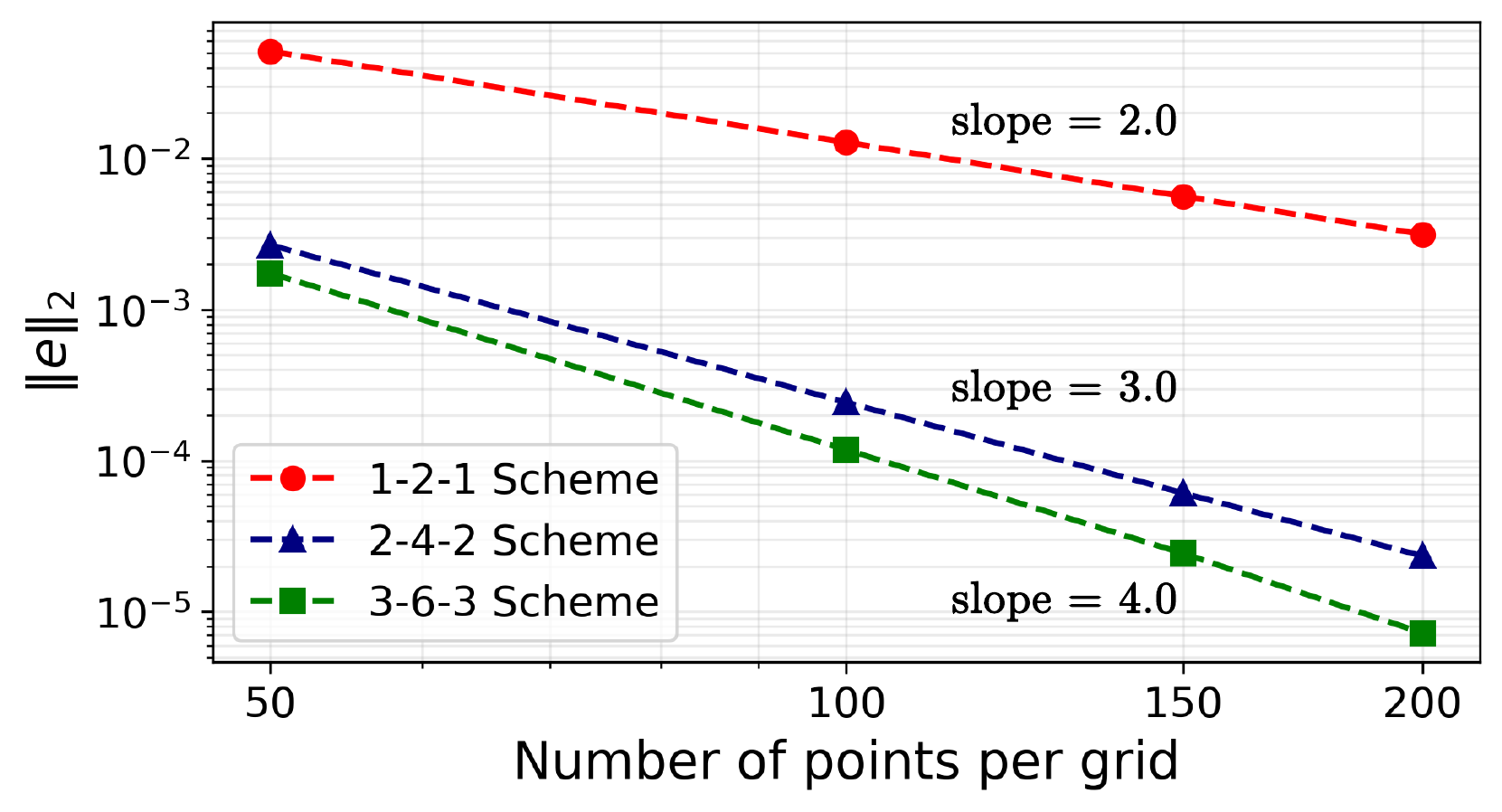}
\par\end{centering}
\begin{centering}
\qquad{}(a)\qquad{}\qquad{}\qquad{}\qquad{}\qquad{}\qquad{}\qquad{}\qquad{}\qquad{}\qquad{}\qquad{}(b)
\par\end{centering}
    \centering
    \caption{(a) Solution from the $3-6-3$ scheme at three times; the vertical dashed lines mark the moving overset interface. 
(b) Convergence plot verifying the global accuracy.}
    \label{fig:sol_accuracy}
\end{figure}

To evaluate the order of accuracy, Fig.~\ref{fig:sol_accuracy}(b) presents the $L_2$-norm of the error with grid refinement, where the errors are calculated at $t=1$, when the initial pulse has traversed the moving middle grid to the right grid, as shown in Fig.~\ref{fig:sol_accuracy}(a). For this convergence analysis, the number of grid points is kept the same in all three grids. A global $p+1$ order of accuracy is observed for various schemes, consistent with the theoretical result of \cite{gustafsson1975convergence}.

To examine the effect of the direction of interface
coupling, the weak interface treatment in
\eqref{eq:left_adv_match}--\eqref{eq:right_adv_match}
is compared with a two-way weak interface treatment.
The latter retains the treatment at the incoming
boundaries and additionally imposes interpolated data
at the right boundaries of the left and middle grids.
This additional treatment is considered only for
the present scalar-advection comparison.

Let $\boldsymbol{I}_{M\to L}$ contain the interpolation
weights from the middle grid to the right boundary
node $u_m$ of the left grid, and let
$\boldsymbol{I}_{R\to M}$ contain the interpolation
weights from the right grid to the right boundary
node $v_n$ of the middle grid.
The additional terms on the right-hand sides of
\eqref{eq:left_adv_match} and
\eqref{eq:middle_adv_match} are, respectively,
\begin{equation}
-\mathbf{p}_L
\left(u_m-\boldsymbol{I}_{M\to L}^{T}\mathbf{v}\right)
\quad \mathrm{and} \quad
-\mathbf{p}_M
\left(v_n-\boldsymbol{I}_{R\to M}^{T}\mathbf{w}\right),
\label{eq:additional_reverse_SAT}
\end{equation}
where
\begin{equation}
\mathbf{p}_L
=\sigma_L\mathcal{X}_L H_L^{-1}\mathbf{e}_m^L,
\quad \mathrm{and} \quad
\mathbf{p}_M
=\sigma_M\mathcal{X}_M H_M^{-1}\mathbf{e}_n^M.
\label{eq:reverse_penalty_vectors}
\end{equation}
Here, $\sigma_L$ and $\sigma_M$ are the additional
penalty parameters, and $\mathbf{e}_m^L$ and
$\mathbf{e}_n^M$ select the right boundary nodes
of the left and middle grids, respectively.
Under the assumed conditions $c>0$ and
$c-x_\tau>0$, these are outgoing boundaries.
The right-grid equation
\eqref{eq:right_adv_match} is unchanged.

Using the same $\mathbf{\Phi}$ and $\mathbf{b}$
as in \eqref{eq:semi_disc} and \eqref{eq:E0_mat},
the modified semi-discrete system is
\begin{equation}
\frac{d\mathbf{\Phi}}{d\tau}
=\hat{M}\mathbf{\Phi}+\mathbf{b},
\label{eq:semi_disc_two_way}
\end{equation}
with
\begin{equation}
\hat{M}=M+
\begin{bmatrix}
-\mathbf{p}_L(\mathbf{e}_m^L)^T
&
\mathbf{p}_L\boldsymbol{I}_{M\to L}^{T}
&
0
\\
0
&
-\mathbf{p}_M(\mathbf{e}_n^M)^T
&
\mathbf{p}_M\boldsymbol{I}_{R\to M}^{T}
\\
0 & 0 & 0
\end{bmatrix},
\label{eq:M_matrix_two_way}
\end{equation}
where $M$ is defined in \eqref{eq:M_matrix}.
The additional interface terms modify the diagonal
blocks and introduce coupling blocks above the
diagonal.

For the eigenvalue comparison, both treatments use
$80$, $100$, and $120$ grid points on the left,
middle, and right grids, respectively.
The grid motion, first-derivative approximations,
and interpolation procedures are identical
between the two treatments. 
The penalty parameters are
$\tau_L=\tau_M=\tau_R=1$ and
$\sigma_L=\sigma_M=1$.
The matrices are evaluated at $t=0.5$.


\begin{figure}
\begin{centering}
\includegraphics[width=0.5\textwidth]{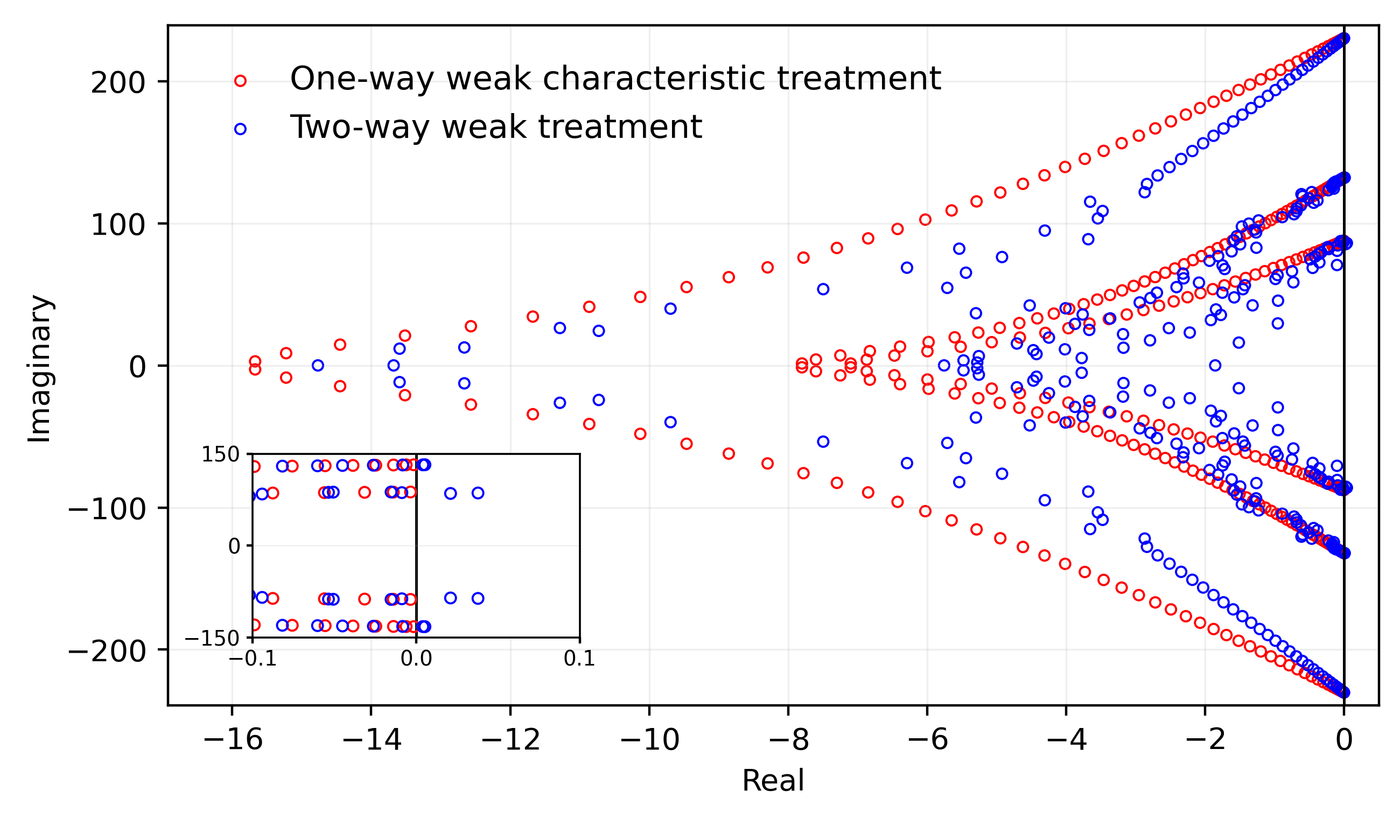}\includegraphics[width=0.5\textwidth]{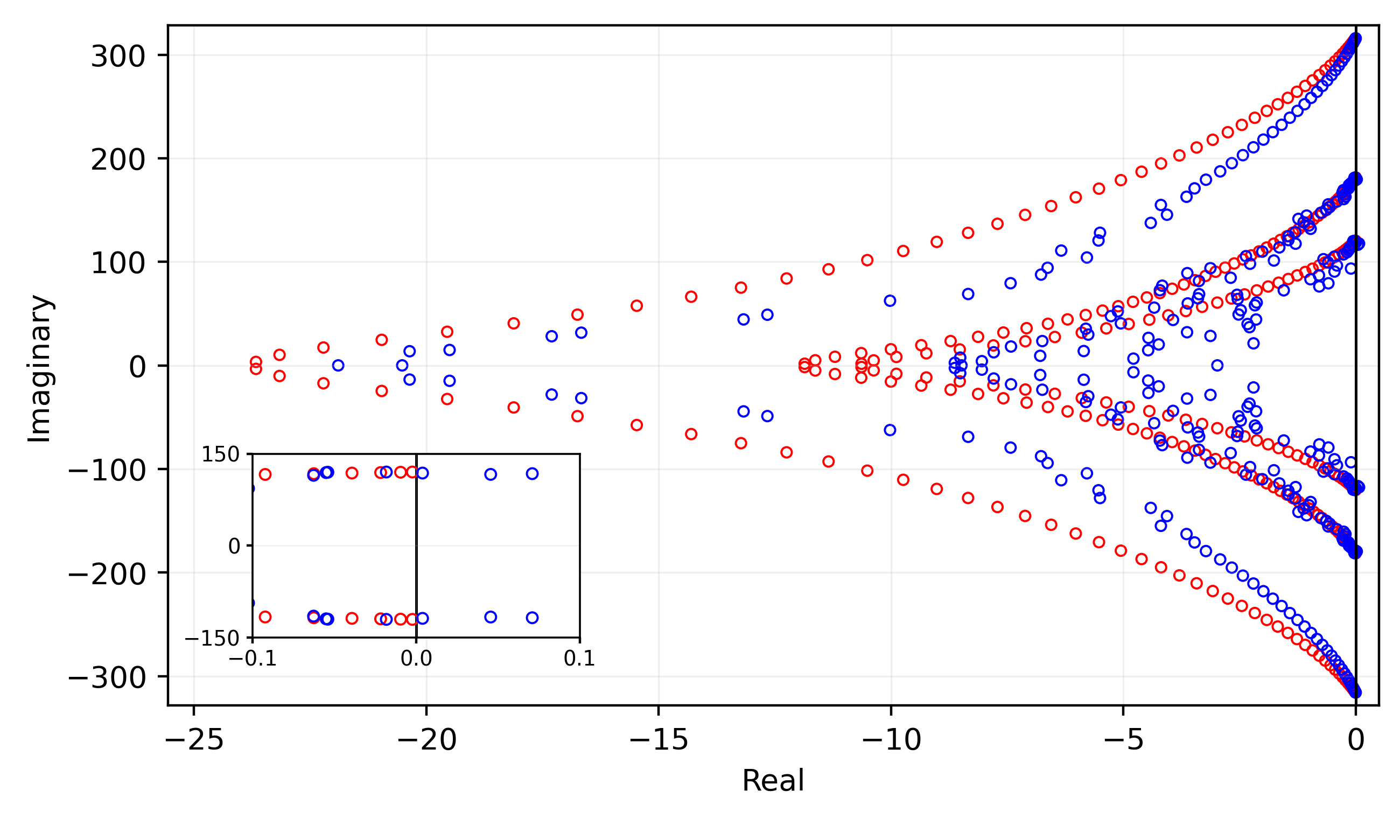}
\par\end{centering}
\begin{centering}
\qquad{}(a)\qquad{}\qquad{}\qquad{}\qquad{}\qquad{}\qquad{}\qquad{}\qquad{}\qquad{}\qquad{}\qquad{}(b)
\par\end{centering}
\begin{centering}
        \includegraphics[width=0.5\textwidth]{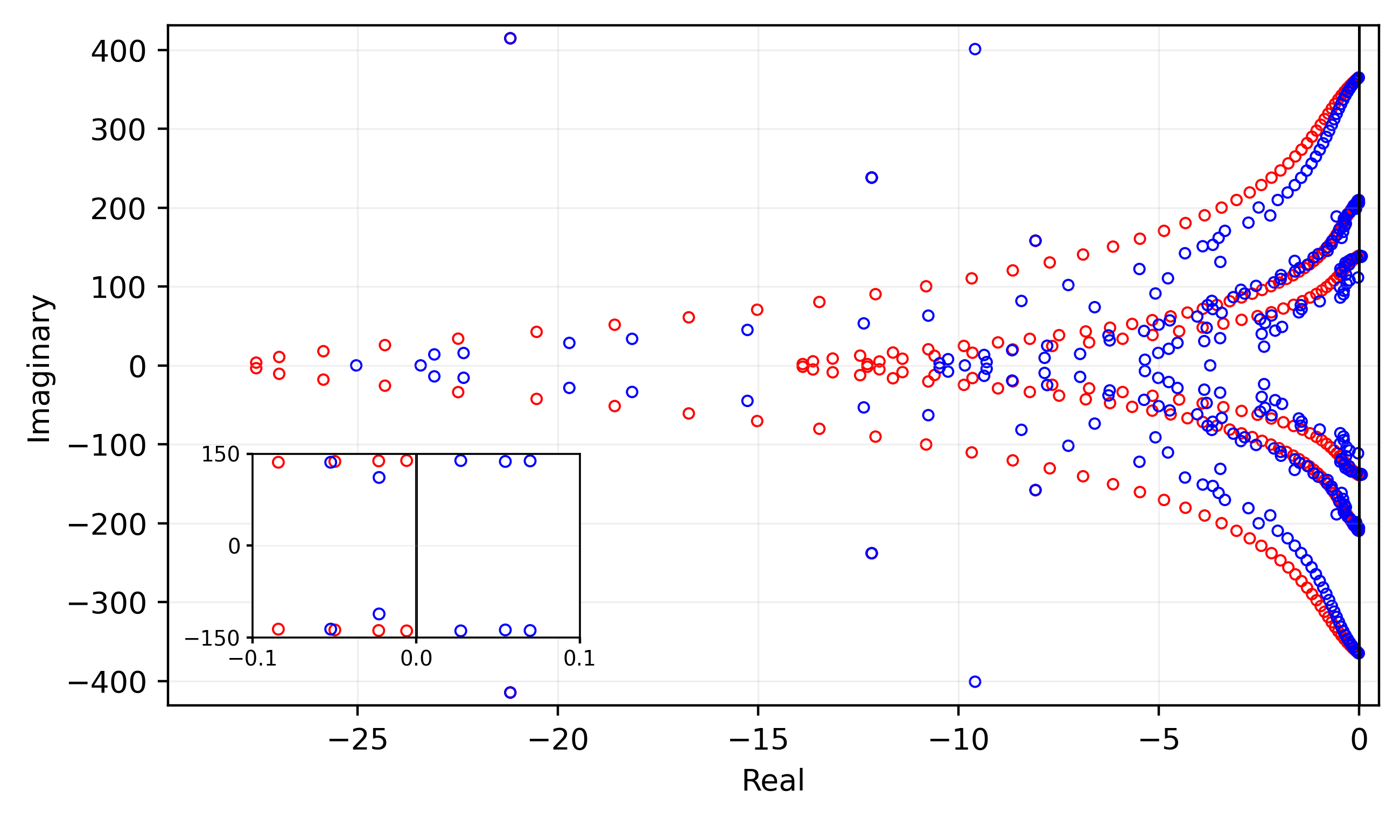}
\par\end{centering}
\begin{centering}
(c)
\par\end{centering}
    \centering
\caption{Eigenvalue spectra of the system matrix
\eqref{eq:M_matrix}, from the characteristic interface
treatment (red circles), and the system matrix
\eqref{eq:M_matrix_two_way}, from the two-way interface
treatment (blue circles), evaluated at $t=0.5$ for
the (a) $1-2-1$, (b) $2-4-2$, and (c) $3-6-3$ schemes.
The maximum real parts of the eigenvalues for the
characteristic (one-way) and two-way interface treatments,
respectively, are
(a) $-1.54792\times10^{-3}$ and $3.77579\times10^{-2}$,
(b) $-1.02826\times10^{-3}$ and $7.09709\times10^{-2}$,
and (c) $-2.27053\times10^{-3}$ and $6.96463\times10^{-2}$.
The insets magnify the region
$-0.1\leq\operatorname{Re}(\lambda)\leq0.1$ and
$-150\leq\operatorname{Im}(\lambda)\leq150$,
showing eigenvalues of the two-way treatment to the
right of the imaginary axis, while those of the
characteristic treatment remain to its left.
Far-left eigenvalues of the two-way treatment lie
outside the displayed limits of the main plots.}

    \label{fig:1D_eigen}
\end{figure}

Figure~\ref{fig:1D_eigen} compares the eigenvalue spectra
of the two interface treatments at $t=0.5$. For all three
schemes, the eigenvalues of the characteristic (one-way)
interface treatment have negative real parts, whereas the
two-way interface treatment exhibits eigenvalues with positive real
parts. Thus, imposing interpolated data at the outgoing
boundaries introduces growing modes when the system
coefficients are held fixed at this instant, for the
configurations and penalty parameters considered.

Under grid motion, the system matrix varies with time due
to changes in grid velocity, overset connectivity, and
interpolation weights. For the characteristic interface treatment, the frozen-time eigenvalues have negative real parts at every sampled time step. Although these spectra establish asymptotic stability of the corresponding frozen-coefficient systems, they do not establish stability of the time-varying system \eqref{eq:semi_disc}. Therefore, we assess perturbation growth under
the prescribed grid motion using the fully discrete propagator
introduced in Section~\ref{subsec:1-D Scalar Advection}.

\begin{figure}
    \centering
    \begin{minipage}[t]{0.6\textwidth}
        \centering
        \includegraphics[width=\linewidth]
            {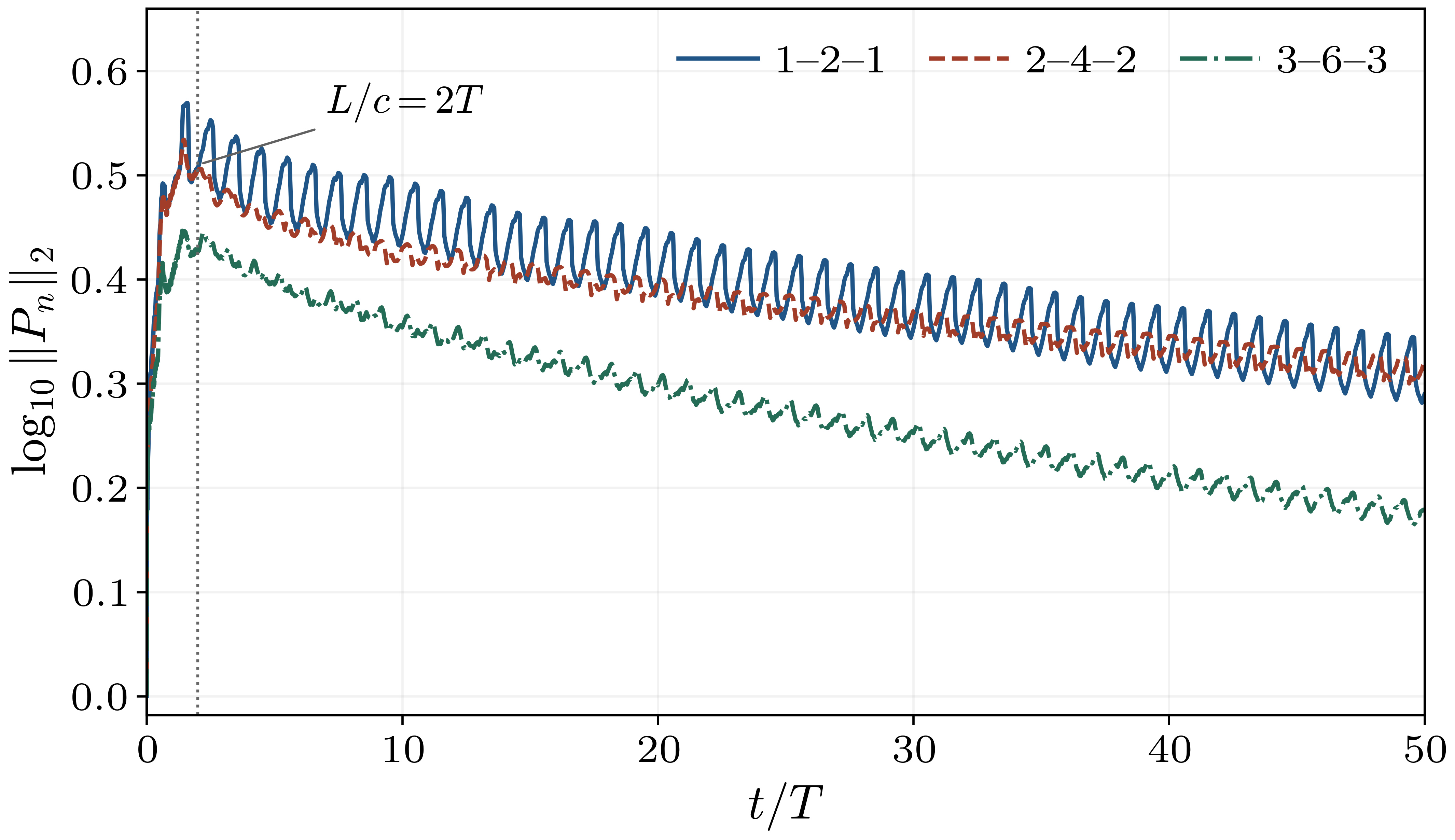}
        \par\smallskip
        (a)
    \end{minipage}%
    \hspace{0.025\textwidth}%
    \begin{minipage}[t]{0.33\textwidth}
        \centering
        \includegraphics[width=\linewidth]
            {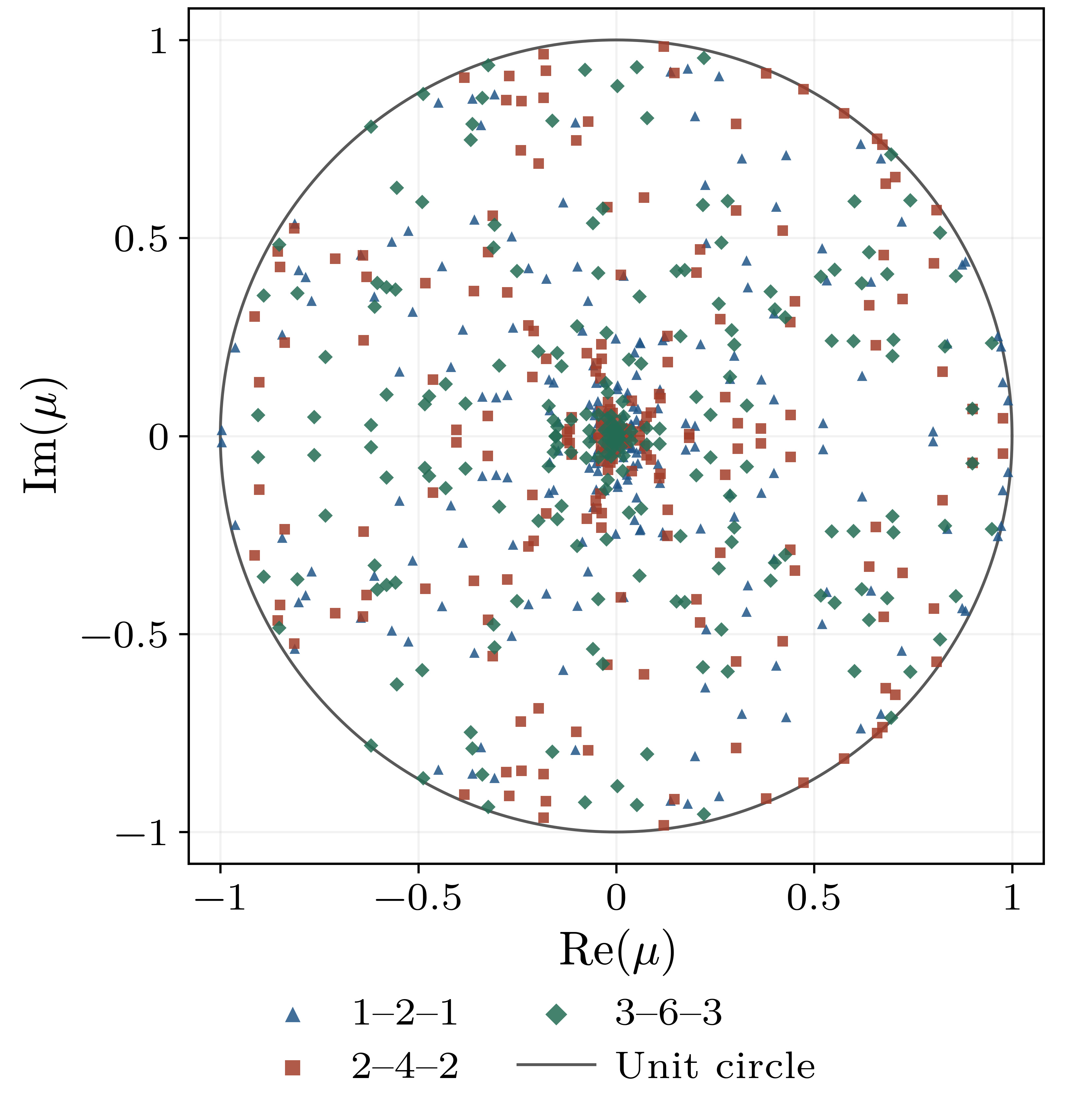}
        \par\smallskip
        (b)
    \end{minipage}
    \caption{Fully discrete stability diagnostics for the characteristic
(one-way) interface treatment
\eqref{eq:left_adv_match}--\eqref{eq:right_adv_match} under the
periodic grid motion \eqref{eq:adv_motion_used}, computed with RK4,
$\Delta t=T/1040$, $80$/$100$/$120$ grid points, and zero inflow data.
(a) $\log_{10}\|P_n\|_2$ over 50 grid-motion periods; the dotted line
marks the domain transit time $L/c=2T$.
(b) Floquet multipliers $\mu$ (eigenvalues of the one-period
propagator $P_T$) and the unit circle; the spectral radii are
$0.99842$, $0.99877$, and $0.99725$ for the $1$--$2$--$1$,
$2$--$4$--$2$, and $3$--$6$--$3$ schemes, respectively.}
    \label{fig:one_way_propagator}
\end{figure}

Figure~\ref{fig:one_way_propagator} reports the fully discrete
diagnostics introduced in Section~\ref{subsec:1-D Scalar Advection}
for the characteristic (one-way) interface treatment. Owing to the block lower-triangular structure of the characteristic discretization \eqref{eq:M_matrix}, the fully discrete propagator retains the same block structure. Consequently, the Floquet multipliers are determined by its diagonal blocks, whereas the propagator norm also captures the transient amplification associated with the off-diagonal interpolation coupling. The grids
($80$, $100$, and $120$ points on the left, middle, and right grids),
the grid motion \eqref{eq:adv_motion_used} with $A=0.1$ and $T=1$,
and the penalty parameters are the same as those used for
Fig.~\ref{fig:1D_eigen}. The inflow data are set to zero
($\mathbf{b}=\mathbf{0}$), and the system is advanced with RK4 using
$\Delta t=T/1040$ (CFL $\approx 0.136$ based on $c$ and the
smallest grid spacing), so that the one-period propagator is
$P_T=P_{1040}$.

In Fig.~\ref{fig:one_way_propagator}(a), $\|P_n\|_2$ increases from
unity to a maximum of approximately $3.7$, $3.4$, and $2.8$ for the
$1$--$2$--$1$, $2$--$4$--$2$, and $3$--$6$--$3$ schemes,
respectively, at $t\approx1.5T$, i.e., before the domain transit time
$L/c=2T$ (dotted line). This bounded transient amplification is
consistent with the non-normality of the discretization: the one-way
interface coupling makes $M$ block lower triangular, so a perturbation
on an upstream grid is passed through the interpolation operators to
the downstream grids, and the Euclidean norm used here differs from
the weighted norm in which the individual grid operators are
dissipative. After the peak, the envelope of $\|P_n\|_2$ decreases
over the remaining $48$ periods (\textit{i.e.}, $24$ domain transit times).
The superposed oscillations at the grid-motion period reflect the
periodic variation of $M(\tau)$ through the grid velocity, the
interface location, and the interpolation weights.

All Floquet multipliers lie strictly inside the unit circle in Fig.~\ref{fig:one_way_propagator}(b). The spectral radii of $P_T$ are $\rho(P_T)=0.99842$, $0.99877$, and $0.99725$ for the $1$--$2$--$1$, $2$--$4$--$2$, and $3$--$6$--$3$ schemes, respectively. Since $\|P_{kN}\|_2\geq\rho(P_T)^k$, these values set the long-time decay rate, which is slow (e-folding over approximately $630$, $810$, and $360$ periods); the faster decrease of the envelope in Fig.~\ref{fig:one_way_propagator}(a) over 50 periods reflects the decay of more strongly damped modes. Halving the time step gives $\rho(P_T)=0.99845$, $0.99897$, and $0.99772$, which remain below unity, indicating that the decay is not an artifact of the temporal discretization. Thus, under the prescribed periodic grid motion, the computed Floquet multipliers indicate that perturbations remain bounded for all time and decay asymptotically for all three schemes. Moreover, the maximum perturbation amplification remains below four over the 50 grid-motion periods examined. In contrast, the two-way treatment already exhibits frozen-time eigenvalues with positive real parts (Fig.~\ref{fig:1D_eigen}). 

We emphasize that the model problem here has one-way interface coupling. For the Euler and Navier--Stokes equations at subsonic overset interfaces, characteristics cross the interface in both directions, so the characteristic treatment \eqref{eq:SAT_overset_NS} couples the grids in both directions and the resulting system matrix is not block triangular. The 1-D analysis therefore does not establish stability of the compressible-flow interface treatment, which is instead assessed through the long-time simulations in Sections~\ref{subsec:2D_Isentropic_Vortex}--\ref{subsec:pitching_plunging_finite_wing}.

\subsection{2-D Isentropic Vortex Convection \label{subsec:2D_Isentropic_Vortex}}
\begin{figure}
\begin{centering}
\includegraphics[width=0.5\textwidth]{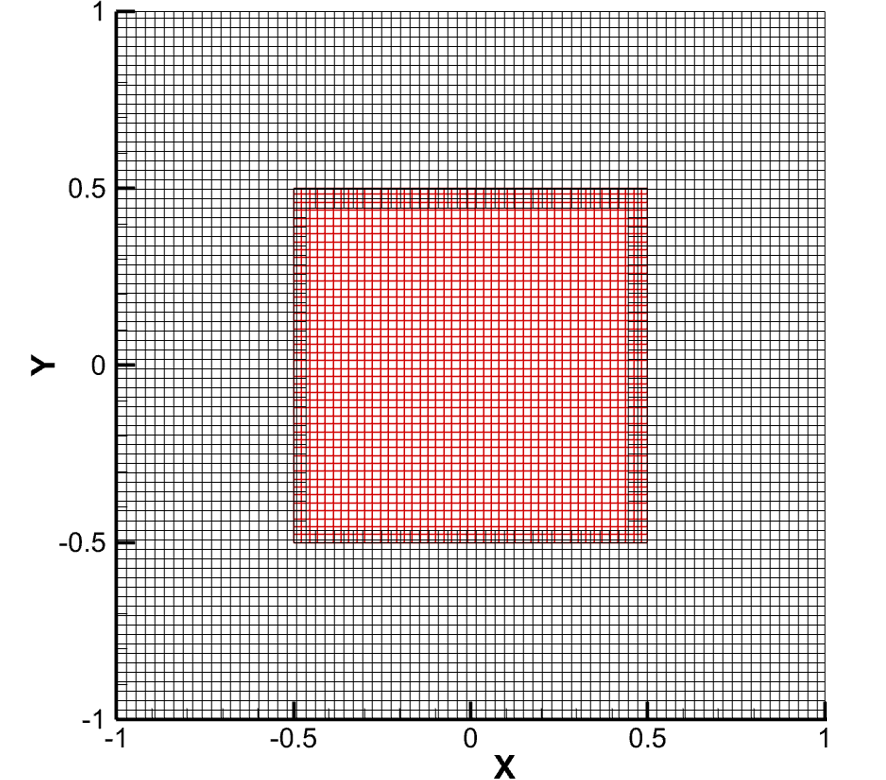}\includegraphics[width=0.5\textwidth]{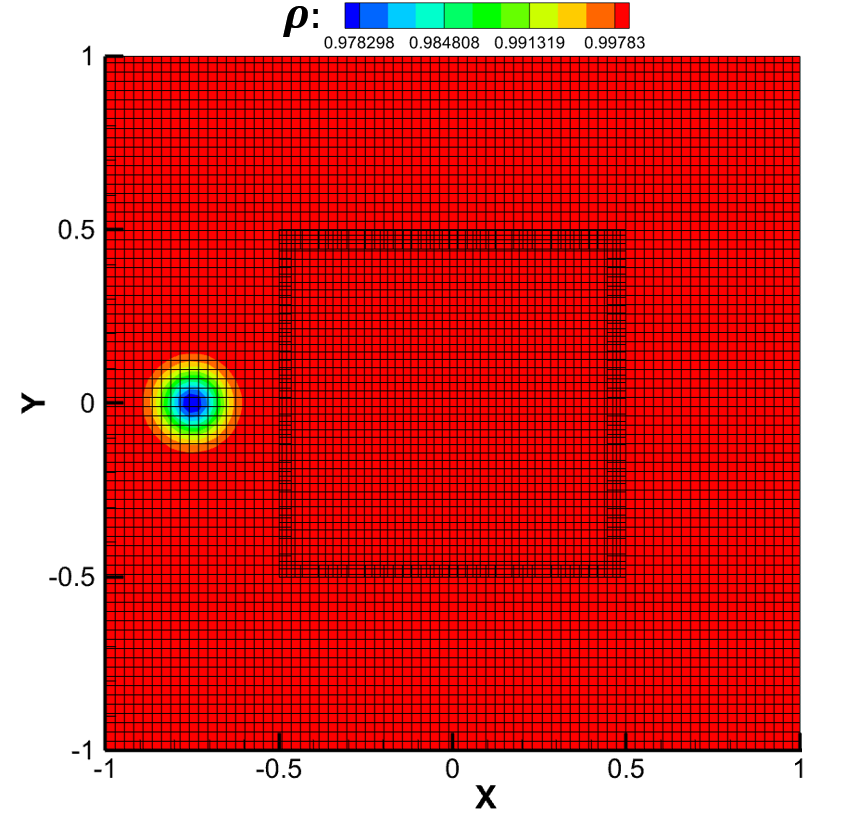}
\par\end{centering}
\begin{centering}
\qquad{}(a)\qquad{}\qquad{}\qquad{}\qquad{}\qquad{}\qquad{}\qquad{}\qquad{}\qquad{}\qquad{}\qquad{}(b)
\par\end{centering}
    \centering
    \caption{2-D vortex convection initial grid: (a) a square background grid (black) overlapping a smaller grid (red) which rotates with time, and (b) the initial density contour.}
    \label{fig:vort_initial}
\end{figure}

To verify the overset interface treatment for an inviscid flow under grid motion, a benchmark problem with a known exact solution is considered.
The grid consists of two overlapping 2-D Cartesian grids, as shown in Fig.~\ref{fig:vort_initial}(a). The smaller square grid
(shown in red) undergoes a prescribed rigid-body rotation about its center (0, 0). Referring to Eq.~\eqref{eq:rigid_map}, this configuration corresponds to zero translation, \(x_p(t)=y_p(t)=z_p(t)=0\), and a pure rotation about the \(z\)-axis, with \(\phi(t)=0\) and \(\theta(t)=0\) in the rotation matrix \eqref{eq:R_full}, and
\begin{equation}
\psi(t)=\psi_A\,\sin\!\left(\frac{2\pi t}{T}\right),
\label{eq:bench_rot}
\end{equation}
where \(\psi_A=\pi/6\) is chosen as the motion amplitude, \(T=5\) is the time period, and \(t\) denotes the non-dimensional solution time. 
Using Eq.~\eqref{eq:grid_vel}, the grid velocity in physical coordinates is then given by
\begin{equation}
x_t=-\dot{\psi}\left(x_0\sin{\psi}+y_0\cos{\psi}\right)=-\dot{\psi}\,y,\qquad
y_t=\dot{\psi}\left(x_0\cos{\psi}-y_0\sin{\psi}\right)=\dot{\psi}\,x,\qquad
z_t=0.
\label{eq:bench_gridvel}
\end{equation}
The grid velocity enters the governing equations through the time-metric derivatives defined in Eq.~\eqref{eq:metric_time}, given for a two-dimensional configuration by
\begin{equation}
\xi_t=-(x_t\xi_x+y_t\xi_y),\qquad
\eta_t=-(x_t\eta_x+y_t\eta_y),\qquad
\zeta_t=0.
\label{eq:bench_timemetrics}
\end{equation}
For the smaller grid, \eqref{eq:bench_gridvel} is substituted in \eqref{eq:bench_timemetrics} to incorporate grid motion, while the background grid remains static, which implies $x_t=y_t=z_t=0$ and hence $\xi_t=\eta_t=\zeta_t=0$.

The vortex convection velocity is prescribed as $(u_0, v_0) = (U_\infty, 0)$ so that the vortex translates only in the x-direction. With periodic boundaries, this choice causes the vortex to re enter the domain and encounter the rotating square overset grid repeatedly, allowing an assessment of the framework's accuracy over multiple passes. Information exchange from one grid to another uses Lagrange bicubic interpolation. The governing equations for this problem are the Euler equations (\textit{i.e.}, \eqref{eq:ns}--\eqref{eq:contravariant_vel} with $\mathbf{F}_v=\mathbf{G}_v=\mathbf{H}_v=\mathbf{0}$), and the exact density solution is given by

\begin{equation}
\rho(x, y, t) = \left[ 1 - \frac{\epsilon^2 (\gamma - 1)}{8\pi^2 \gamma}
e^{\,1 - s^2\left((x - x_0 - u_0t)^2 + (y - y_0 - v_0t)^2\right)} \right]^{\frac{1}{\gamma - 1}},
\label{eq:vortex_density}
\end{equation}
where $\epsilon$ determines the strength of the density perturbation,
$\gamma$ is the specific heat ratio (isentropic exponent),
$s$ controls the spatial decay rate of the exponential function, and
$(x_0, y_0)$ is the initial location of the vortex center. Fig.~\ref{fig:vort_initial}(b) shows the initial density contours, where $(x_0, y_0)=(-0.75, 0.0)$. The solutions to other flow variables for this problem can be found in \cite[Section 5.2]{sharan2018time}.
\vspace{0.0cm}

\begin{figure}
\begin{centering}
\includegraphics[width=0.5\textwidth]{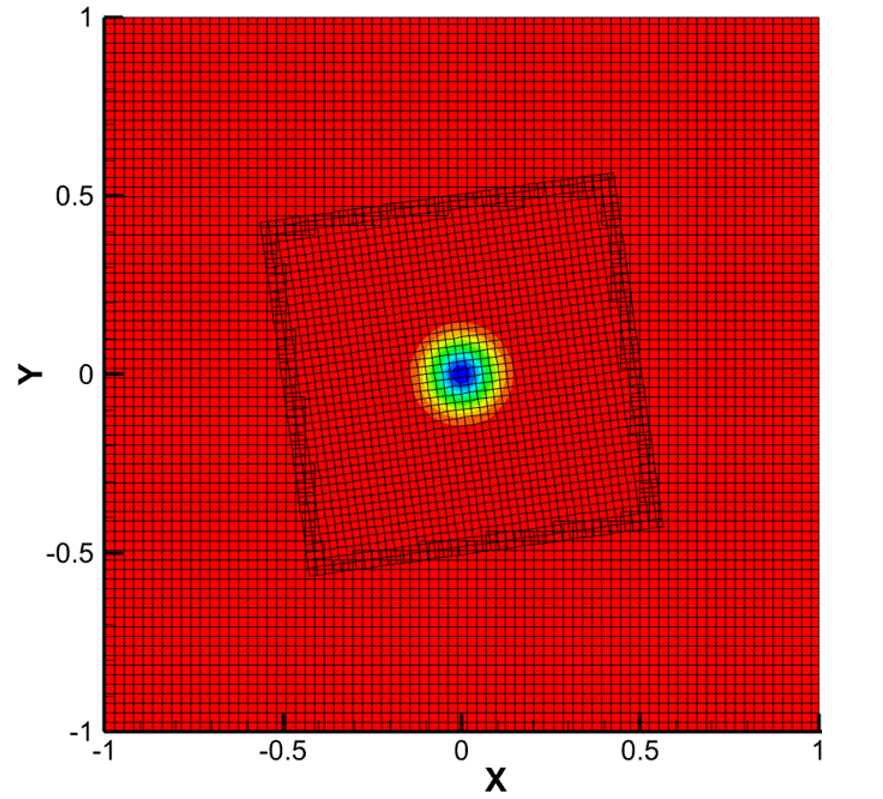}\includegraphics[width=0.5\textwidth]{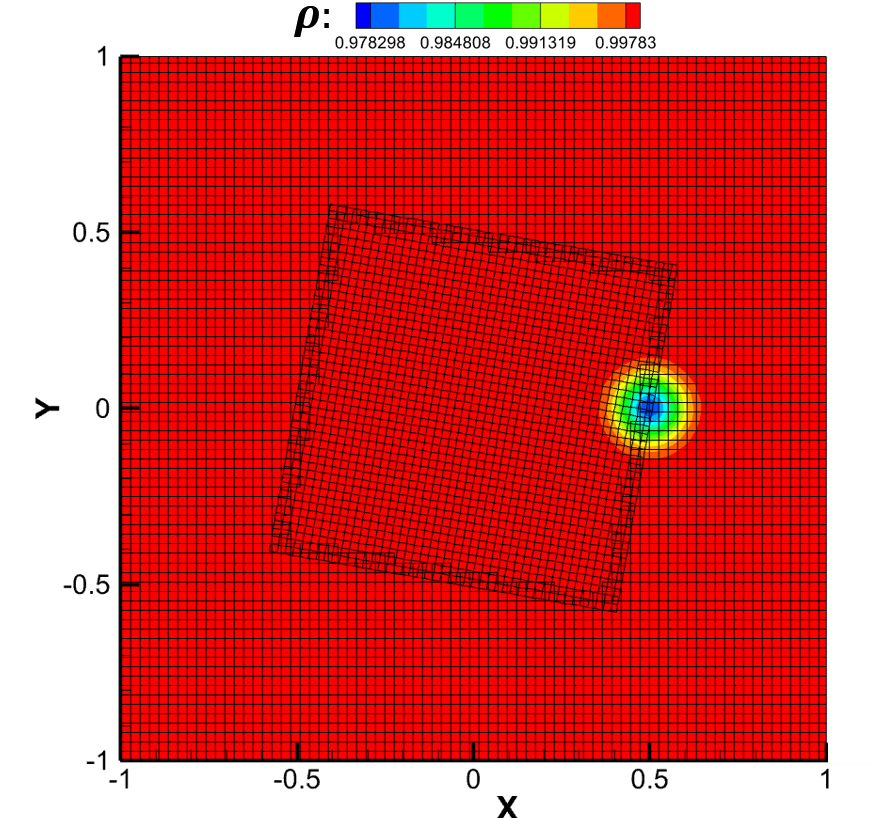}
\par\end{centering}
\begin{centering}
\qquad{}(a)\qquad{}\qquad{}\qquad{}\qquad{}\qquad{}\qquad{}\qquad{}\qquad{}\qquad{}\qquad{}\qquad{}(b)
\par\end{centering}
    \centering
\caption{Density contours at two different time instants for the 2-D isentropic vortex convection on the moving overset configuration. 
}
    \label{fig:vort_history}
\end{figure}
Fig.~\ref{fig:vort_history} shows the density contours at two different time instants as the vortex traverses the rotating square overset grid. To evaluate stability, the simulation is carried out for a long duration and the density error, \(\varepsilon_{\rho}(\mathbf{x},t)\), is examined by calculating the difference between the exact and numerical solution at each time step. 
The grid resolutions considered for the moving overset grid cases are shown in Table \ref{tab:vortex_convection_2D}, where \(N_{\mathrm{sg}}\) and \(N_{\mathrm{bg}}\) denote the number of grid points in each direction for the smaller and background grids, respectively. For simplicity, an equal number of grid points is used in each coordinate direction. Therefore, the smaller grid contains \(N_{\mathrm{sg}}\times N_{\mathrm{sg}}\) points, while the background grid contains \(N_{\mathrm{bg}}\times N_{\mathrm{bg}}\) points. Numerical results that confirm the accuracy of the proposed approach for a static grid using various $p-2p-p$ schemes are discussed in~\ref{sec:appdxB}.

\begin{figure}
\begin{centering}
\includegraphics[width=0.51\textwidth]{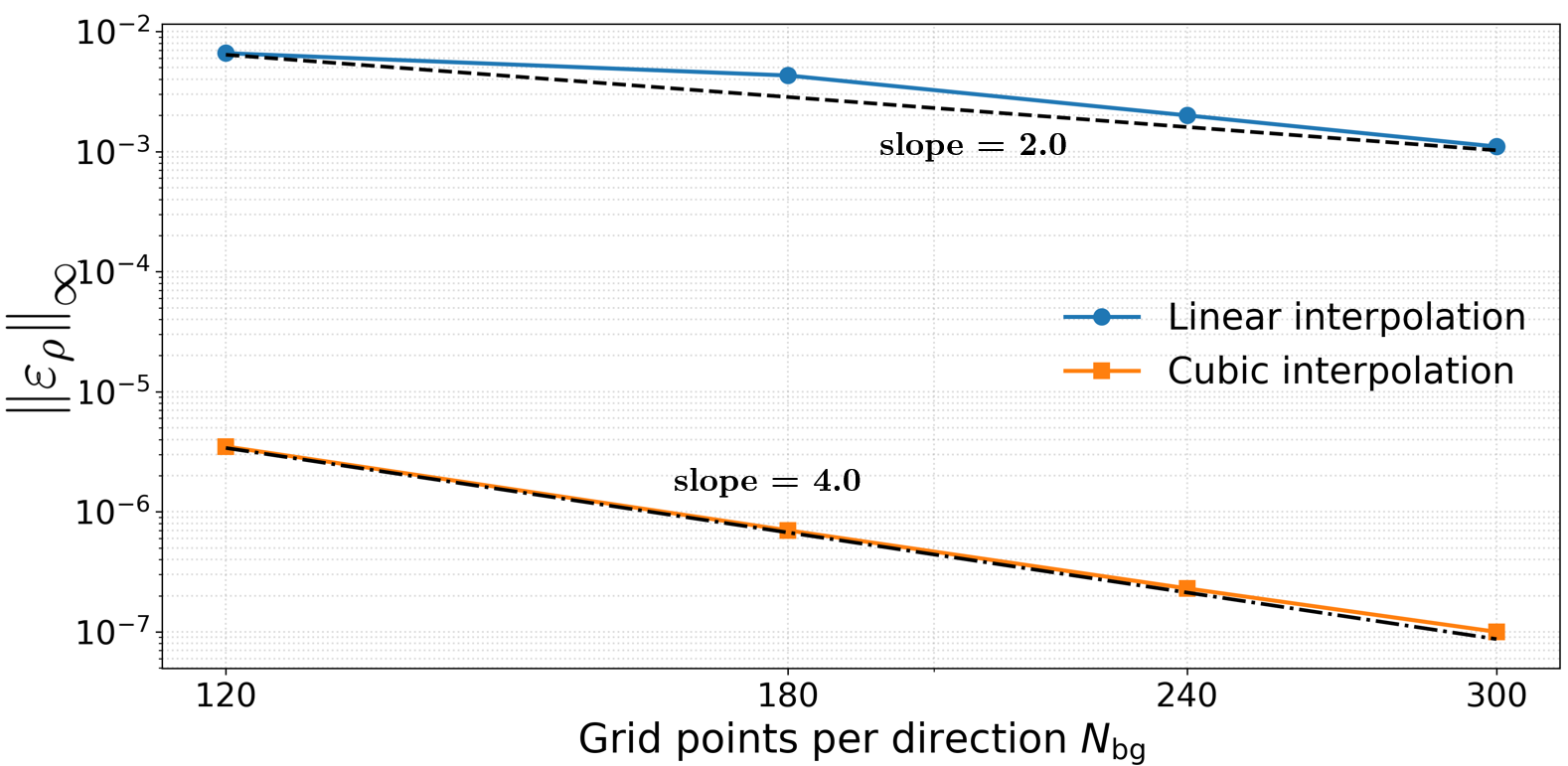}\includegraphics[width=0.47\textwidth]{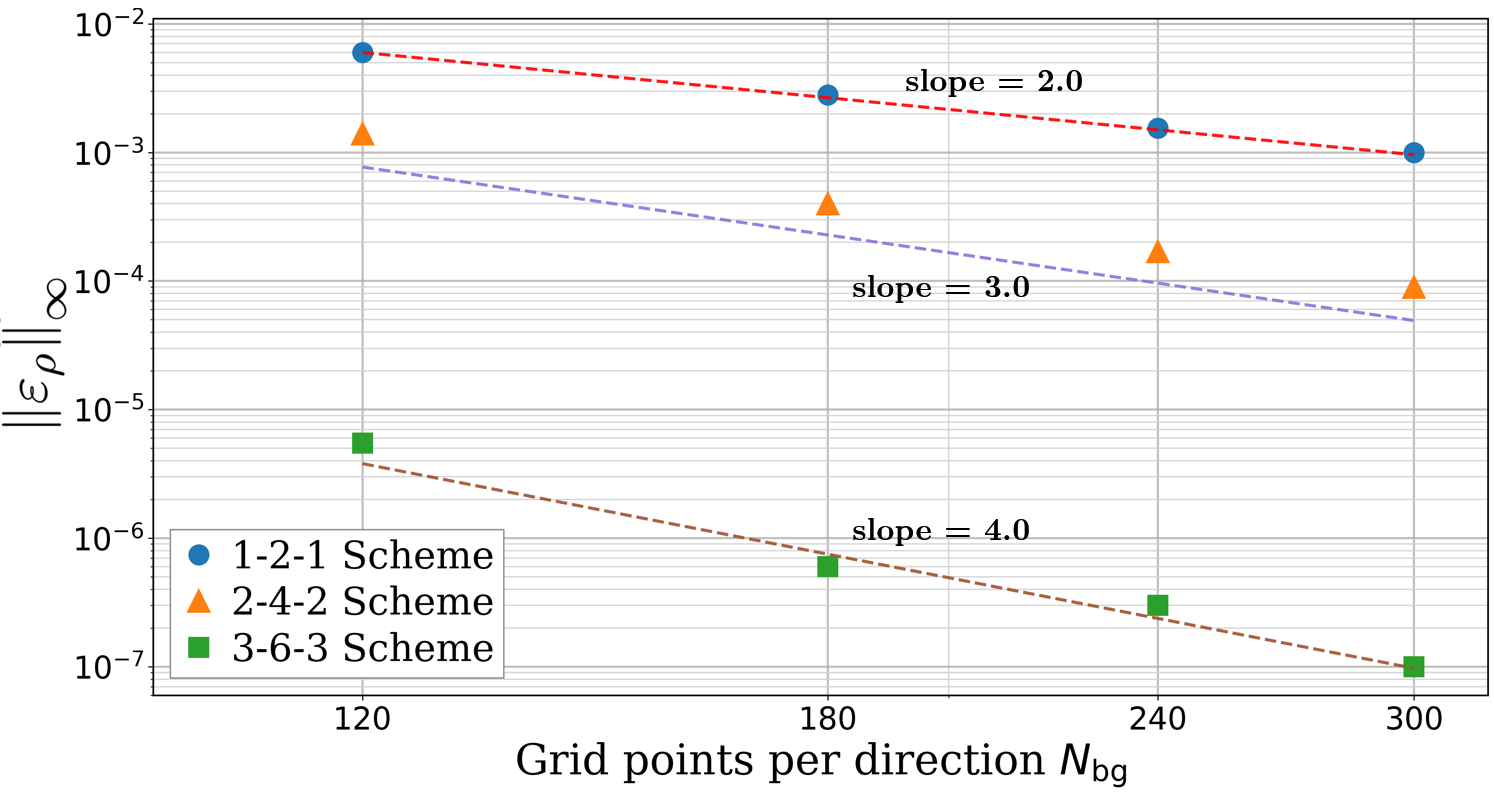}
\par\end{centering}
\begin{centering}
\qquad{}(a)\qquad{}\qquad{}\qquad{}\qquad{}\qquad{}\qquad{}\qquad{}\qquad{}\qquad{}\qquad{}\qquad{}(b)
\par\end{centering}
    \centering   
    \caption{Log--log convergence plots of \(\left\|\varepsilon_{\rho}\right\|_{\infty}\). (a) Comparison of the errors from linear and cubic interpolation with $3-6-3$ scheme. (b) Comparison of the errors from $1-2-1$, $2-4-2$, and $3-6-3$ schemes using Lagrange cubic interpolation.}
    \label{fig:lin_cu_global}
\end{figure}

\begin{figure}
\begin{centering}
\includegraphics[width=0.34\textwidth]{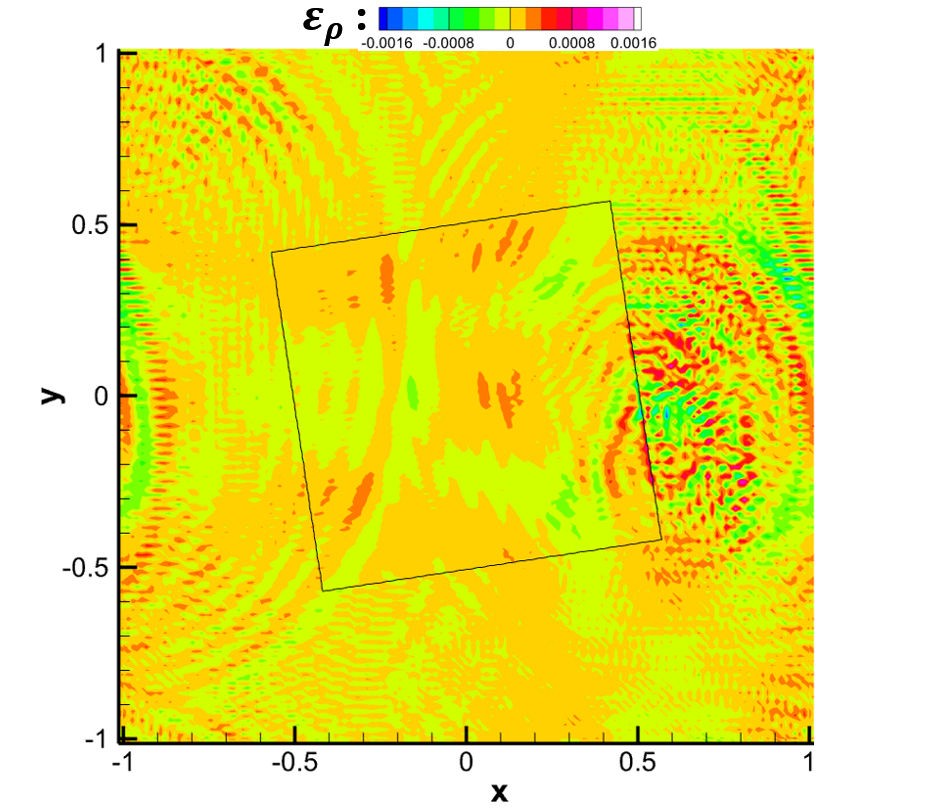}\includegraphics[width=0.34\textwidth]{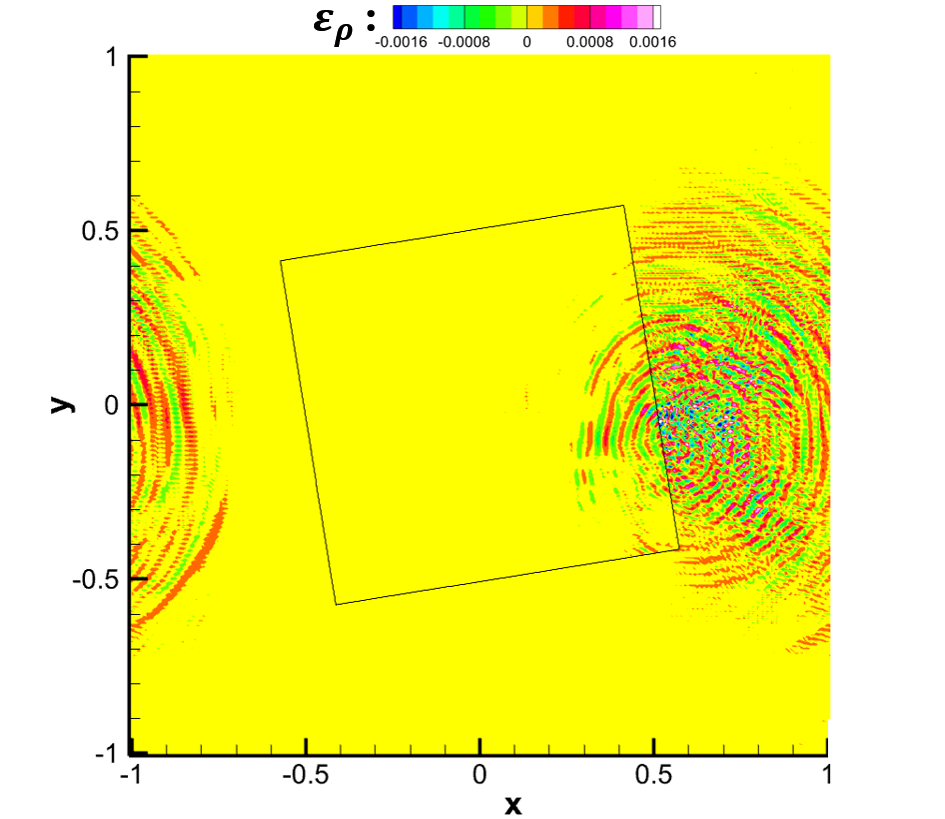}\includegraphics[width=0.305\textwidth]{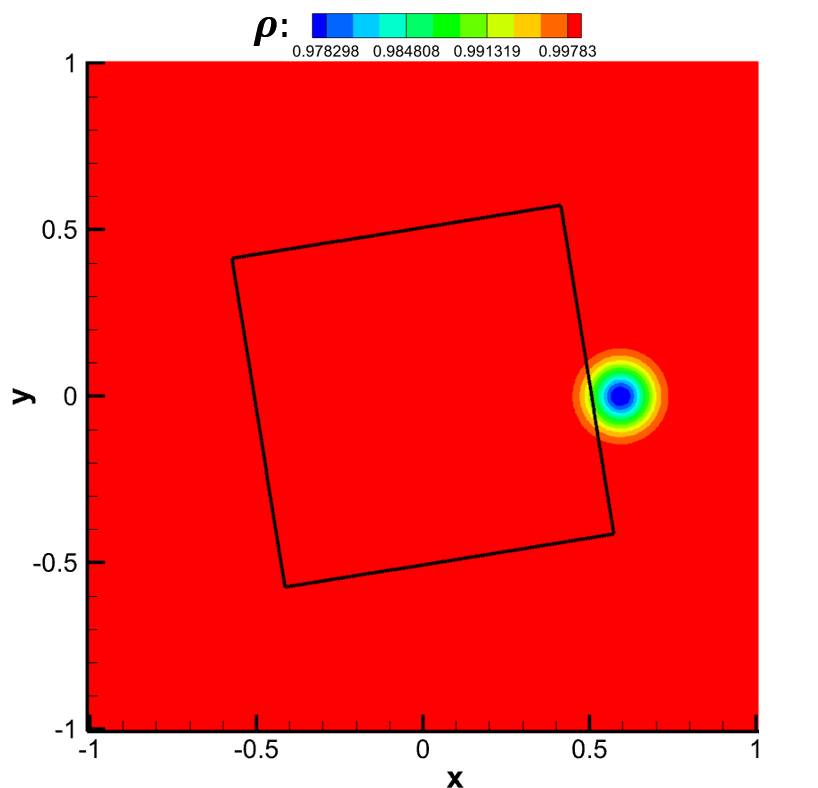}
\par\end{centering}
\begin{centering}
\qquad{}(a)\qquad{}\qquad{}\qquad{}\qquad{}\qquad{}\qquad{}\qquad{}\qquad{}(b)\qquad{}\qquad{}\qquad{}\qquad{}\qquad{}\qquad{}\qquad{}(c)
\par\end{centering}
    \centering
\caption{Density error contours using the $3-6-3$ scheme on (a) Resolution 1 and (b) Resolution 4 gird. (c) Density contours showing the location of the isentropic vortex at that solution time.}
    \label{fig:error_2D}
\end{figure}

The $L_{\infty}$-norm of the density error, \(\left\|\varepsilon_{\rho}\right\|_{\infty}\), at a non-dimensional time of 10 for the four grid resolutions listed in Table \ref{tab:vortex_convection_2D} is shown in a log-log plot in Fig. \ref{fig:lin_cu_global}. The $3-6-3$ scheme is used in Fig. \ref{fig:lin_cu_global}(a) with linear and cubic Lagrange interpolations. As expected, linear interpolation yields second-order accuracy, compromising the accuracy of the underlying scheme, whereas cubic interpolation provides global fourth-order accuracy. Fig. \ref{fig:lin_cu_global}(b) compares \(\left\|\varepsilon_{\rho}\right\|_{\infty}\) for various schemes, confirming $p+1$ order of accuracy for the $p-2p-p$ scheme, consistent with the theoretical result of \cite{gustafsson1975convergence}. 
The effect of grid refinement on error reduction is also shown in Fig.~\ref{fig:error_2D} through the density error contours at a non-dimensional time of 8 from $3-6-3$ scheme. Fig.~\ref{fig:error_2D}(a) shows the errors for Resolution 1, whereas Fig.~\ref{fig:error_2D}(b) shows the errors for Resolution 4; for a one-to-one comparison the same contour levels are used in both plots.\vspace{0.0cm}
\begin{table}
\centering
\resizebox{\textwidth}{!}{
\begin{tabular}{|c|c|c|c|c|}
    \hline
    \textbf{Parameters} 
        & \textbf{Resolution 1} 
        & \textbf{Resolution 2} 
        & \textbf{Resolution 3} 
        & \textbf{Resolution 4} \\
    \hline

    \begin{tabular}[c]{@{}c@{}}\textbf{Grid 1 (smaller,  \(N_{\mathrm{sg}}\times N_{\mathrm{sg}}\))} \\ \textbf{Grid 2 (background,  \(N_{\mathrm{bg}}\times N_{\mathrm{bg}}\))}\end{tabular}
    & \begin{tabular}[c]{@{}c@{}}$50 \times 50$\\$120 \times 120$\end{tabular}
    & \begin{tabular}[c]{@{}c@{}}$75 \times 75$\\$180 \times 180$\end{tabular}
    & \begin{tabular}[c]{@{}c@{}}$100 \times 100$\\$240 \times 240$\end{tabular}
    & \begin{tabular}[c]{@{}c@{}}$125 \times 125$\\$300 \times 300$\end{tabular} \\
    \hline
\end{tabular}
}
\caption{Various grid resolutions for the moving overset grid configuration shown in Fig.~\ref{fig:vort_initial}(a) for isentropic vortex convection tests.}
\label{tab:vortex_convection_2D}
\end{table}

\begin{figure}
    \centering
    \includegraphics[width=0.5\textwidth]{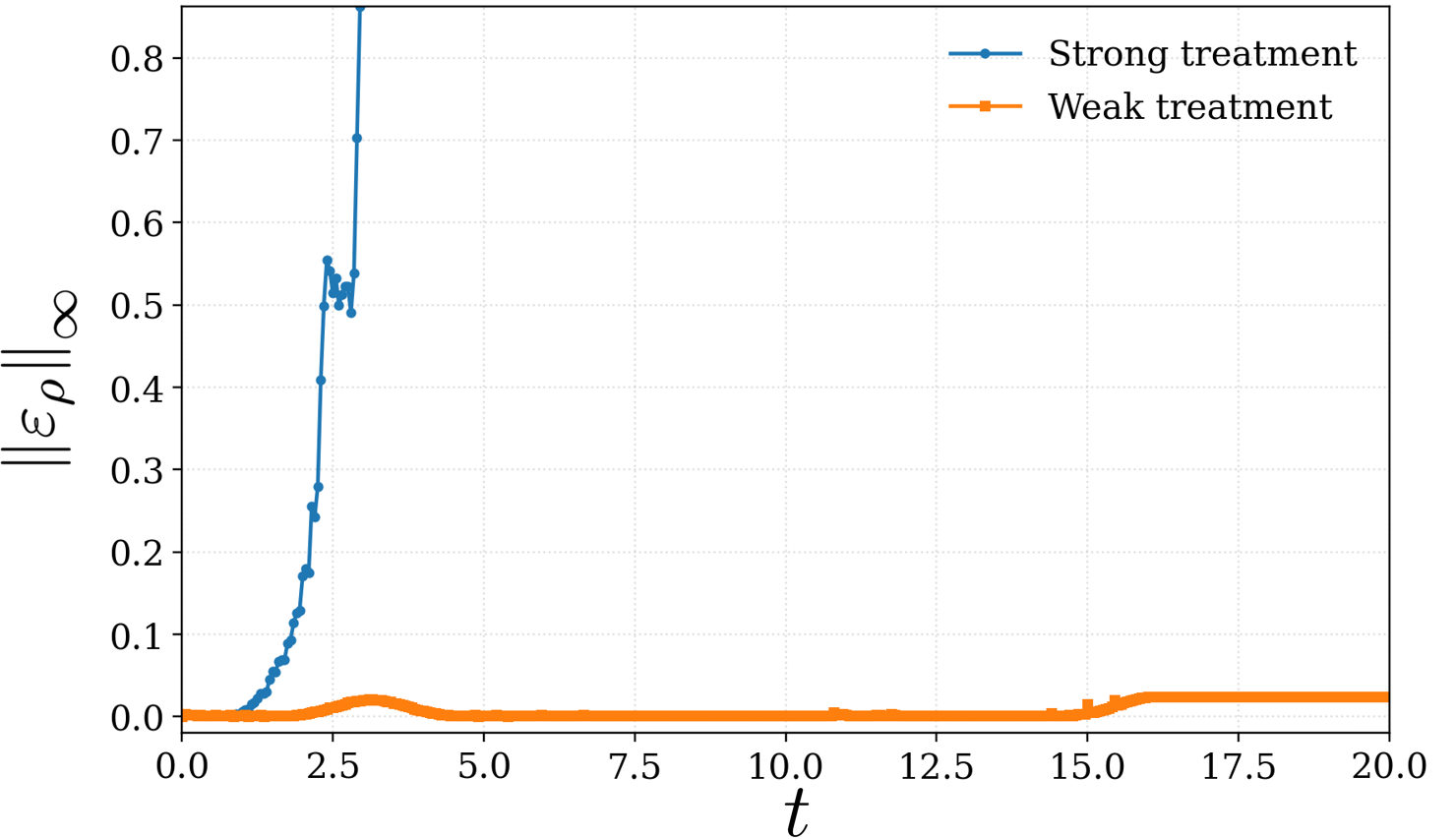}
    \caption{A comparison of \(\left\|\varepsilon_{\rho}\right\|_{\infty}\) from the strong and weak overset treatment applied to solve the 2-D vortex convection problem on a moving overset configuration shown in Fig. \ref{fig:vort_initial}.}
    \label{fig:error_time}
\end{figure}

To compare the performance of the two moving overset interface treatments, strong and weak imposition, as described in Sections~\ref{subsec:strong_treatment} and \ref{subsec:weak_treatment}, a time history of \(\left\|\varepsilon_{\rho}\right\|_{\infty}\) from the two treatments is evaluated. The commonly used strong treatment (\textit{e.g.}, \cite{Alabi2007,Brazell2016JCP}) injects the interpolated data, overwriting the solution at the receiver points, whereas the proposed weak treatment includes a characteristic decomposition and imposes the interpolated incoming characteristic variables weakly. Figure~\ref{fig:error_time} presents the \(\left\|\varepsilon_{\rho}\right\|_{\infty}\) time history for both approaches on the Resolution 4 grid with the grid motion described above. In the strong imposition run, the error increases steadily and then shoots up sharply, leading to a solution blow-up around a non-dimensional time of 2.6; the instability appears as an abrupt spike in the error curve. In contrast, the weak imposition remains stable over long durations for this moving overset inviscid problem, with the simulation continuing robustly even as the vortex interacts with the rotating smaller grid multiple times.

\subsection{2-D Pitching Airfoil Flow}\label{subsec:2d_pitching_airfoil}
\begin{figure}
\begin{centering}
\includegraphics[width=0.48\textwidth]{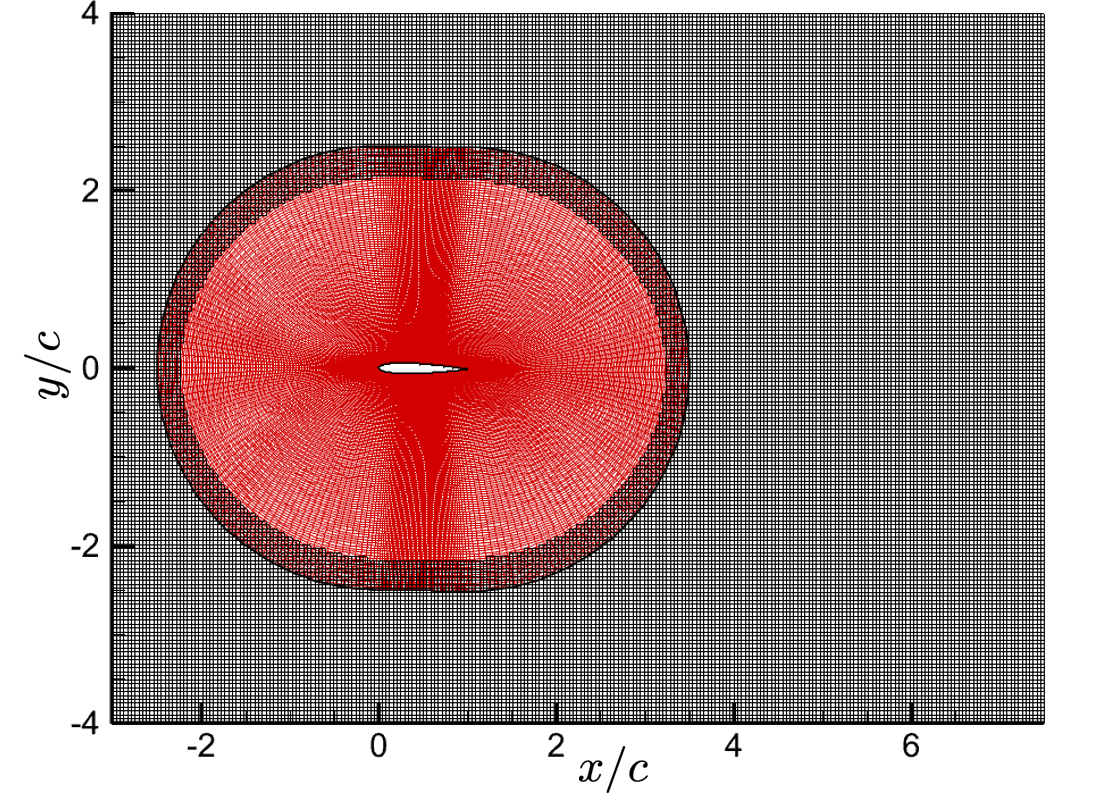}\includegraphics[width=0.47\textwidth]{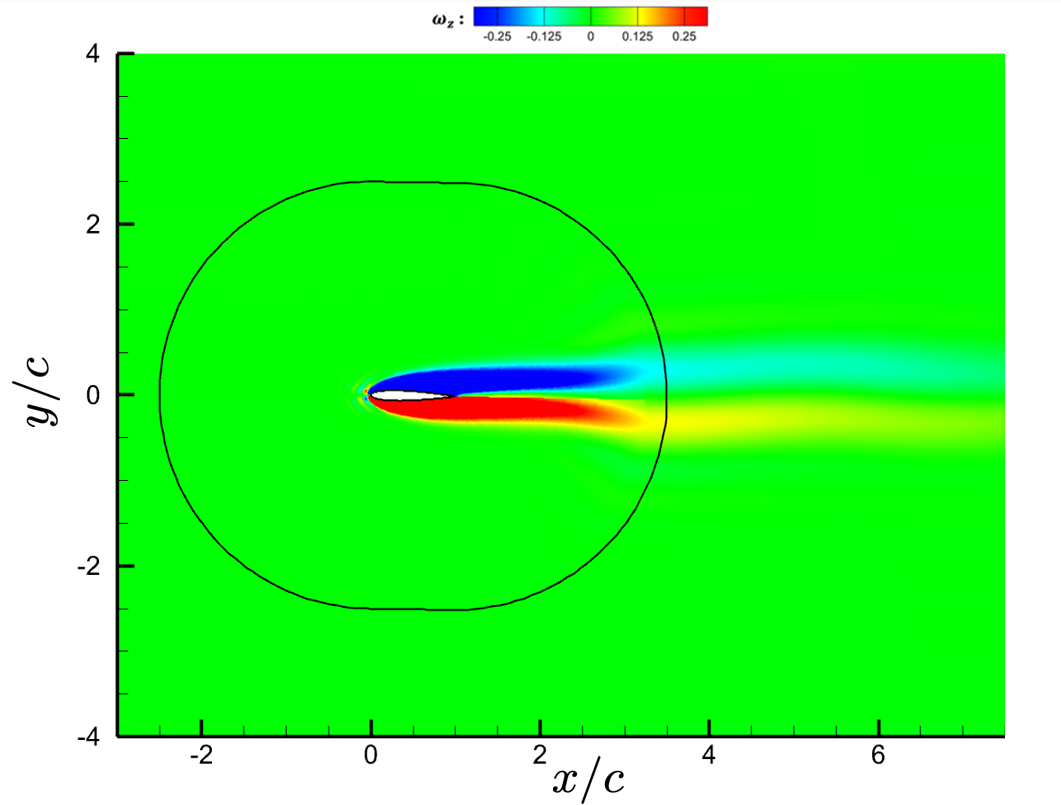}
\par\end{centering}
\begin{centering}
\qquad{}(a)\qquad{}\qquad{}\qquad{}\qquad{}\qquad{}\qquad{}\qquad{}\qquad{}\qquad{}\qquad{}\qquad{}(b)
\par\end{centering}
    \centering
\caption{2-D pitching NACA~0012 airfoil for $\alpha_0=0^\circ$: (a) moving overset grid configuration and (b) instantaneous $z$-vorticity field with contour levels in the range $\omega_z\in[-0.3,\,0.3]$.}
    \label{fig:2D_0}
\end{figure}

The curvilinear moving overset framework described in Section~\ref{subsec:Compressible Navier-Stokes equations} is validated by simulating the flow over a pitching 2-D NACA 0012 airfoil at a freestream Mach number $M_\infty=0.2$ and a Reynolds number (based on the airfoil chord length) $Re_c=1000$. A body-conforming O-grid is used around the airfoil; it overlaps with a Cartesian background grid, as shown in Fig. \ref{fig:2D_0}(a). The O-grid moves with respect to the background grid to model the pitching motion. This problem provides a rigorous evaluation of the proposed overset treatment for a viscous flow problem on moving grids. As a first step, a static airfoil configuration was validated against existing results by comparing the aerodynamic coefficients, as discussed in~\ref{sec:appdxC}.

\begin{figure}
    \centering
    \includegraphics[width=0.42\textwidth]{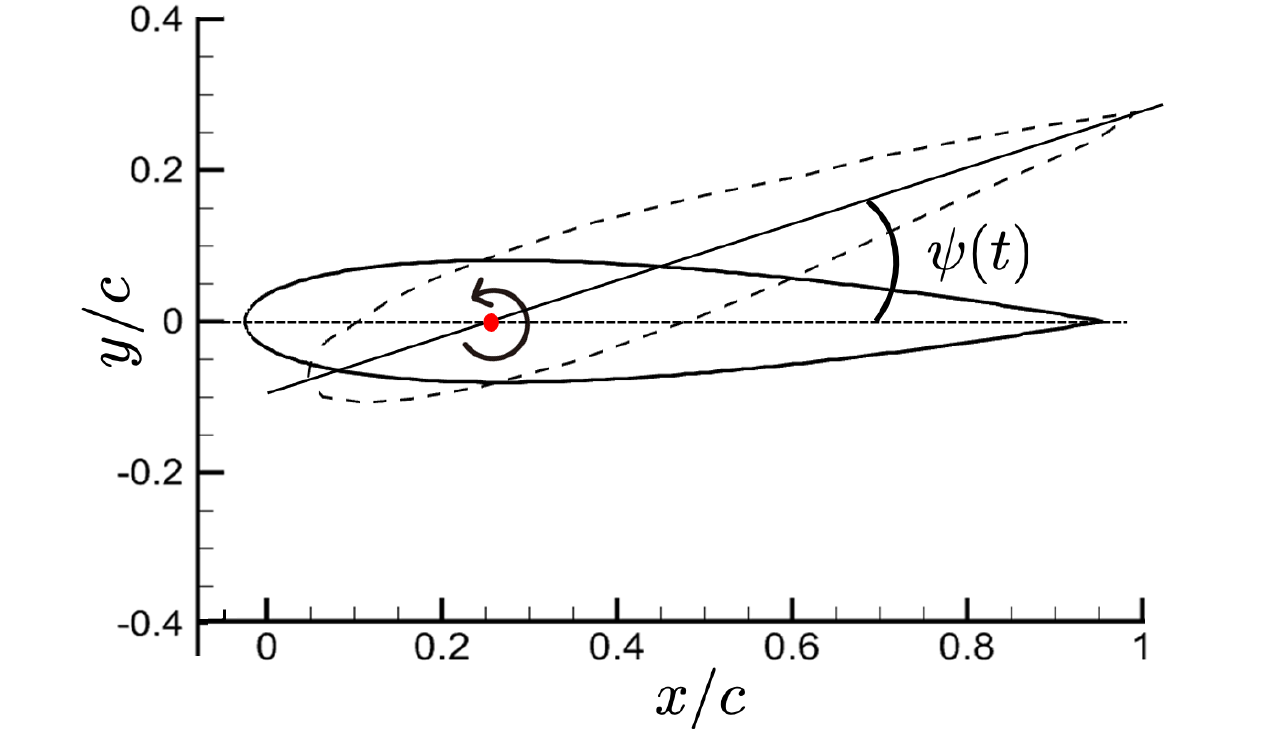}
    \caption{Schematic of the prescribed 2-D pitching motion of the NACA~0012 airfoil about the quarter-chord pivot, $(x_p,y_p)=(0.25c,0)$. The solid outline indicates the initial position for the mean pitch angle $\alpha_0=0^\circ$ and the dashed outline shows an instantaneous rotated position with respect to $\alpha_0$, given by grid rotation $\psi(t)=-[\alpha(t)-\alpha_0]$. 
    }
    \label{fig:pitch_schematics}
\end{figure}

For a rigid-body motion described by Eq.~\eqref{eq:rigid_map}, a 2-D pitching airfoil motion in the \(x\)--\(y\) plane corresponds to
pure rotation about the \(z\)-axis, as depicted in Fig.~\ref{fig:pitch_schematics}. Therefore, the motion is represented in
\eqref{eq:grid_vel} by setting \(\dot{x}_p(t)=\dot{y}_p(t)=\dot{z}_p(t)=0\) and \(\phi(t)=\theta(t)=0\), allowing rotation only about the \(z\)-axis by the angle \(\psi(t)\).
Following \cite{kurtulus2019unsteady}, the pitch angle \(\alpha(t)\) is defined positive nose-up, which, for a free stream in the \(+x\) direction, corresponds to a clockwise rotation when viewed from the positive \(z\)-axis toward the origin. It is prescribed as
\begin{equation}
\alpha(t)=\alpha_0+\alpha_A\sin(2\pi f t),
\label{eq:naca_pitch}
\end{equation}
where \(\alpha_0=\alpha(0)\) is the mean pitch angle, \(\alpha_A\) is the pitching amplitude, and \(f\) is the pitching frequency; angles are converted to radians before use. The O-grid is generated at the initial pitch angle \(\alpha(0)=\alpha_0\), so that \(\mathbf{x}_0\) in Eq.~\eqref{eq:rigid_map} is the grid at \(t_0=0\). Because \(\psi\) in Eq.~\eqref{eq:R_full} is positive counterclockwise, the rotation angle measured from this initial configuration is
\begin{equation}
\psi(t)=-\left[\alpha(t)-\alpha(0)\right]=-\alpha_A\sin(2\pi f t),
\qquad
\dot{\psi}(t)=-2\pi f\,\alpha_A\cos(2\pi f t),
\label{eq:naca_psi}
\end{equation}
which gives \(\psi(0)=0\) and hence \(\mathbf{R}(\phi(0),\theta(0),\psi(0))=\mathbf{I}\), consistent with Eq.~\eqref{eq:rigid_map} at \(t=t_0\). The airfoil rotates about a pivot located at the
quarter-chord,
$(x_p,y_p)=(x_{p,0},y_{p,0})=(0.25c,0)$, so that Eq.~\eqref{eq:rigid_map} reduces (in 2-D) to a rotation about $(x_p,y_p)$. The grid-velocity components, given by Eq.~\eqref{eq:grid_vel}, then simplifies to
\begin{equation}
x_t=-\dot{\psi}\left(y-y_p\right),\qquad
y_t=\dot{\psi}\left(x-x_p\right),\qquad
z_t=0.
\label{eq:naca_gridvel}
\end{equation}
These grid-velocity components are incorporated into the governing equations using Eq.~\eqref{eq:metric_time}, which for a 2-D motion simplifies to Eq.~\eqref{eq:bench_timemetrics}.
For validation, the pitching amplitude and frequency are chosen to match~\cite{kurtulus2019unsteady}: \(\alpha_A=\;1^\circ\) and \(f=0.20\) (based on \(a_\infty^*\) and \(c^*\); equivalently \(f^*c^*/U_\infty^*=1.00\)). Two cases are considered with $\alpha_0=0^\circ$ and $\alpha_0=11^\circ$.

\begin{figure}
\begin{centering}
\includegraphics[width=0.48\textwidth]{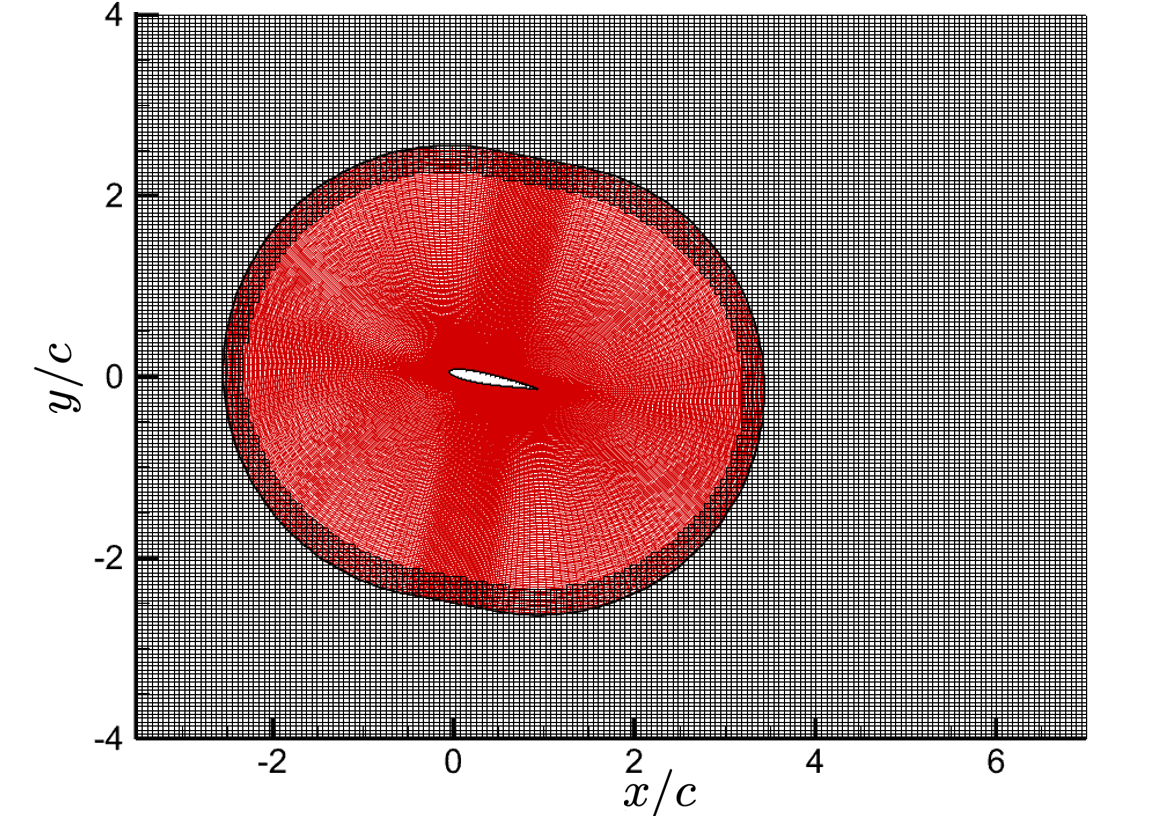}\includegraphics[width=0.47\textwidth]{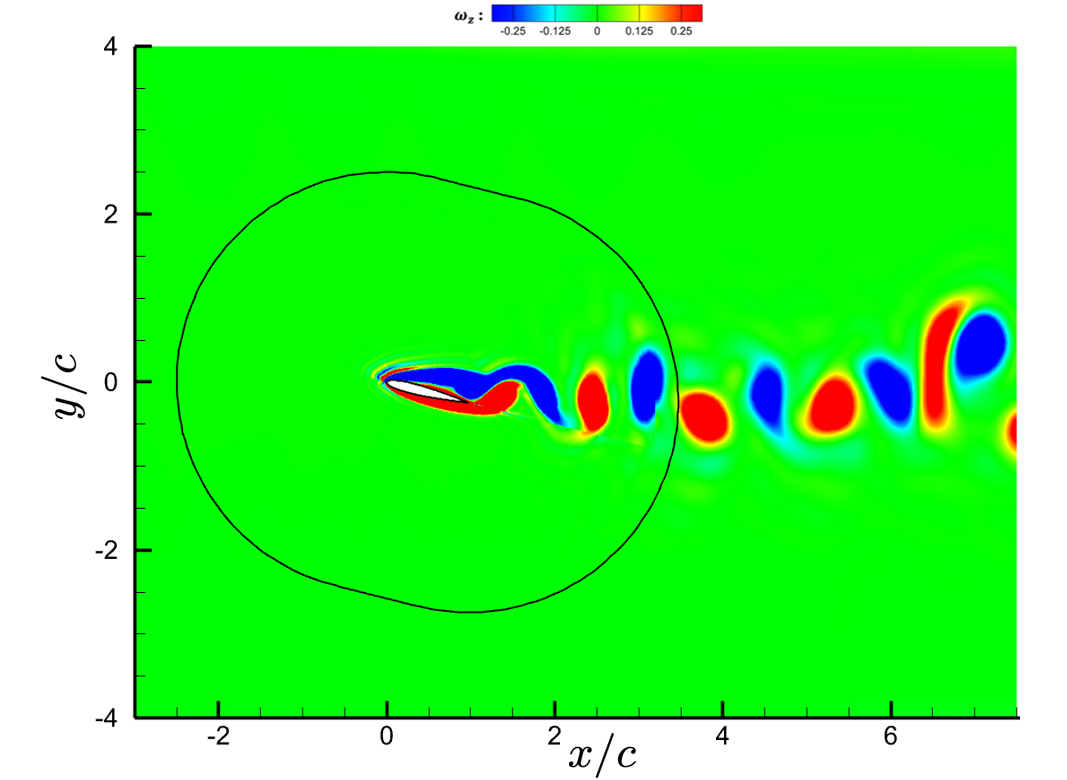}
\par\end{centering}
\begin{centering}
\qquad{}(a)\qquad{}\qquad{}\qquad{}\qquad{}\qquad{}\qquad{}\qquad{}\qquad{}\qquad{}\qquad{}\qquad{}(b)
\par\end{centering}
    \centering
\caption{2-D pitching NACA~0012 airfoil for $\alpha_0=11^\circ$: (a) moving overset grid configuration and (b) instantaneous $z$-vorticity field with contour levels in the range $\omega_z\in[-0.3,\,0.3]$.}
    \label{fig:2D_15}
\end{figure}

\begin{figure}
\begin{centering}
\includegraphics[width=0.485\textwidth]{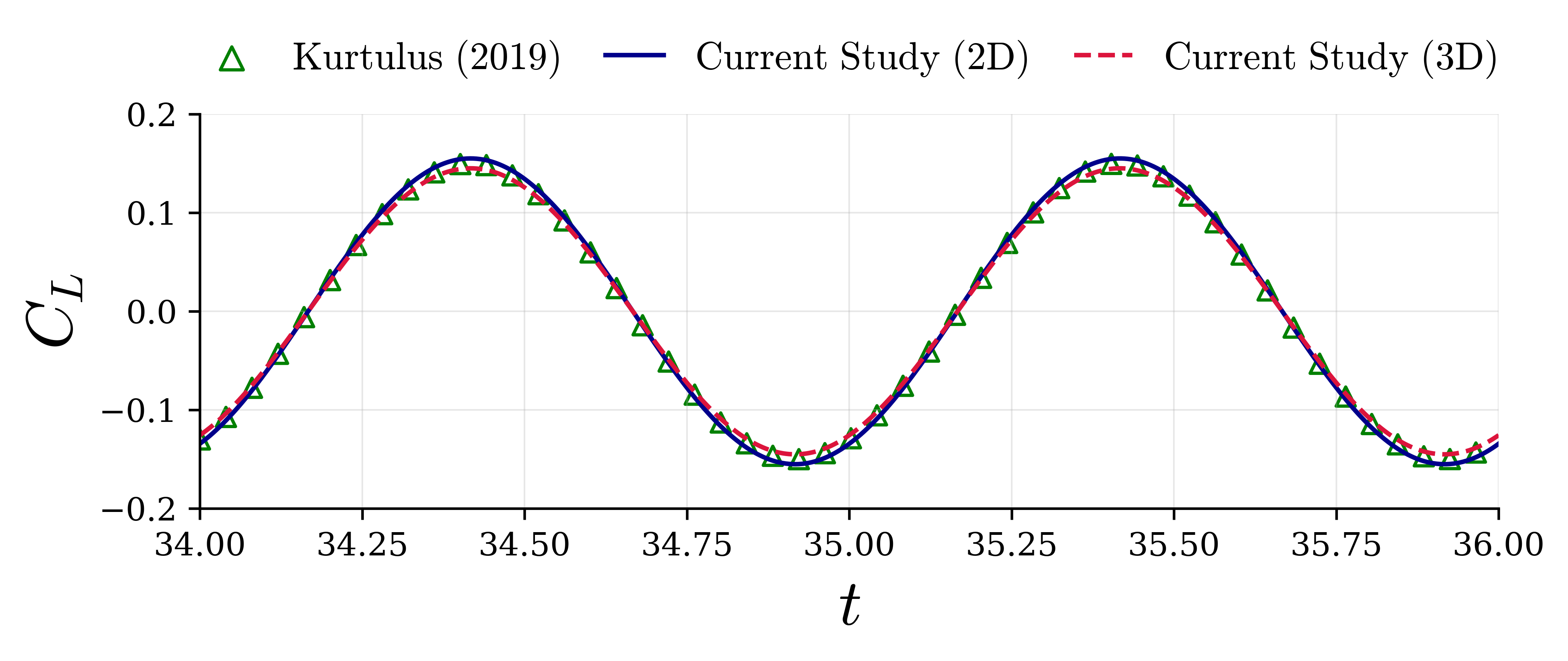}\includegraphics[width=0.52\textwidth]{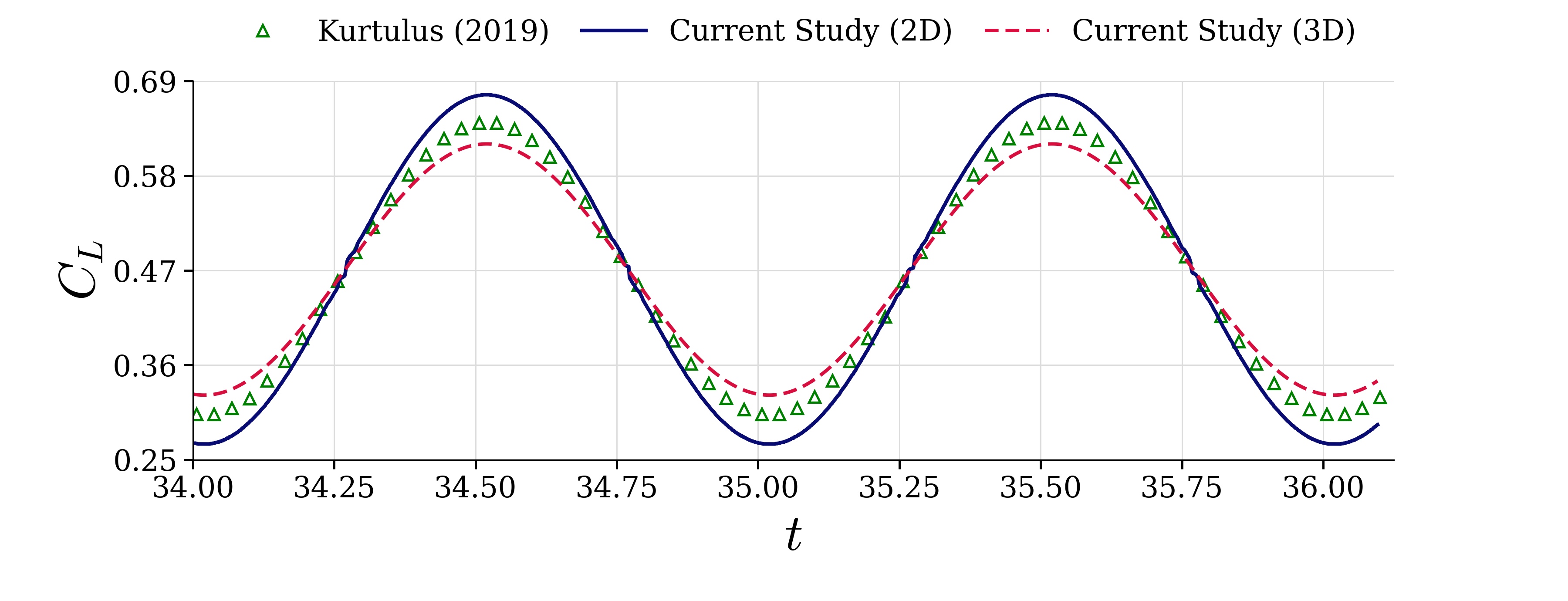}
\par\end{centering}
\begin{centering}
\qquad{}(a)\qquad{}\qquad{}\qquad{}\qquad{}\qquad{}\qquad{}\qquad{}\qquad{}\qquad{}\qquad{}\qquad{}(b)
\par\end{centering}
    \centering
    \caption{Validation of the lift coefficient $C_L$ for the (a) $\alpha_0=0^\circ$ and (b) for $\alpha_0=11^\circ$ pitching airfoil/wing flow cases against the results of \cite{kurtulus2019unsteady}. The time axis is renormalized using the freestream velocity and airfoil chord length to facilitate direct comparison with the reference data.}
    \label{fig:validation_airfoil_cl}
\end{figure}

The overset grid setup and $z$-vorticity after few cycles of motion for the \(\alpha_0 = 0^\circ\) and \(\alpha_0 = 11^\circ\) flow cases are shown in Figs.~\ref{fig:2D_0} and \ref{fig:2D_15}, respectively. The lift coefficients \(C_L\) for the two cases are compared with the reference data in Fig.~\ref{fig:validation_airfoil_cl}, which shows good agreement. 
The simulation is run for a sufficiently long duration to ensure that the unsteady aerodynamic response reached a time-periodic state after the initial transient effects. The favorable agreement with the reference data in Fig.~\ref{fig:validation_airfoil_cl} and the long time simulations confirm the accuracy and the numerical robustness of the proposed moving overset treatment.

\subsection{3-D Pitching Wing Flow}
\begin{figure}
\begin{centering}
\includegraphics[width=0.47\textwidth]{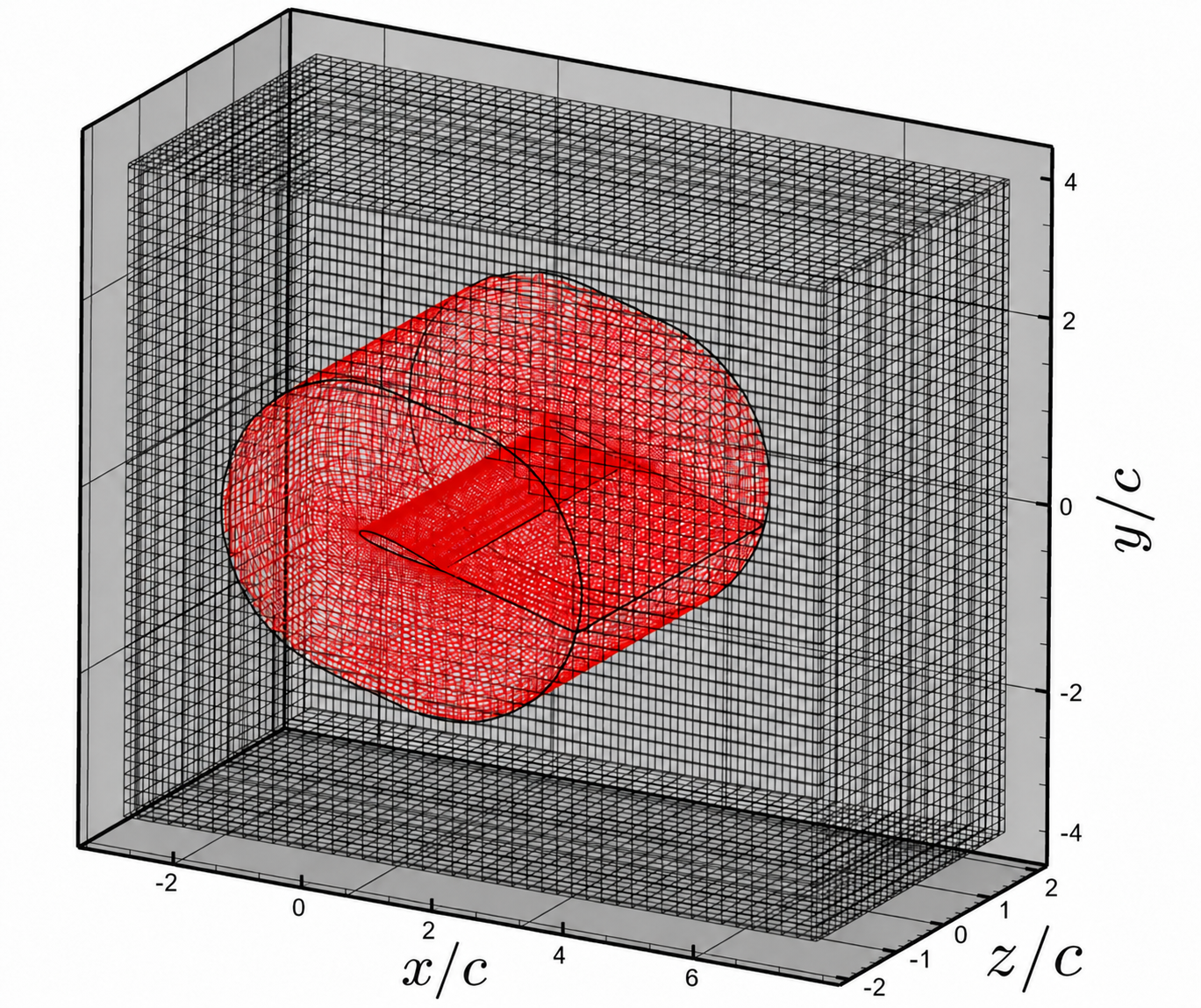}\includegraphics[width=0.52\textwidth]{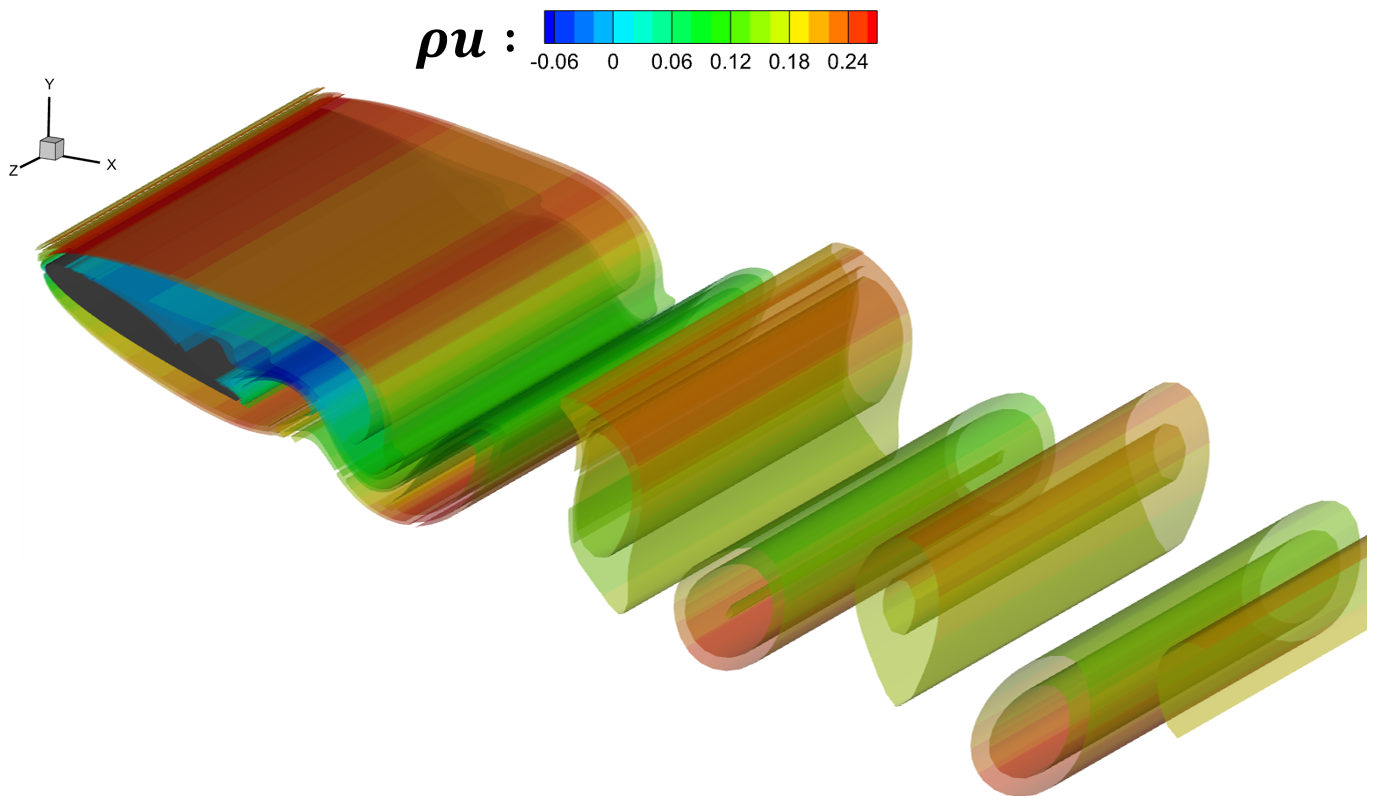}
\par\end{centering}
\begin{centering}
\qquad{}(a)\qquad{}\qquad{}\qquad{}\qquad{}\qquad{}\qquad{}\qquad{}\qquad{}\qquad{}\qquad{}\qquad{}(b)
\par\end{centering}
    \centering
\caption{Flow past a pitching 3-D NACA~0012 wing for $\alpha_0=11^\circ$: (a) moving overset grid configuration and (b) iso-surfaces of vorticity magnitude at $|\boldsymbol{\omega}|=0.3,\ 0.5,\ 1.0$, colored by $x$-momentum.}
    \label{fig:3D_15}
\end{figure}

To verify the consistency of the three-dimensional implementation, pitching wing simulations are conducted for $\alpha_0=0^\circ$ and $\alpha_0=11^\circ$, extending the pitching airfoil configurations considered in the previous section to a spanwise-periodic domain. Since geometry, imposed motion, and initial conditions are uniform in the spanwise direction, and the low Reynolds number $Re_c=1000$ does not promote spanwise instability, the solution is expected to remain nearly two-dimensional. This case therefore serves as a consistency check, confirming that the three-dimensional solver reproduces the corresponding two-dimensional result on a three-dimensional mesh, rather than as a test of genuinely three-dimensional interface behavior, which is assessed with the finite-span wings in Sections~\ref{subsec:plunging_finite_wing} and~\ref{subsec:pitching_plunging_finite_wing}.
The computational grid consists of a body-fitted
O-grid of $250\times350\times200$ points in the
radial, circumferential, and spanwise directions,
respectively, and a background grid of
$400\times400\times200$ points in the $x$, $y$,
and $z$ directions. The spanwise domain length
is $L_z/c=4$. The lift-coefficient time history, $C_L(t)$, is compared with the same reference study \cite{kurtulus2019unsteady}. By enforcing periodic boundary conditions in the spanwise direction, the 3-D configuration approximates an infinite wing and removes finite span/tip effects. As shown in Fig.~\ref{fig:3D_15}, the spanwise periodic configuration yields wake dynamics that for a low $Re_c$ of 1000 considered here 
can be directly compared with the 2-D pitching airfoil case. 

The close match of the lift-coefficient time history, $C_L(t)$, in Fig.~\ref{fig:validation_airfoil_cl} with the reference data demonstrates that the proposed approach accurately preserves the phase and amplitude of the unsteady aerodynamic response during grid motion in three dimensions. The close agreement between the two- and three-dimensional lift histories is consistent with the nearly two-dimensional character of this spanwise-periodic flow. Unlike strong overset interface treatments, stable long-time force predictions are obtained without introducing any filtering or numerical dissipation at the overset interface.

\subsection{2-D Forced Oscillation of a Circular Cylinder}\label{subsec:2-D Forced Oscillation of a Circular Cylinder}
The curvilinear moving-overset treatment discussed in Section \ref{subsec:Compressible Navier-Stokes equations} is applied here to study the flow past a circular cylinder undergoing prescribed transverse (heaving) oscillations. The Reynolds number based on the cylinder diameter ($D$) is set to $Re_D=185$, the oscillation amplitude to $A_o/D=0.2$, and the oscillation frequency to $f_o/f_{vs}=1.0$, where $f_{vs}$ denotes the vortex shedding frequency for a stationary circular cylinder. The non-dimensional frequency $f_o$, based on the reference velocity scale ($=a_{\infty}^{*}$, the ambient speed of sound) and the reference length scale ($=D^{*}$, the cylinder diameter), where the superscript $*$ denotes a dimensional variable, as discussed in Section \ref{subsec:Compressible Navier-Stokes equations}, is given by
\begin{equation}
f_o = \frac{f^*_oD^*}{a^*_{\infty}}=\left(\frac{f^*_oD^*}{U^*_{\infty}}\right)\left(\frac{U^*_{\infty}}{a^*_{\infty}}\right)=St\,M_\infty,
\label{eq:shedding_frequency}
\end{equation}
where the Strouhal number, \(St\), for a stationary circular cylinder is $\sim 0.2$ and the free-stream Mach number considered here is \(M_\infty=0.2\) yielding $f_o=0.04$. 
These values are chosen to match the configuration reported in \cite{GUILMINEAU2002773}. 

\begin{figure}
    \centering
    \includegraphics[width=1.0\textwidth]{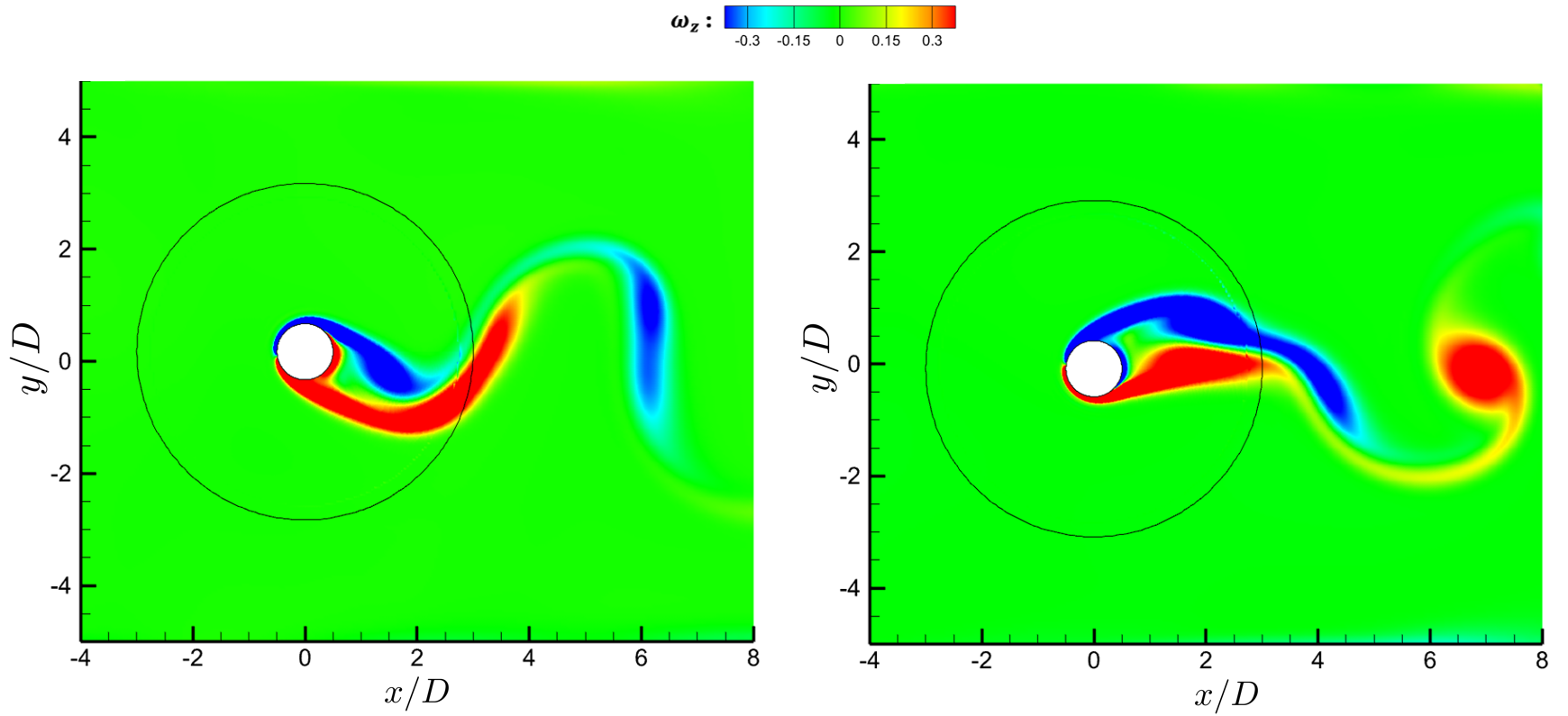}
    \caption{Instantaneous $z$-vorticity contours at two different time instants in the flow past a 2-D heaving circular cylinder.}
    \label{fig:heave_1}
\end{figure}

This two-dimensional case corresponds to pure translation in the transverse ($y$) direction with no rotation, therefore, \(x_p(t)=z_p(t)=0\) and $\phi(t)=\theta(t)=\psi(t)=0$ in Eq.~\eqref{eq:rigid_map}. The prescribed heaving motion is
\begin{equation}
y_p(t)=A_o\sin\!\left(2\pi f_o t\right).
\label{eq:heave_motion}
\end{equation}
The grid velocity components in Eq.~\eqref{eq:grid_vel} are then given by
\begin{equation}
x_t=0,\qquad
y_t=\dot y_p=2\pi f_o A_o \cos\!\left(2\pi f_o t\right),\qquad
z_t=0.
\label{eq:heave_gridvel}
\end{equation}
The grid-velocity components are incorporated into the curvilinear governing equations \eqref{eq:ns} through the time-metric derivatives defined in Eq.~\eqref{eq:metric_time}. Accordingly, the moving overset grid time metrics for the present
case are evaluated as
\begin{equation}
\xi_t=-\left(y_t\,\xi_y\right),\qquad
\eta_t=-\left(y_t\,\eta_y\right),\qquad
\zeta_t=0.
\label{eq:heave_timemetrics_2d}
\end{equation}
Figure~\ref{fig:heave_1} illustrates two cylinder positions during the motion and the associated $z$-vorticity contours. The corresponding lift-coefficient time series, $C_L(t)$, is provided in Fig.~\ref{fig:ratio_1}. The lift-coefficient time series $C_L(t)$ obtained using the present framework agrees fairly well with the reference data from \cite{GUILMINEAU2002773}.

\subsection{3-D Forced Oscillation of a Circular Cylinder}
\begin{figure}
    \centering
    \includegraphics[width=0.9\textwidth]{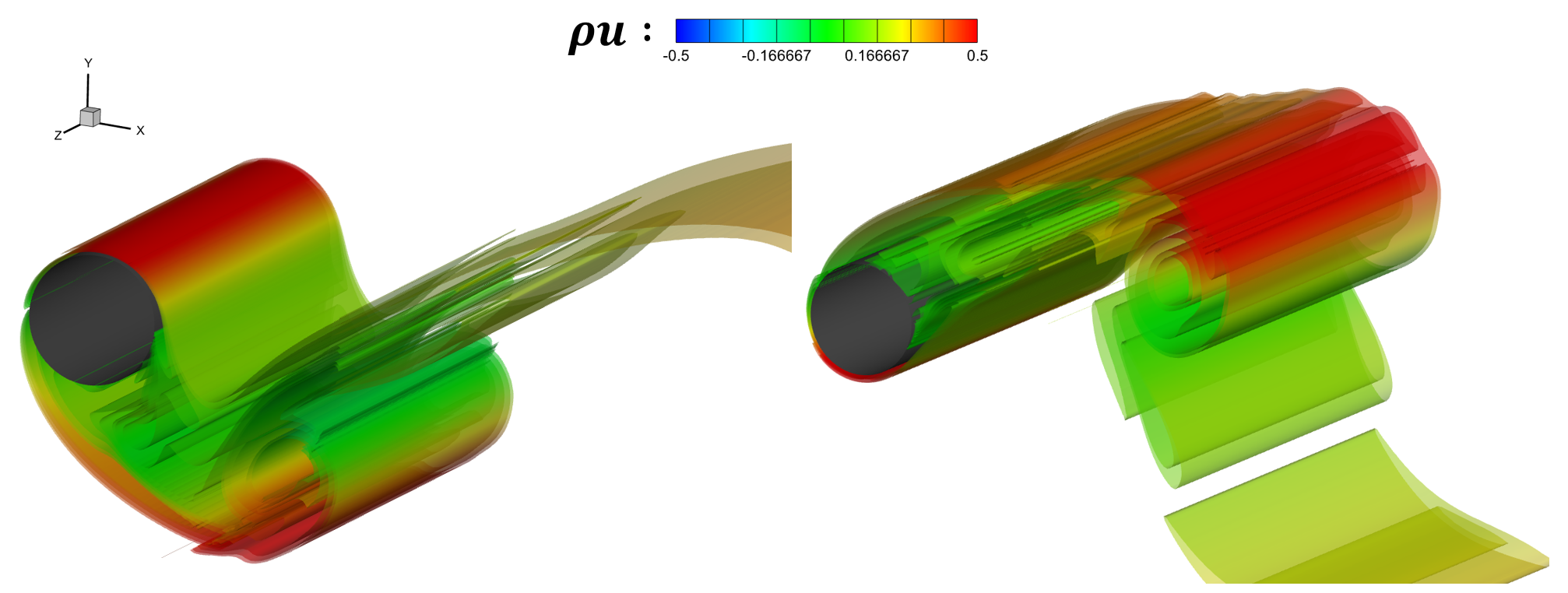}
    \caption{3-D wake visualization for a transversely oscillating (heaving) circular cylinder at $Re_D=185$ and $f_o/f_{vs}=1.0$. Iso-surfaces of vorticity magnitude, $|\boldsymbol{\omega}|=0.3,\ 0.5,\ 1.0$, are shown and colored by $x$-momentum.}
    \label{fig:3D_heave}
\end{figure}

\begin{figure}
    \centering
    \includegraphics[width=0.5\textwidth]{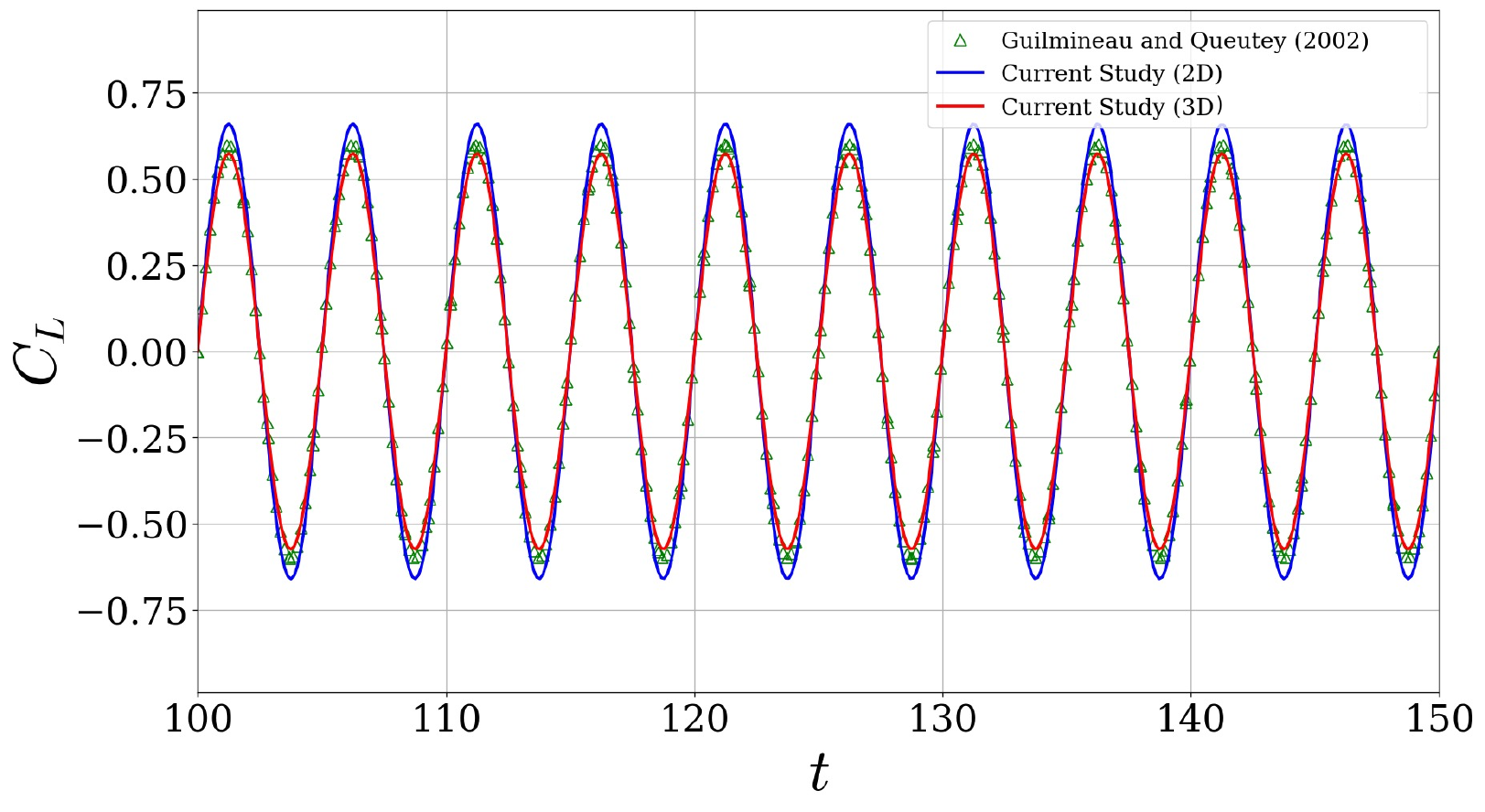}
    \caption{Lift coefficient ($C_L$) response of a heaving circular cylinder at $f_o/f_{vs} = 1.0$. Time is renormalized using the freestream velocity and cylinder diameter for consistency with the reference study \cite{GUILMINEAU2002773}.}
    \label{fig:ratio_1}
\end{figure}

As for the spanwise-periodic pitching wing, this configuration has spanwise-uniform geometry, motion, and initial conditions, and is used as a consistency check of the three-dimensional implementation against the corresponding two-dimensional case. Genuinely three-dimensional effects, including tip flow, non-aligned curved interfaces between multiple overlapping grids, and spanwise variations of the wake, are examined in Sections~\ref{subsec:plunging_finite_wing} and~\ref{subsec:pitching_plunging_finite_wing}.
The computational grid consists of a body-fitted
O-grid containing $250\times350\times200$ points in the
radial, circumferential, and spanwise directions,
respectively, and a background grid of
$400\times400\times200$ points in the $x$, $y$,
and $z$ directions. The spanwise domain length
is $L_z/D=4$.
The Reynolds number, oscillation amplitude, and frequency are chosen to be the same as in the previous section to allow direct comparisons with the results of \cite{GUILMINEAU2002773}. At a low Reynolds number, $Re_D=185$, the spanwise inhomogeneity is expected to be minimal, enabling  comparison with the 2-D results discussed in Section ~\ref{subsec:2-D Forced Oscillation of a Circular Cylinder}. The wake dynamics from the 3-D simulations is shown in Fig.~\ref{fig:3D_heave} using isosurfaces of vorticity magnitude, colored by $x$-momentum.
To eliminate initial transients from the starting of the cylinder motion, the simulation was advanced for many cycles until the lift response, $C_L$, reached a time-periodic behavior. The $C_L$ time series is compared with the results of \cite{GUILMINEAU2002773} in Fig.~\ref{fig:ratio_1}. For the spanwise-periodic 3-D configuration, the lift profile is nearly indistinguishable from the two-dimensional results, consistent with the expected spanwise-homogeneous dynamics at low $Re_D$ and in the absence of end effects.

The favorable agreement with the reference data demonstrates the accuracy of the proposed moving overset treatment for viscous problems. Moreover, the preservation of coherent vortical structures across the moving overset interface demonstrates the ability of the non-dissipative scheme, coupled with the weak interface treatment, to resolve unsteady flow features with minimal numerical degradation.
\subsection{2-D Rotating Circular Cylinder}\label{subsec:2-D Rotating Circular Cylinder}
The curvilinear moving overset framework is further applied to simulate flow past a circular cylinder rotating at a constant angular velocity in a uniform free stream. Following the standard definition for this configuration, the Reynolds number is given by $Re_D=U^*_{\infty}D^*/\nu^*$, where $D^*$ is the cylinder diameter, $U^*_{\infty}$ is the free-stream velocity, and $\nu^*$ is the kinematic viscosity. The superscript * denotes dimensional variables. The rotation rate is defined as $\alpha=D^*\omega^*/(2U^*_{\infty})$, where $\omega^*$ denotes the angular velocity of the cylinder about its axis. For this test case, the Reynolds number is set to $Re_D=200$, and the non-dimensional rotation rate is prescribed as $\alpha=1$. The flow conditions and non-dimensional parameters are matched to the reference study~\cite{MITTAL_KUMAR_2003}, enabling a direct comparison of the lift-coefficient ($C_L$) time history. This flow problem assesses the capability of the present moving overset treatment in handling large rigid-body rotational motions.

\begin{figure}[H]
\begin{centering}
\includegraphics[width=0.4\textwidth]{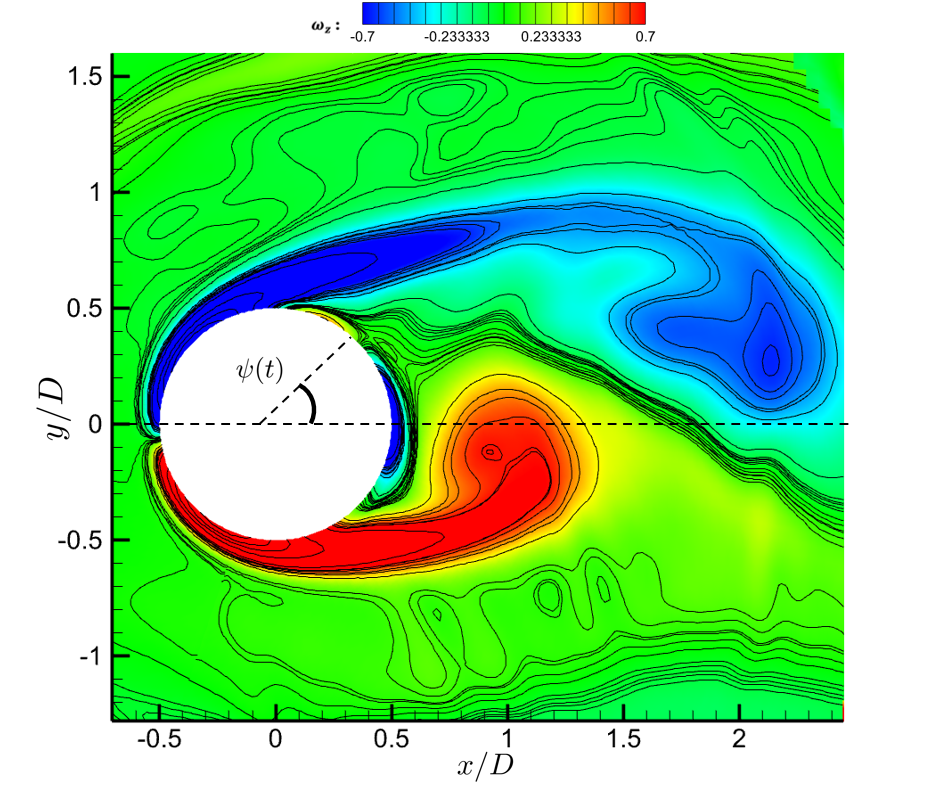}\includegraphics[width=0.57\textwidth]{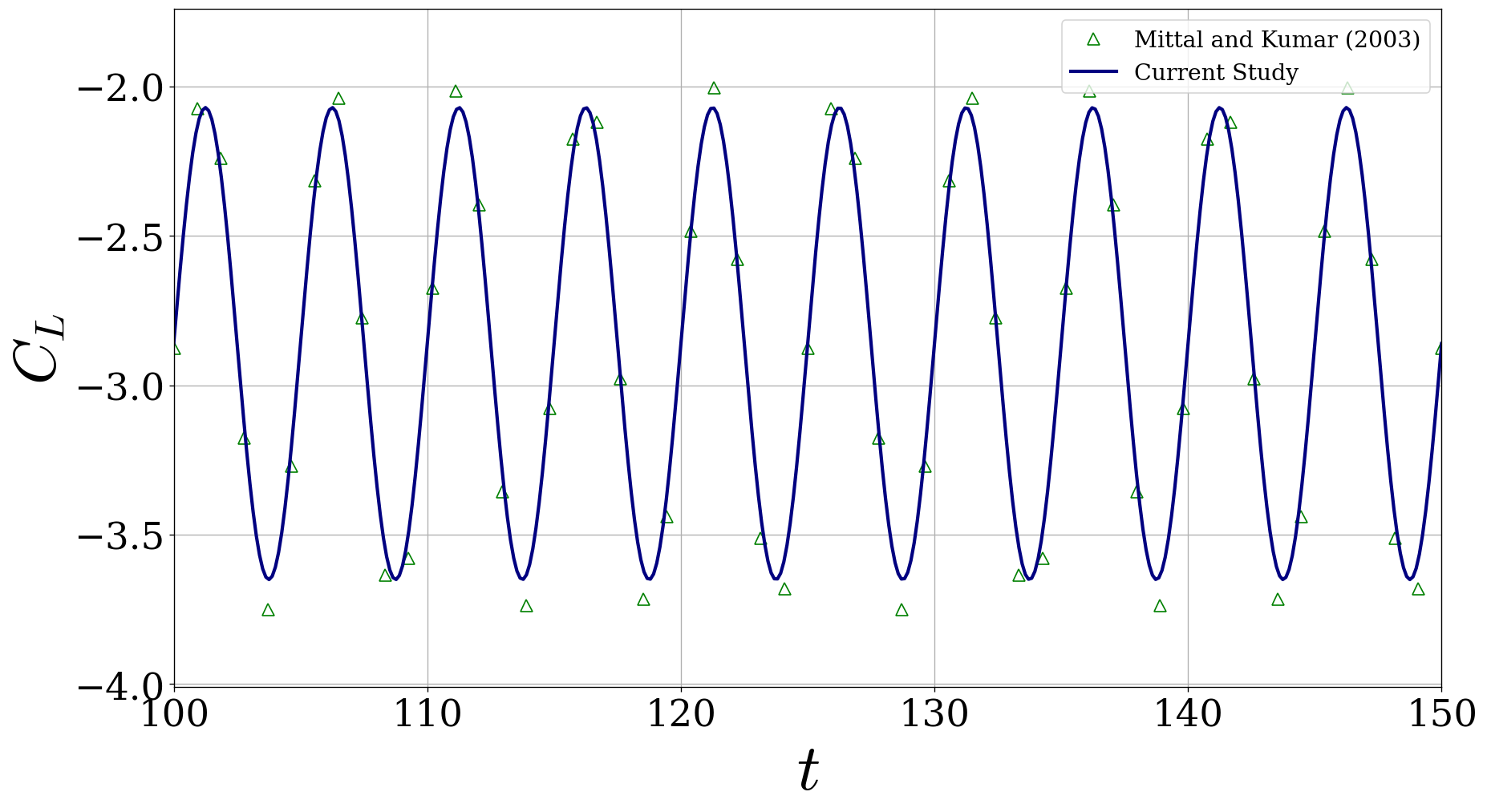}
\par\end{centering}
\begin{centering}
\qquad{}(a)\qquad{}\qquad{}\qquad{}\qquad{}\qquad{}\qquad{}\qquad{}\qquad{}\qquad{}\qquad{}\qquad{}(b)
\par\end{centering}
    \centering
\caption{Flow past a rotating circular cylinder at $Re_D=200$: (a) zoomed in instantaneous $z$-vorticity contours around the overset interface illustrating the asymmetric wake induced by cylinder rotation; (b) time history of the lift coefficient $C_L$ compared with reference results. The time axis in subfigure (b) is renormalized using the freestream velocity and cylinder diameter to facilitate comparison with \cite{MITTAL_KUMAR_2003}.}
    \label{fig:rotating_cyl}
\end{figure}

Here, the grid motion is restricted to a rigid-body rotation about the $z$-axis with no translational motion. Therefore, in Eq. \eqref{eq:rigid_map}, \(x_p(t)=y_p(t)=z_p(t)=0\) and $\phi(t)=\theta(t)=0$. The non-dimensional rotation rate $\alpha=D^*\omega^*/(2U^*_{\infty})$ is defined using $U^*_{\infty}$ for validation against the results of~\cite{MITTAL_KUMAR_2003}; however, as discussed in Section \ref{subsec:Compressible Navier-Stokes equations}, the present code implementation uses the ambient speed of sound, $a_{\infty}^{*}$, as the velocity scale and the cylinder diameter, $D^*$, as the length scale. Therefore, for the calculation of the grid velocity components in Eq. \eqref{eq:grid_vel}, which reduces to Eq.~\eqref{eq:naca_gridvel} for this pure rotation case,
\begin{equation}
\dot{\psi}=\alpha_{\text{code}}=\frac{D^*\omega^*}{a^*_{\infty}}=\left(\frac{D^*\omega^*}{2U^*_{\infty}}\right)\left(\frac{2U^*_{\infty}}{a^*_{\infty}}\right)=2\alpha\,M_{\infty}.
\end{equation}
For a constant rotation rate, assuming $\psi(t=0)=0$, see Fig.~\ref{fig:rotating_cyl},
\begin{equation}
\psi(t)=2\alpha\,M_{\infty}\,t,
\end{equation}
where the freestream Mach number, $M_{\infty}$, is chosen as 0.2. The grid velocity components are then given by Eq.~\eqref{eq:naca_gridvel} with time metric terms defined by Eq.~\eqref{eq:bench_timemetrics}.
Figure~\ref{fig:rotating_cyl} compares the lift-coefficient time history, $C_L$, with the reference results of \cite{MITTAL_KUMAR_2003}.
The favorable match between the computed and reference results demonstrates the accuracy of the proposed framework in capturing the unsteady force response for rotating-body flow. The simulations in this and the previous sections are performed using only one layer of fringe points, indicating that the weak overset interface treatment produces accurate and stable results without additional fringe layers.

\subsection{3-D Plunging Finite Wing}
\label{subsec:plunging_finite_wing}

A rectangular finite wing with a NACA 0012 cross-section
undergoing pure plunging motion is considered to test the
moving overset framework for three-dimensional tip flow
and spanwise variations absent from the preceding
spanwise-uniform cases.
The configuration follows the pure-plunging case of
~\cite{doi:10.2514/6.2011-1316}, whose aerodynamic results provide
the reference for comparison. The reference configuration has
a root-to-tip span of four chord lengths, with a chord-based
Reynolds number $Re_c=2000$ and freestream Mach number
$M_\infty=0.1$. The computational domain extends over $x/c\in[-5,16]$,
$y/c\in[-10,10]$, and $z/c\in[-11,11]$, with the origin
at the midspan leading edge of the unpitched wing at
zero plunge displacement.
The domain is discretized using four overlapping grids:
a body-fitted O-grid of $200\times300\times200$ points,
two wing-tip cap grids of $100\times150\times150$ points each,
and a background grid of $200\times300\times300$ points.
The four grids contain a total of $34.5$ million points.

\begin{figure}[htbp]
    \centering
    \begin{minipage}[t]{0.32\textwidth}
        \centering
        \includegraphics[
            width=\linewidth,
            trim=2pt 2pt 2pt 2pt,
            clip
        ]{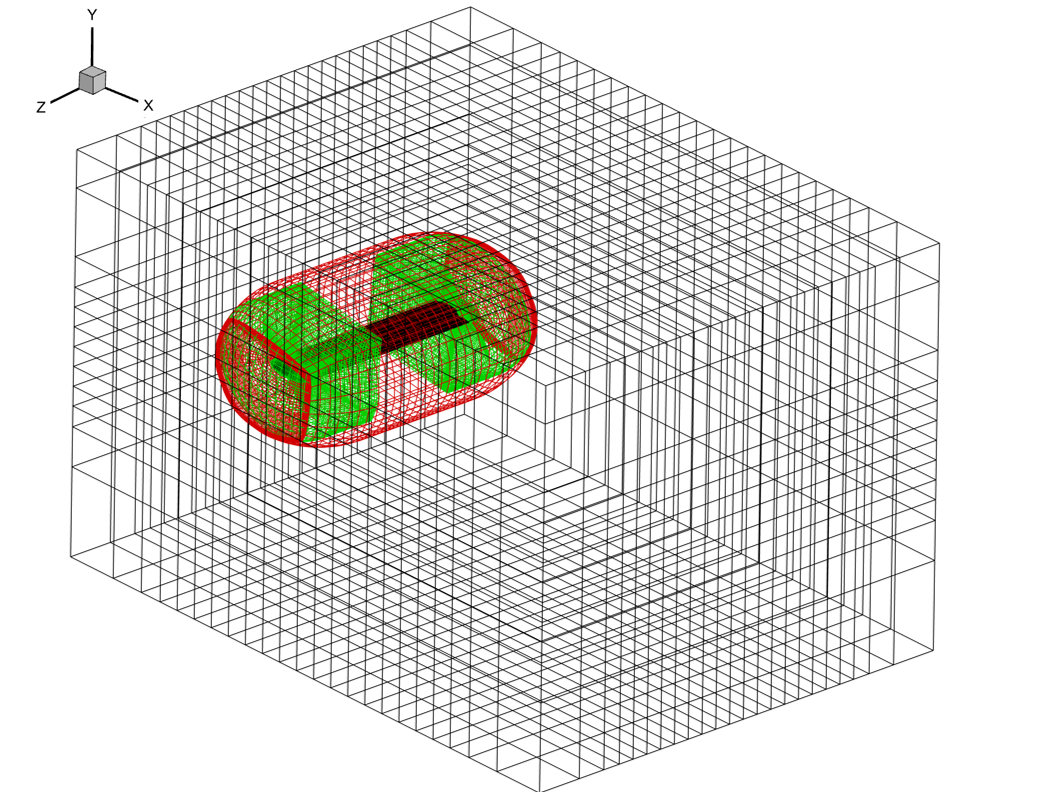}
        \par\smallskip
        (a)
    \end{minipage}
    \hfill
    \begin{minipage}[t]{0.32\textwidth}
        \centering
        \includegraphics[
            width=\linewidth,
            trim=22pt 20pt 20pt 20pt,
            clip
        ]{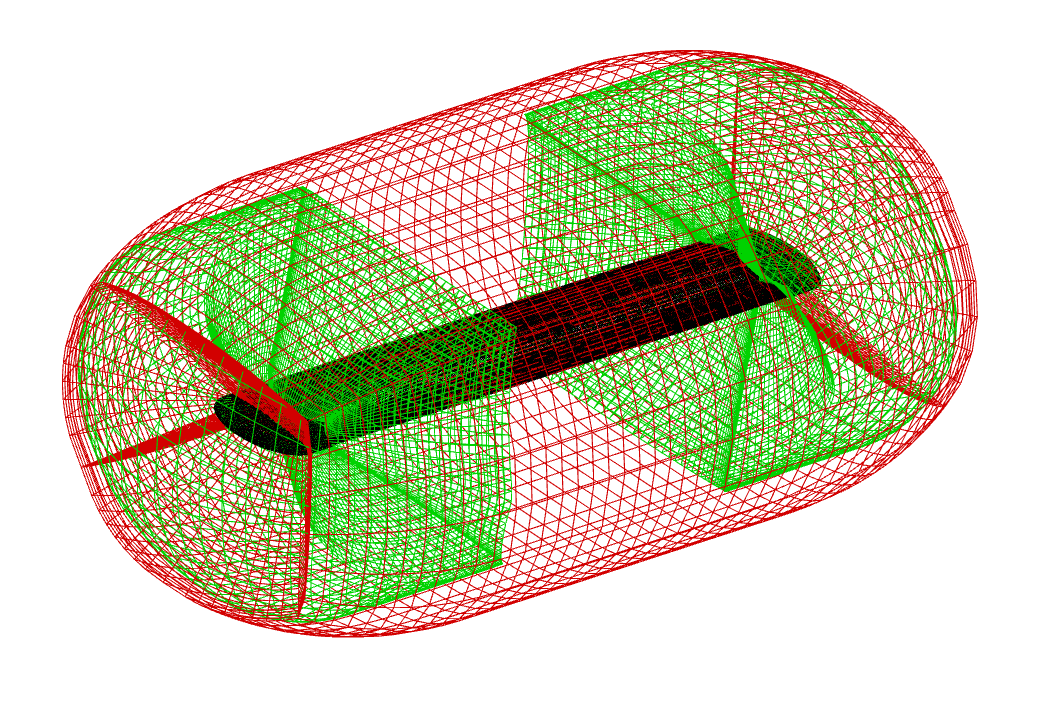}
        \par\smallskip
        (b)
    \end{minipage}
    \hfill
    \begin{minipage}[t]{0.32\textwidth}
        \centering
        \includegraphics[
            width=\linewidth,
            trim=22pt 72pt 22pt 22pt,
            clip
        ]{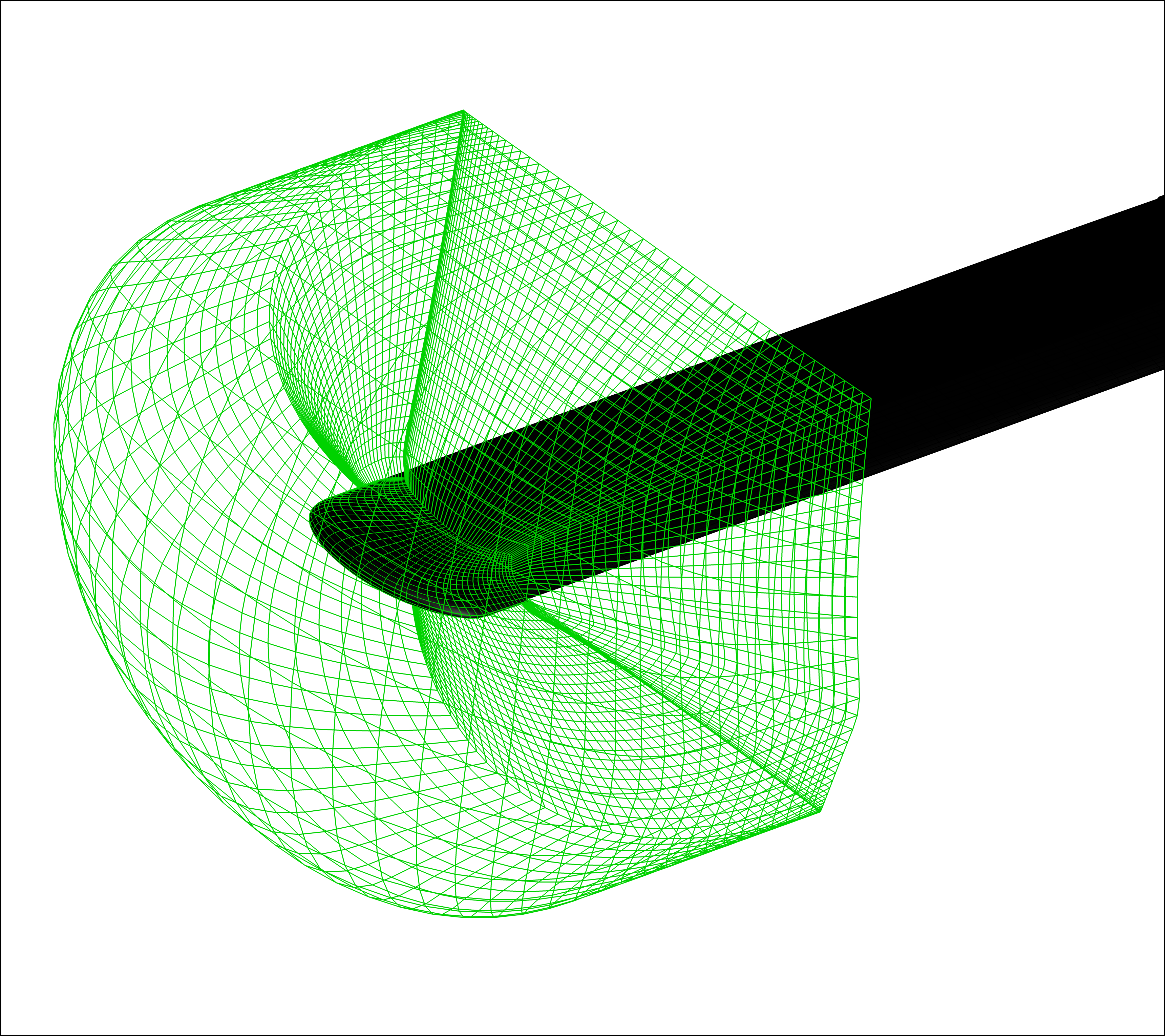}
        \par\smallskip
        (c)
    \end{minipage}
    \caption{Overset grids for the finite wing:
    (a) complete arrangement showing the O-grid (red),
    wing-tip cap grids (green), and background grid
    (black); (b) close-up view of the overlap between
    the O-grid and cap grids; and (c) a cap grid
    conforming to the wing-tip geometry. The wing
    surface is shaded in black in all subfigures.}
    \label{fig:finite_wing_grids}
\end{figure}

Figure~\ref{fig:finite_wing_grids}(a) shows the complete
overset grid arrangement, with the body-fitted O-grid
in red, the wing-tip cap grids in green, and the
background grid in black. The close-up view in Fig.~\ref{fig:finite_wing_grids}(b)
shows sufficient overlap between the O-grid and cap
grids to apply the weak characteristic interface
treatment and accurately resolve flow features
across the interfaces.
Figure~\ref{fig:finite_wing_grids}(c) shows one cap
grid conforming smoothly to the wing-tip geometry.
The wing surface is shaded in black in all panels.

Unlike the spanwise-periodic cases, this configuration exercises the genuinely three-dimensional components of the moving overset framework. The overset interfaces between the O-grid, the tip-cap grids, and the background grid are curved surfaces that are not aligned with one another or with any coordinate direction, so receiver points are served by fully three-dimensional (tensor-product Lagrange) donor stencils, Eq.~\eqref{eq:lagrange_interp_tensor}, whose donor cells and interpolation weights change in all three directions as the grids move. Many receiver points on the tip-cap grids lie on edges and corners of the grid blocks, where the interface terms are applied separately in each normal direction and summed, as described following Eq.~\eqref{eq:SAT_overset_NS}. The cap--O-grid interfaces translate with the wing, whereas the O-grid--background and cap--background interfaces move relative to the static background grid. Finally, because the domain is decomposed across 480 MPI ranks in all three directions, donor and receiver points across these interfaces generally reside on different ranks, so the interpolated solution and viscous fluxes are exchanged through spanwise as well as in-plane inter-grid communication at each time step.

The wing undergoes rigid-body translation in the
$y$-direction. In the notation of Eq.~\eqref{eq:rigid_map},
$x_p=z_p=0$ and $\phi=\theta=\psi=0$, while
\begin{equation}
    y_p(t)=h_A\sin(2\pi f_o t),
    \qquad h_A=0.75.
    \label{eq:finite_wing_plunge}
\end{equation}
Lengths are normalized by the chord length. Following the
reference definition, the Strouhal number is based on
the single-sided plunge amplitude,
$St=f^*h_A^*/U_\infty^*=0.2$. The non-dimensional
frequency is
\begin{equation}
    f_o=\frac{f^*c^*}{a_\infty^*}
       =\frac{St\,M_\infty}{h_A}
       \approx 0.02667.
    \label{eq:finite_wing_frequency}
\end{equation}
 The velocity of
the translating grids is
\begin{equation}
    x_t=z_t=0,
    \qquad y_t=2\pi f_o h_A\cos(2\pi f_o t).
    \label{eq:finite_wing_velocity}
\end{equation}

\begin{figure}
    \centering
    \begin{minipage}[t]{0.48\textwidth}
        \centering
        \includegraphics[width=\linewidth]{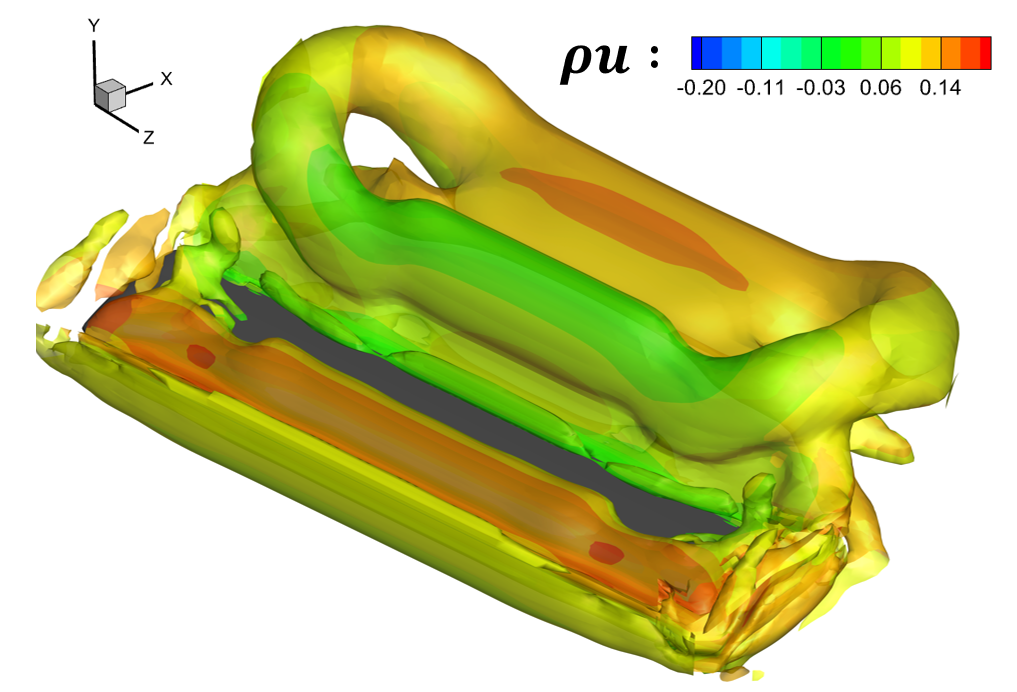}
        \par\smallskip
        (a)
    \end{minipage}
    \hfill
    \begin{minipage}[t]{0.48\textwidth}
        \centering
        \includegraphics[width=\linewidth]{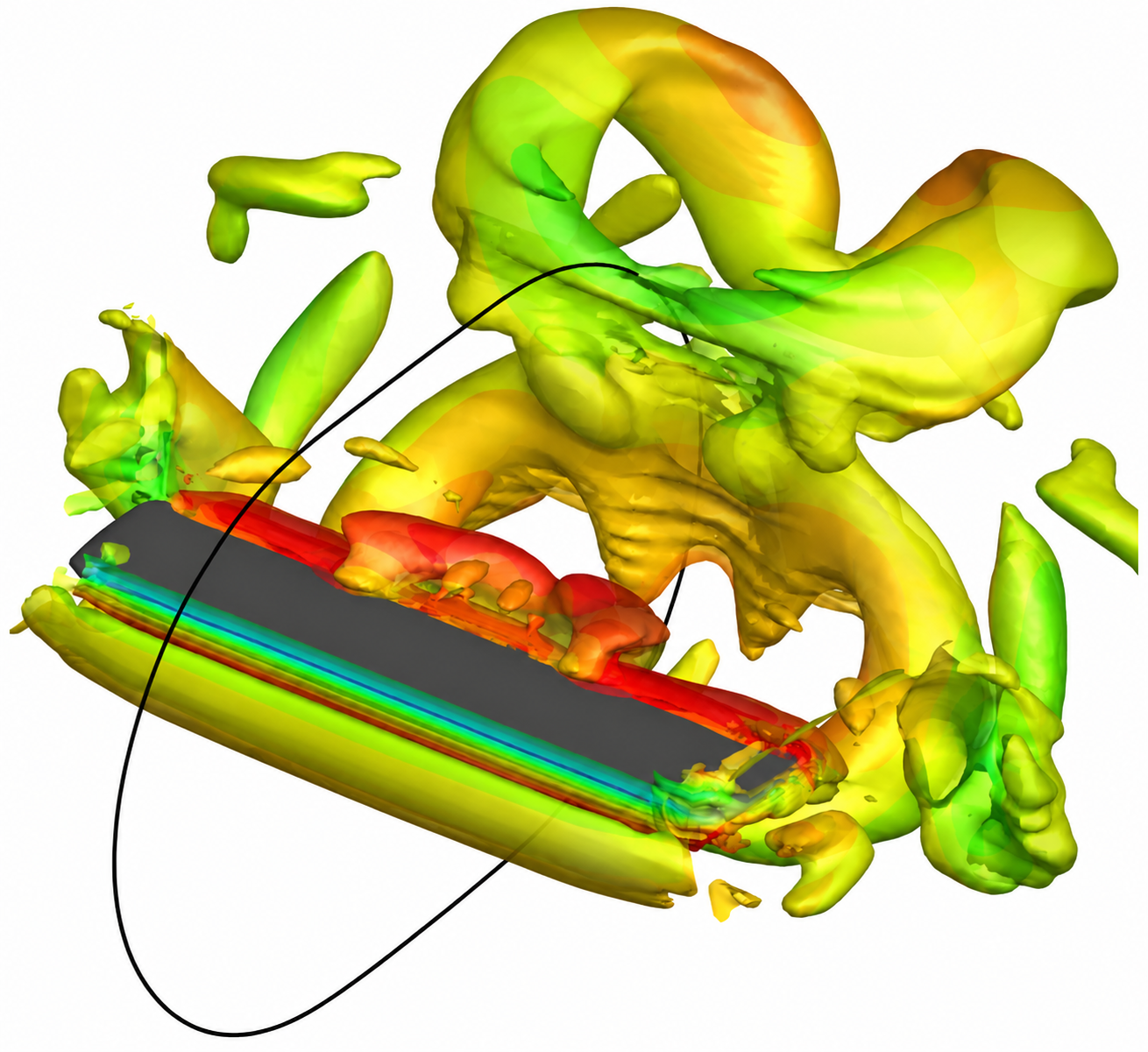}
        \par\smallskip
        (b)
    \end{minipage}
    \caption{Instantaneous vortex structures around the plunging
    finite wing at (a) $t \approx 17$ and (b) $t \approx 37$,
    visualized using iso-surfaces of the $Q$-criterion
    at $Q=0.003$ and colored by $x$-momentum.
    In sub figure (b), the black line indicates the overset interface between the moving O-grid and the background grid at the wing mid-span.}
    \label{fig:plunging_wing_Q}
\end{figure}

Figure~\ref{fig:plunging_wing_Q} shows the spanwise
variation and deformation of the vortical structures.
At $t=17$, elongated structures extend across the central
portion of the wing and curve around its tips. At $t=37$,
large deformed vortex loops extend into the wake,
accompanied by more pronounced spanwise variations
and smaller structures near the tips. In
Fig.~\ref{fig:plunging_wing_Q}(b), the black line marks
the interface between the moving O-grid and the
background grid at the wing mid-span. The vortical structures remain
smooth as they pass through this interface, with
no apparent discontinuities in the iso-surfaces,
providing further evidence of the robustness of the
proposed moving-grid interface treatment. Together
with the multi-grid arrangement shown in
Fig.~\ref{fig:finite_wing_grids}, these results also
demonstrate a three-dimensional wake that is not
represented by the preceding spanwise-uniform cases.

\begin{figure}
    \centering
    \begin{minipage}[t]{0.48\textwidth}
        \centering
        \includegraphics[width=\linewidth]{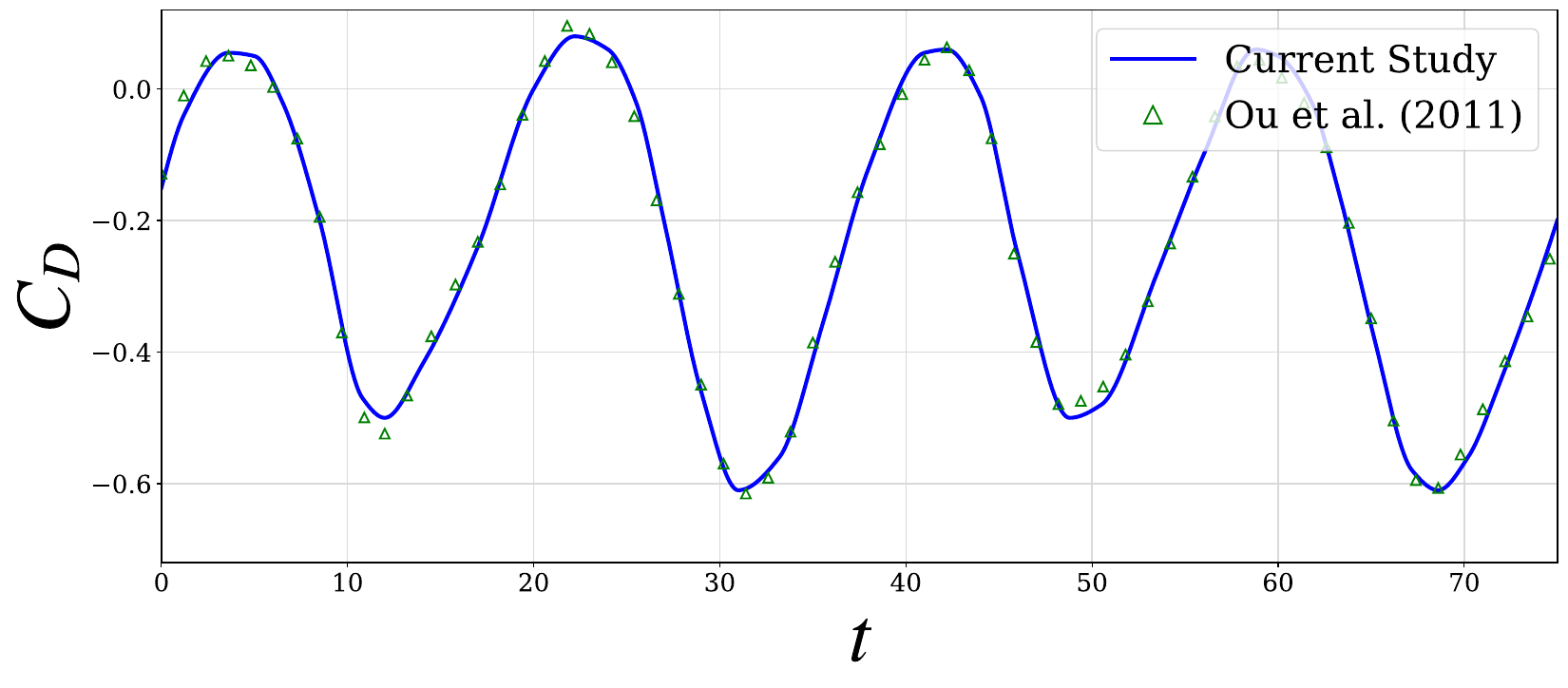}
        \par\smallskip
        (a)
    \end{minipage}
    \hfill
    \begin{minipage}[t]{0.48\textwidth}
        \centering
        \includegraphics[width=\linewidth]{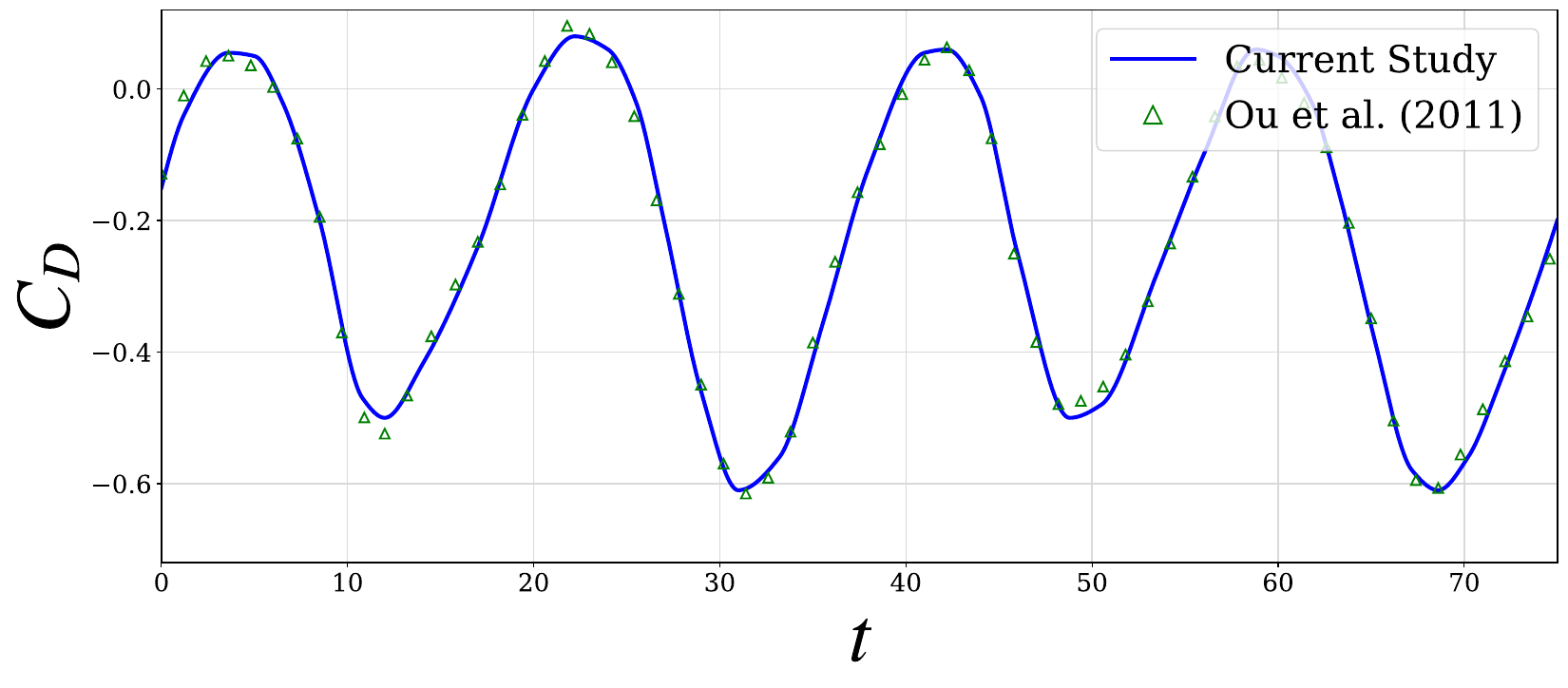}
        \par\smallskip
        (b)
    \end{minipage}
    \caption{Time histories of (a) the lift coefficient, $C_L$,
    and (b) the drag coefficient, $C_D$, for the plunging
    finite wing, compared with the reference results.}
    \label{fig:plunging_wing_forces}
\end{figure}

Figure~\ref{fig:plunging_wing_forces} compares the lift
and drag histories with the results of ~\cite{doi:10.2514/6.2011-1316} over two plunging
cycles. Both coefficients show good agreement
with the reference results.

\begin{figure}[htbp]
    \centering
    \includegraphics[width=0.98\textwidth]
        {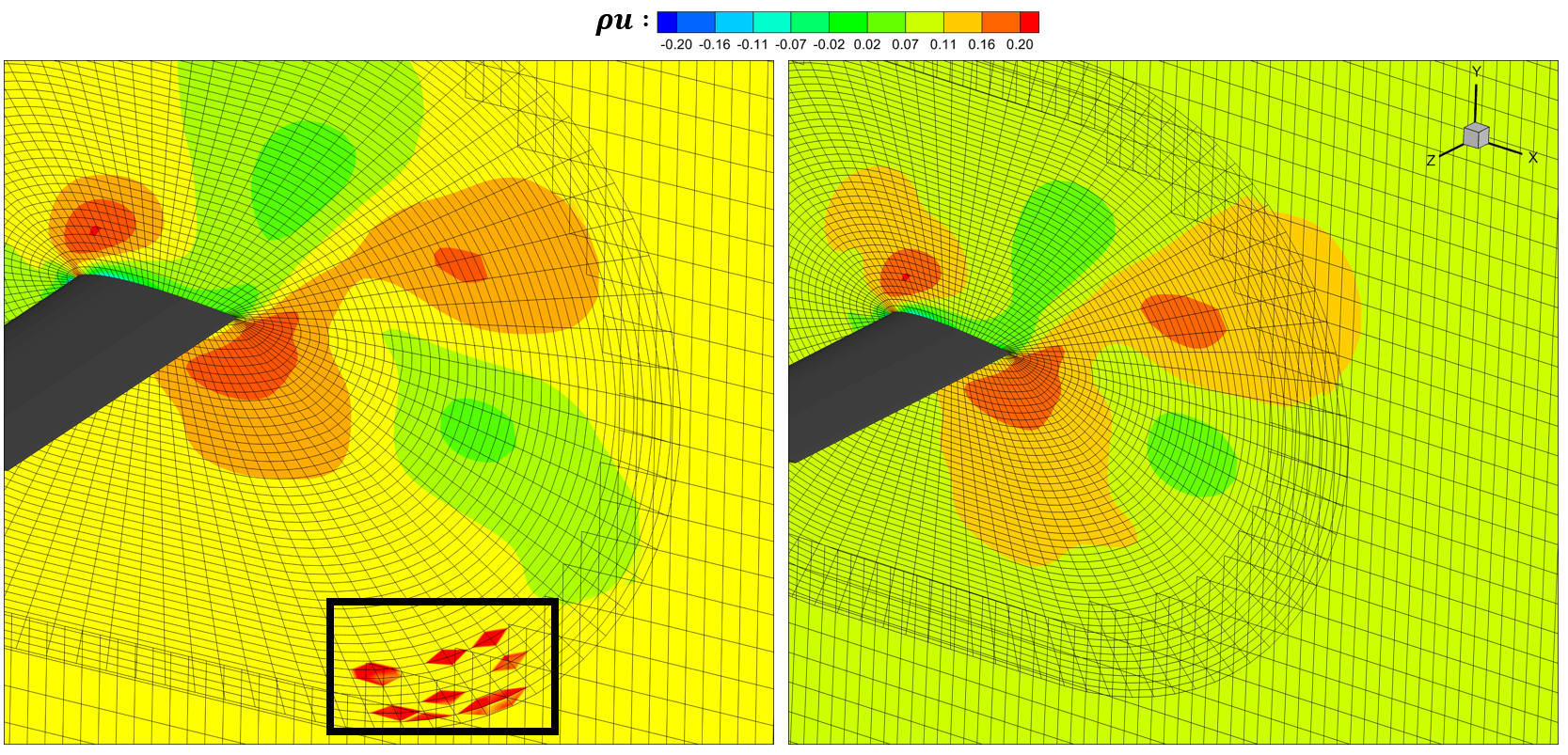}
    \caption{Streamwise momentum, $\rho u$, on a
    constant-$z$ plane near midspan at $t \approx 15.6$ of the plunging finite wing using the strong injection treatment (left) and the weak characteristic treatment (right). The wing surface is shaded in dark gray. The region within the black rectangle in the left subfigure highlights numerical artifacts near the overset interface in the strong injection
    interface treatment.}
    \label{fig:plunging_interface_comparison}
\end{figure}

\begin{figure}[htbp]
    \centering
    \includegraphics[width=0.65\textwidth]
        {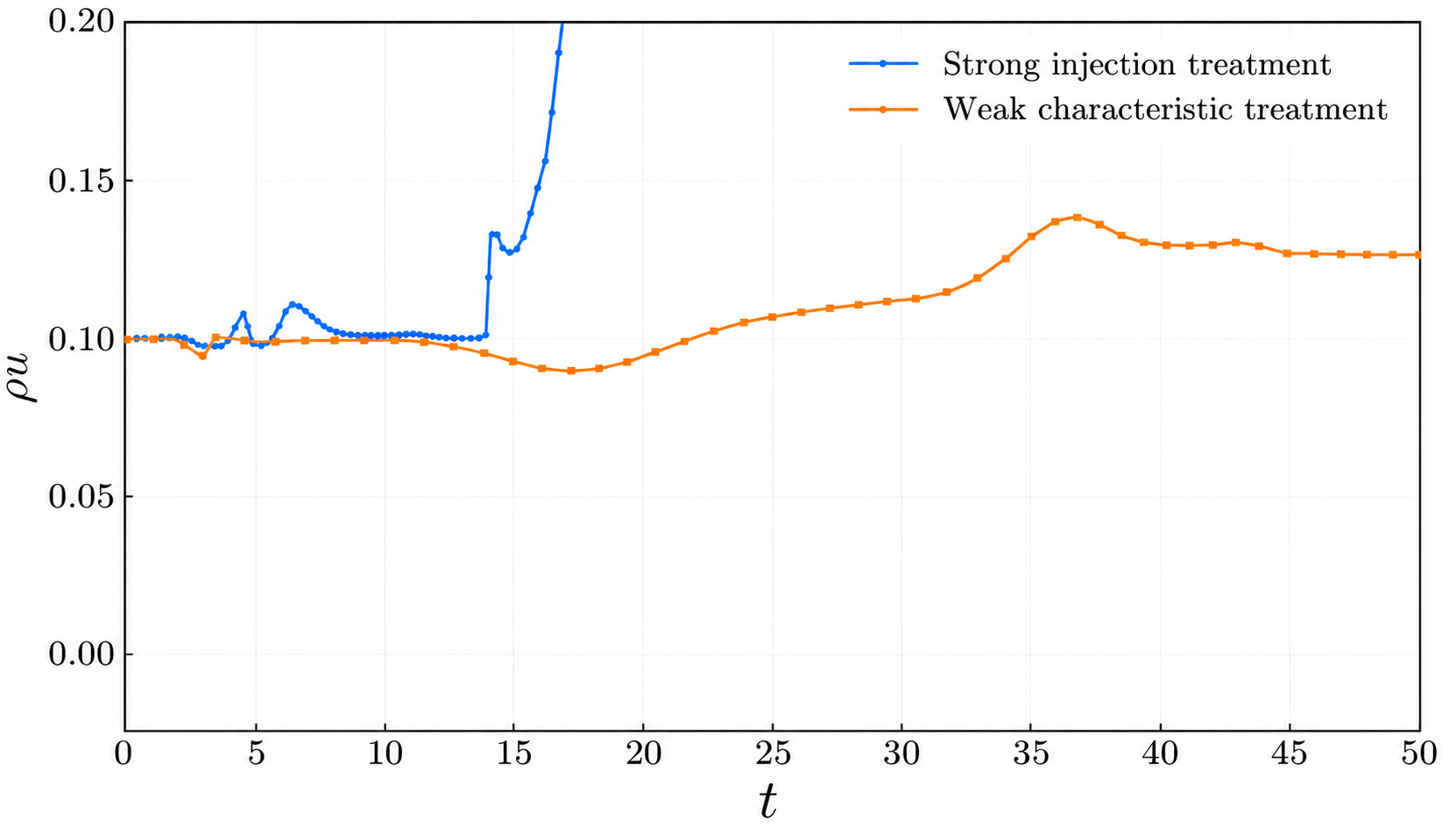}
    \caption{Time histories of streamwise momentum, $\rho u$, at an interface point on the moving O-grid at its interface with the background grid for
the plunging finite wing. The histories compare the strong
injection and weak characteristic treatments.}
    \label{fig:plunging_momentum_history}
\end{figure}

The comparison between the strong injection and weak
characteristic treatments in
Section~\ref{subsec:2D_Isentropic_Vortex} is extended
here to the three-dimensional plunging-wing case.
Both computations use identical grids, initial and
boundary conditions, time-stepping parameters and
interpolation order; only
the overset interface treatment is changed.
Figure~\ref{fig:plunging_interface_comparison} compares
the resulting streamwise momentum contours.
The strong injection treatment produces localized
numerical artifacts near the overset interface,
highlighted in the inset, whereas the weak characteristic
treatment maintains a smooth solution in the corresponding
region. Shortly after these artifacts appear, the
computation using strong injection becomes unstable
and blows up at the non-dimensional time $t\approx16.1632$.
In contrast, the weak characteristic treatment remains
stable over multiple plunging cycles. To better understand the development of these disturbances,
the time histories of streamwise momentum at an interface
point within the region exhibiting numerical artifacts
are shown in Fig.~\ref{fig:plunging_momentum_history}. This controlled
comparison extends the evidence from the isentropic-vortex convection
test and demonstrates the numerical robustness of the
weak characteristic treatment for the present highly
unsteady, three-dimensional viscous flow with spanwise inhomogeneity.

\subsection{3-D Pitching and Plunging Finite Wing}
\label{subsec:pitching_plunging_finite_wing}

A rectangular finite wing undergoing combined pitching
and plunging motion is considered following the second
NACA 0012 case of~\cite{doi:10.2514/6.2011-1316}.
The wing geometry, flow conditions, computational domain extents and
grids are identical to those used in
Section~\ref{subsec:plunging_finite_wing}.

The prescribed motion combines translation in the
$y$-direction with rotation about a spanwise axis located
one-third of the chord aft of the leading edge of the
unpitched wing. In the notation of Eq.~\eqref{eq:rigid_map},
the rotation angles $\phi$ and $\theta$ are zero. With the
leading edge of the unpitched wing at $x=0$, the fixed
pivot coordinates are $x_p(t)=x_{p,0}=c/3$ and
$z_p(t)=z_{p,0}$. The pivot position in the $y$-direction
is prescribed as
\begin{equation}
    y_p(t)=y_{p,0}+h_A\sin(2\pi f_o t),
    \label{eq:finite_wing_combined_plunge}
\end{equation}
where $y_{p,0}=0$ and $h_A=0.75$.
As in the 2-D pitching case (Section \ref{subsec:2d_pitching_airfoil}), the pitch
angle $\alpha(t)$ is defined positive nose-up (clockwise when viewed from the
positive $z$-axis toward the origin) and is measured from the unpitched
configuration. It is prescribed as
\begin{equation}
    \alpha(t)=\alpha_A\sin(2\pi f_o t+\varphi),
    \label{eq:finite_wing_combined_motion}
\end{equation}
where $\alpha_A=\pi/12$ (=$15^\circ$) is the pitch amplitude
and $\varphi=5\pi/12$ (=$75^\circ$) is the phase lead of
the pitch angle relative to the plunge displacement. Pitching and
plunging occur at the same frequency.

At $t=0$, the plunge displacement is zero and the prescribed
initial pitch is
\begin{equation}
    \alpha(0)=\alpha_A\sin\varphi\approx14.49^\circ.
\end{equation}
In Eq.~\eqref{eq:rigid_map}, the spanwise rotation angle
$\psi(t)$ is set to 
the grid rotation measured from this initial configuration,
\begin{equation}
    \psi(t)=-[\alpha(t)-\alpha(0)],
\end{equation}
so that $\psi(0)=0$ and $\mathbf{R}=\mathbf{I}$ at $t=0$.
Positive $\psi(t)$ corresponds to counterclockwise
rotation about the spanwise $z$-axis when viewed from the
positive $z$-axis toward the origin.


The non-dimensionalization and Strouhal number are
unchanged from the pure-plunging case, giving
$f_o=St\,M_\infty/h_A\approx0.02667$ and a period
$T=37.5$. The translational and angular velocities
are
\begin{equation}
    \dot{y}_p(t)=2\pi f_o h_A\cos(2\pi f_o t),
    \qquad
    \dot{\psi}(t)
    =-2\pi f_o\alpha_A\cos(2\pi f_o t+\varphi).
    \label{eq:finite_wing_combined_velocities}
\end{equation}
The grid-point velocities follow from Eq.~\eqref{eq:grid_vel},
including both the translational and rotational
contributions.

\begin{figure}
    \centering
    \begin{minipage}[t]{0.48\textwidth}
        \centering
        \includegraphics[width=\linewidth]{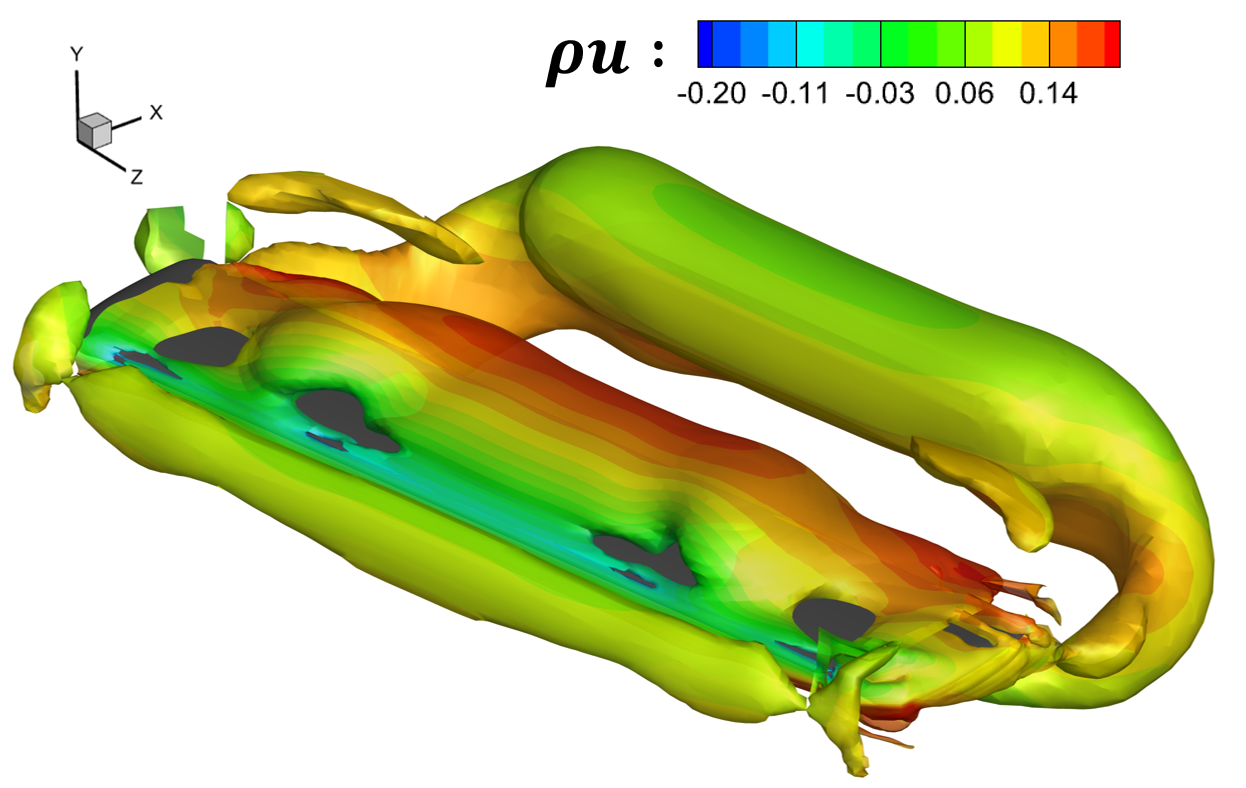}
        \par\smallskip
        (a)
    \end{minipage}
    \hfill
    \begin{minipage}[t]{0.48\textwidth}
        \centering
        \includegraphics[width=\linewidth]{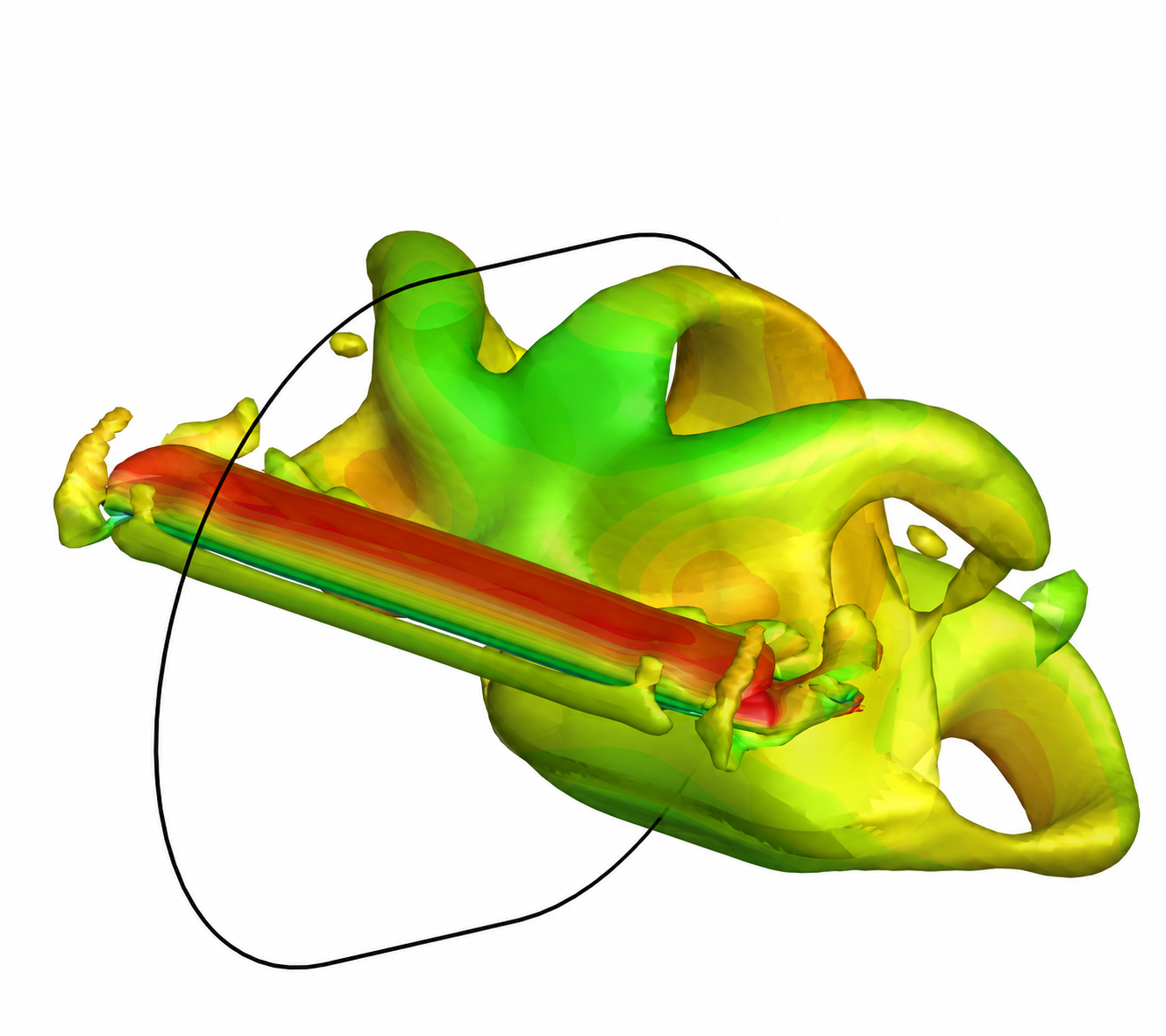}
        \par\smallskip
        (b)
    \end{minipage}
    \caption{Instantaneous vortex structures around the
    pitching and plunging finite wing at (a) $t \approx 17$ and (b) $t \approx 37$, visualized using
    iso-surfaces of the $Q$-criterion at $Q=0.003$
    and colored by $x$-momentum. In sub figure (b), the black
    line indicates the interface between the moving O-grid
    and the background grid.}
    \label{fig:pitching_plunging_wing_Q}
\end{figure}

\begin{figure}
    \centering
    \begin{minipage}[t]{0.48\textwidth}
        \centering
        \includegraphics[width=\linewidth]{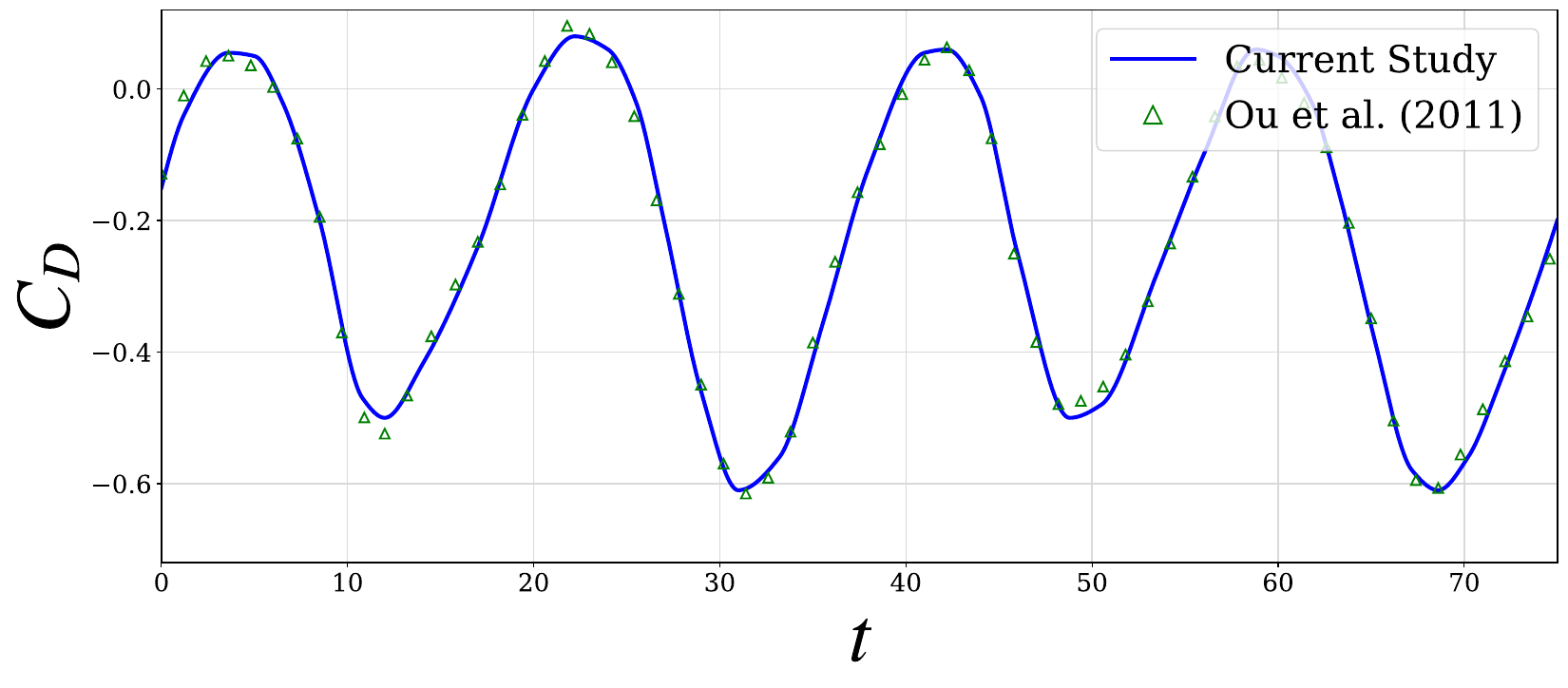}
        \par\smallskip
        (a)
    \end{minipage}
    \hfill
    \begin{minipage}[t]{0.48\textwidth}
        \centering
        \includegraphics[width=\linewidth]{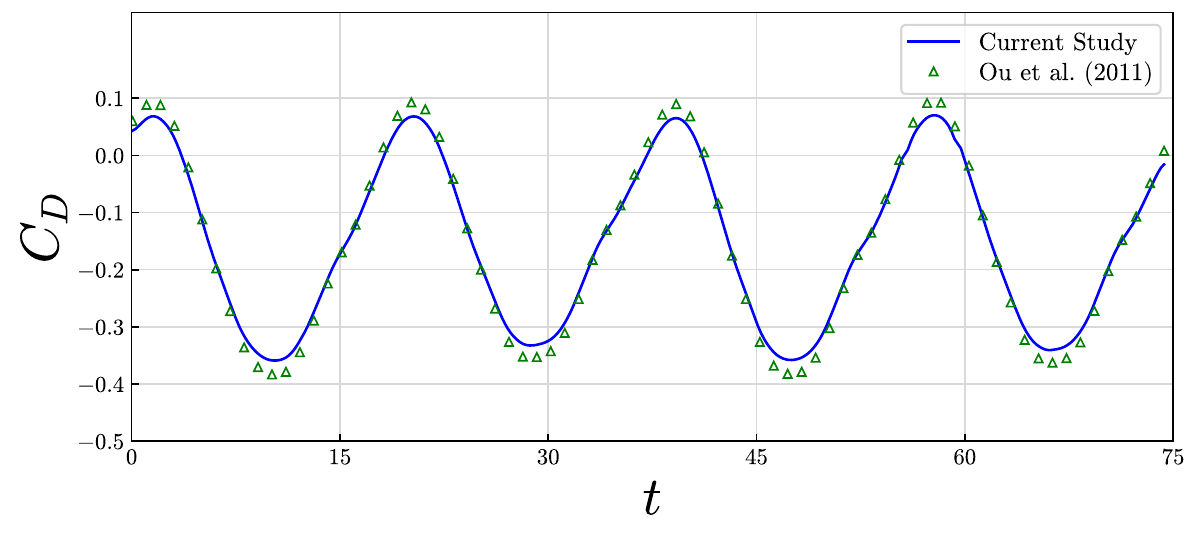}
        \par\smallskip
        (b)
    \end{minipage}
    \caption{Time histories of (a) the lift coefficient,
    $C_L$, and (b) the drag coefficient, $C_D$, for the
    pitching and plunging finite wing, compared with
    the reference results.}
    \label{fig:pitching_plunging_wing_forces}
\end{figure}

Figure~\ref{fig:pitching_plunging_wing_Q} shows marked
variations in streamwise momentum along the vortical
structures.
The addition of pitching changes the wing orientation
relative to the incident flow while retaining the
prescribed plunging displacement, thereby modifying
the unsteady aerodynamic loading. 
For this phase relationship, ~\cite{doi:10.2514/6.2011-1316} reported
lower mean thrust than for pure plunging.
The same trend is observed by comparing
Figs.~\ref{fig:plunging_wing_forces}(b)
and~\ref{fig:pitching_plunging_wing_forces}(b),
with combined pitching and plunging yielding
a less negative cycle-averaged drag coefficient.

\section{Effect of Fringe-Layer Count on Communication Volume and Runtime}
\label{sec:fringe_communication}
Conventional high-order overset formulations typically employ two or more layers of fringe points to preserve the order of accuracy of the interior scheme at the interface (\textit{e.g.}, \cite{Alabi2007,chan2002overgrid}), whereas the present weak characteristic treatment retains high-order accuracy and stability with a single layer. Fewer fringe points reduce the number of receiver points at which interpolated data must be exchanged, but a smaller data volume alone does not guarantee a lower overall parallel cost, since the cost of each exchange also depends on message latency, synchronization, and donor--receiver rank ownership. This section therefore quantifies the effect of the fringe-layer count directly, by measuring the overset data volume, the MPI-exchange time, and the total solver wall-clock time for one, two, and three fringe layers under otherwise identical conditions.

The effect of fringe-layer count is assessed using a
three-dimensional plunging-wing case on four overlapping grids:
a body-fitted O-grid of $200\times300\times200$ points, two
wing-tip cap grids of $100\times150\times150$ points each, and
a background grid of $200\times300\times300$ points, totaling
$34.5$ million grid points. All computations use 10 compute
nodes with 48 processor cores per node, totaling 480 cores,
and are advanced to a non-dimensional time of $t=20$.
The grid dimensions and computational resources are held
fixed while the number of fringe layers is varied from one
to three.

\begin{figure}
    \centering
    \begin{minipage}[t]{0.48\textwidth}
        \centering
        \includegraphics[width=\linewidth]{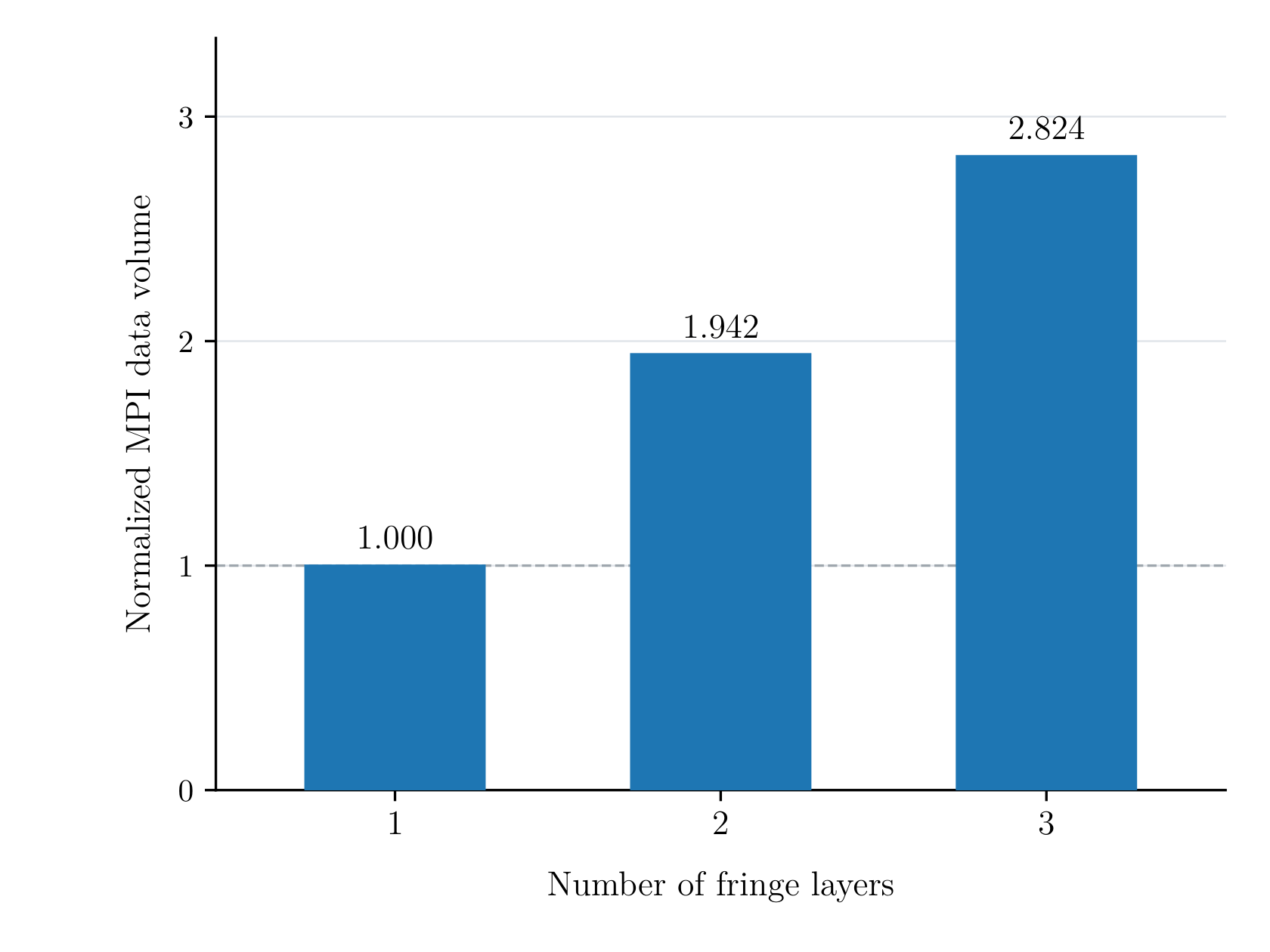}
        \par\smallskip
        (a)
    \end{minipage}
    \hfill
    \begin{minipage}[t]{0.48\textwidth}
        \centering
        \includegraphics[width=\linewidth]{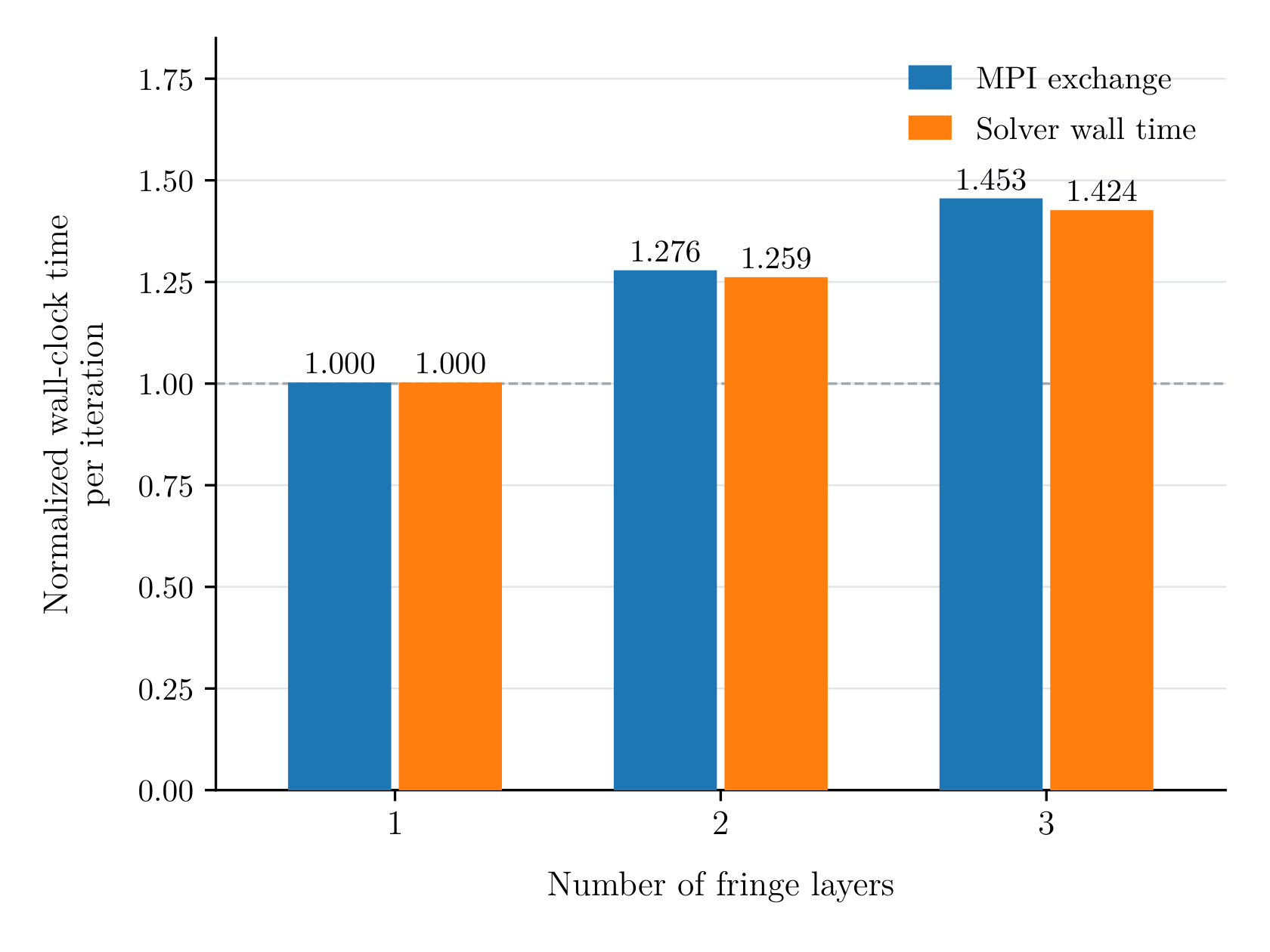}
        \par\smallskip
        (b)
    \end{minipage}
    \caption{Effect of fringe-layer count on (a) MPI data volume
    and (b) MPI-exchange and total solver wall-clock times per
    iteration. Each quantity is normalized by its corresponding
    value for the single-layer configuration.}
    \label{fig:fringe_communication}
\end{figure}

Figure~\ref{fig:fringe_communication}(a) reports the inter-rank
data volume associated with overset data exchanges. These
exchanges transfer interpolated conservative variables and
the three Cartesian viscous-flux vectors from donor ranks to
receiver ranks. The two- and three-layer
configurations exchange approximately 1.942 and 2.824 times
the single-layer volume, respectively. These ratios need not equal the number of fringe layers
because successive layers can contain different numbers
of receiver points. Interface geometry, hole masking, and
donor--receiver rank ownership determine the actual amount
of data transferred between MPI ranks.

The MPI-exchange time in Fig.~\ref{fig:fringe_communication}(b)
includes the time spent transferring interpolated conservative
variables and Cartesian viscous fluxes between donor and receiver
ranks through collective MPI operations, as well as the waiting
time within those operations. For each rank, we divide its
accumulated exchange time by the number of time steps to obtain
its mean exchange time per step. The figure reports the largest
of these rank means. We measure the total solver wall-clock time
separately over the full time-stepping loop.
Relative to the single-layer configuration, two and three
fringe layers increase MPI-exchange time by 27.6\% and
45.3\%, respectively, and total solver time by 25.9\% and
42.4\%. Together with the data-volume comparison in
Fig.~\ref{fig:fringe_communication}(a), these results
demonstrate that the single-layer treatment reduces
overset communication volume and elapsed exchange time,
with an accompanying reduction in overall runtime for
the tested configuration.

\section{Conclusions\label{sec:Conclusions}}

This work develops and validates a non-dissipative, high-order moving overset-grid method for unsteady compressible flows using central finite-difference schemes in the interior. A sixth-order centered interior scheme with summation-by-parts boundary closure achieves fourth-order global accuracy, while the weak overset coupling yields long-time stable solutions in all cases considered without upwind flux stabilization or ad hoc filtering/artificial dissipation. Stability is demonstrated numerically for inviscid and viscous moving overset grid simulations, and is further supported, for a 1-D model problem under periodic grid motion, by the fully discrete propagator, whose norm remains bounded, and by Floquet multipliers inside the unit circle; frozen-coefficient spectra also remain in the left half-plane at all sampled times. A rigorous theoretical stability analysis of the time-dependent moving-grid problem will be the subject of future work. Validation on 1-D scalar advection, 2-D isentropic vortex convection, flow past a 2-D rotating cylinder, and 2-D pitching airfoil and oscillating cylinder flows, together with spanwise-periodic 3-D counterparts that verify consistency of the three-dimensional implementation, confirms the method's accuracy and robustness. Genuinely three-dimensional interface behavior, including tip flow and non-aligned interfaces between the body, tip-cap, and background grids, is demonstrated for finite-span wings undergoing pure plunging and combined pitching--plunging motions. Moreover, in all simulations, stable results are obtained using only one layer of fringe points, and the inviscid convergence studies confirm $p+1$ global order with this single layer. The effect of this choice on parallel cost was quantified for the three-dimensional plunging finite-wing simulation (34.5 million grid points, 480 cores) in Section~\ref{sec:fringe_communication}. Relative to the single-layer configuration, two and three fringe layers increased the inter-rank overset data volume by factors of 1.94 and 2.82, the MPI-exchange time by 27.6\% and 45.3\%, and the total solver time by 25.9\% and 42.4\%, respectively. Thus, for the tested configuration, the single-layer treatment reduced overall runtime by approximately 21\% and 30\% compared to the two- and three-layer configurations.

\section{Acknowledgments}
This work was supported in part by the Defense Established Program to Stimulate Competitive Research (DEPSCoR) grant AF-FA9550-24-1-0170 and the National Science Foundation (NSF) CBET Award\# 2418406 (under the direction of Dr.~Ron Joslin). The computations used the Advanced Cyberinfrastructure Coordination Ecosystem: Services \& Support (ACCESS) resources under grants PHY210037 and PHY240185 and the Auburn University Easley Cluster.

\appendix

\section{2-D Isentropic Vortex Convection On Static Overset Grids}
\begin{figure}
    \centering
    \includegraphics[width=0.5\textwidth]{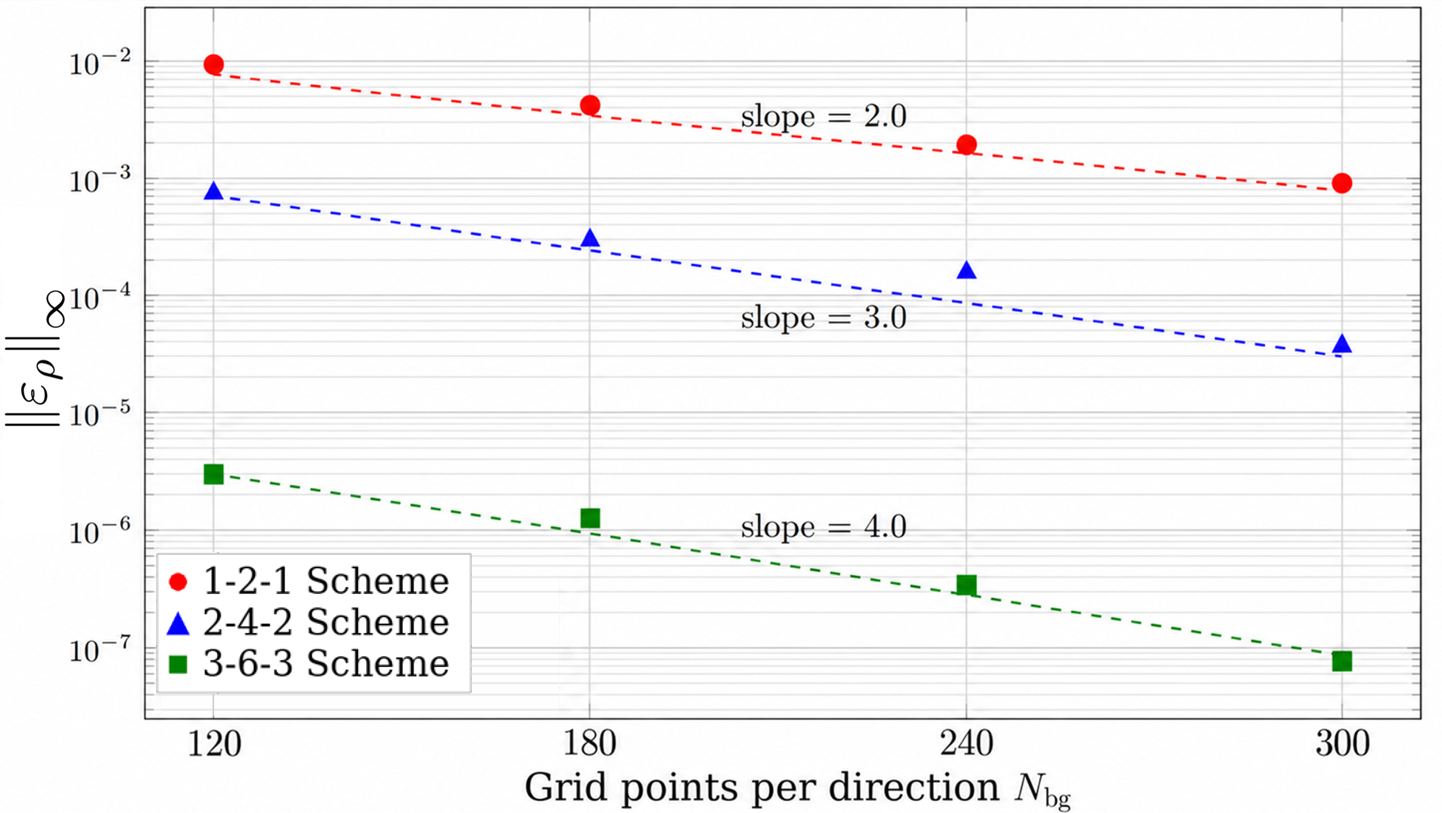}
    \caption{Grid convergence results showing \(\left\|\varepsilon_{\rho}\right\|_{\infty}\) from the proposed overset scheme for the 2-D isentropic vortex convection problem solved over static overset grids.}
    \label{fig:static_vort}
\end{figure}

The inviscid test case discussed in Section~\ref{subsec:2D_Isentropic_Vortex} is validated here on a static overset grid configuration using various $p-2p-p$ schemes. The grid configuration consists of two overlapping Cartesian square grids, as shown in Fig.~\ref{fig:vort_initial}(a), with a smaller square overset grid containing \(N_{\mathrm{sg}}\times N_{\mathrm{sg}}\) points placed inside a background grid containing \(N_{\mathrm{bg}}\times N_{\mathrm{bg}}\) points. In the static case, both grids remain fixed in time. Therefore, the time-metric terms defined in Eq.~\eqref{eq:metric_time} reduce to
\begin{equation}
\xi_t = \eta_t = \zeta_t = 0.
\end{equation}
The $1-2-1$ scheme employs Lagrange linear interpolation, whereas the $2-4-2$ and $3-6-3$ schemes use Lagrange cubic interpolation. Figure ~\ref{fig:static_vort} presents the corresponding grid convergence results for the three schemes over four grid resolutions. A $p+1$ order of accuracy is observed for the $p-2p-p$ scheme, consistent with the theoretical result of \cite{gustafsson1975convergence}. Based on these results, the $3-6-3$ scheme is used for several moving grid simulations discussed in Section \ref{sec:Numerical-results}.
\label{sec:appdxB}

\section{2-D NACA 0012 Airfoil On Static Overset Grids}
\label{sec:appdxC}
\begin{figure}
\begin{centering}
\includegraphics[width=0.42\textwidth]{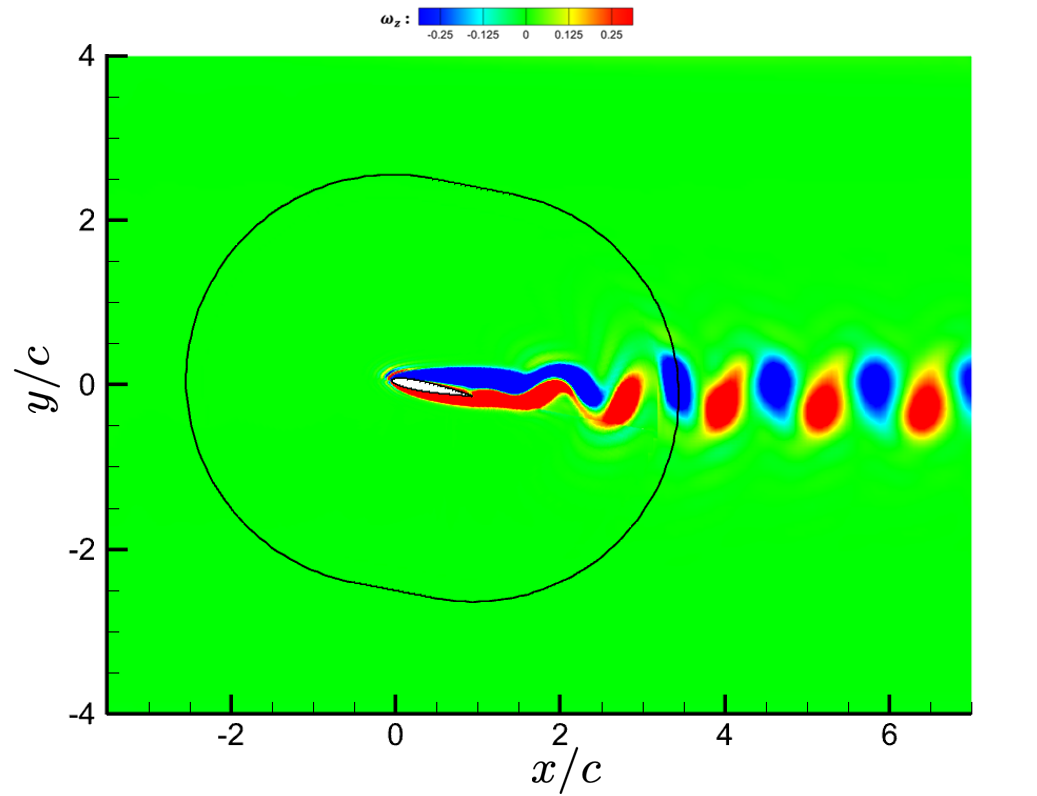}\includegraphics[width=0.5\textwidth]{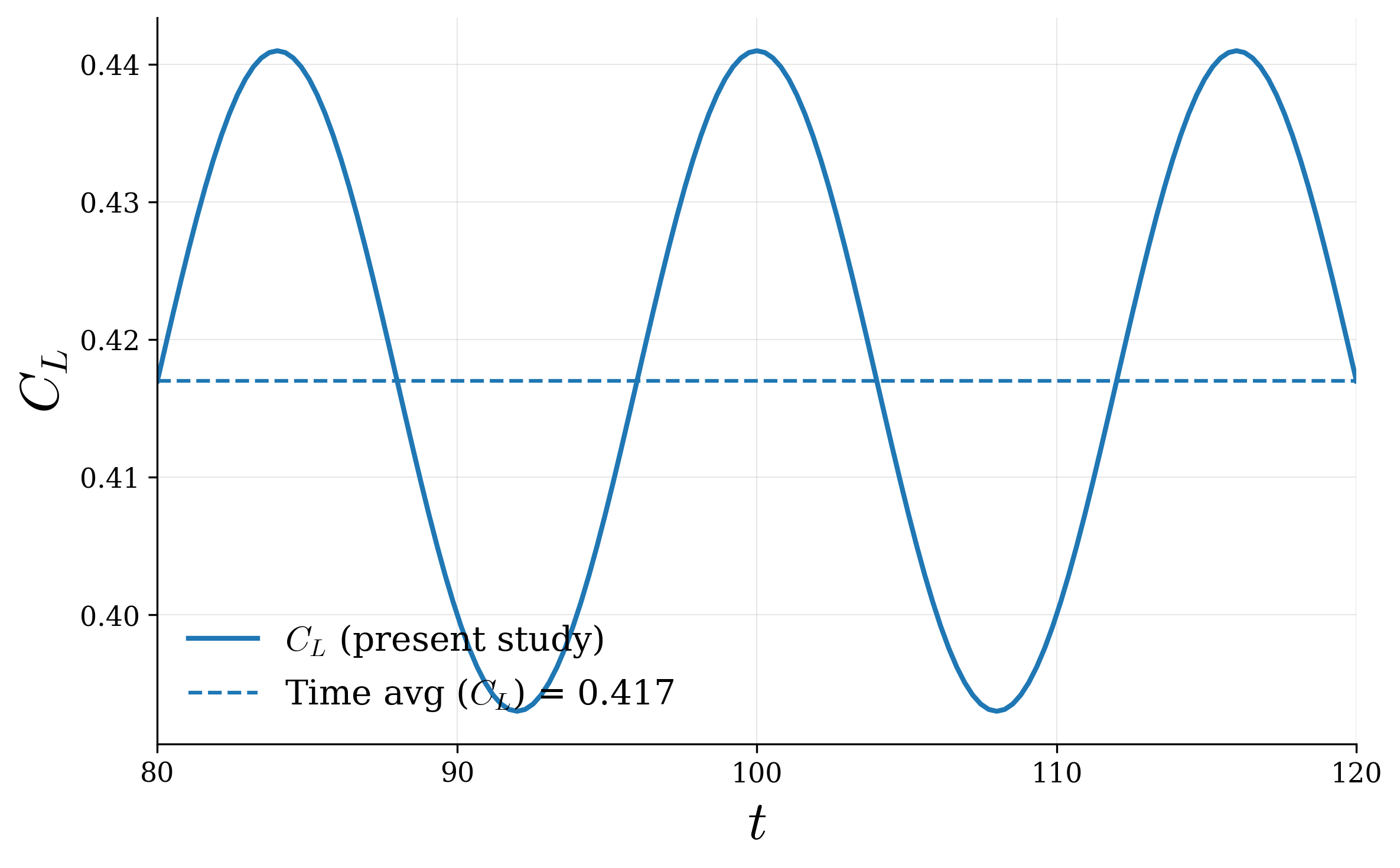}
\par\end{centering}
\begin{centering}
\qquad{}(a)\qquad{}\qquad{}\qquad{}\qquad{}\qquad{}\qquad{}\qquad{}\qquad{}\qquad{}\qquad{}\qquad{}(b)
\par\end{centering}
    \centering
\caption{NACA 0012 airfoil at $\alpha=10^\circ$ and $Re_c=1000$: (a) instantaneous z-vorticity ($\omega_z$) contour and (b) time history of the lift coefficient, $C_L$.}
    \label{fig:air_initial}
\end{figure}

The flow over a static two-dimensional NACA 0012 airfoil at an angle of attack $\alpha=10^\circ$ and a Reynolds number $Re_c=1000$ is simulated using overset grids for validation. A body-conforming O-grid is used around the airfoil with an overset rectangular (background) grid to capture the wake flow. Fig.~\ref{fig:air_initial}(a) shows the $z$-vorticity contours at a time instant and Fig.~\ref{fig:air_initial}(b) presents the temporal evolution of the lift coefficient, $C_L$. The time-averaged value of computed $C_L$ is $0.417$, which is in good agreement with the reference value of \(C_L=0.41\) reported in \cite{MITTAL1994253}. This supports the accuracy of the static overset implementation.

\section{Flow Over a Circular Cylinder On Static Overset Grids}
\label{sec:appdxD}
\begin{figure}[H]
\begin{centering}
\includegraphics[width=0.445\textwidth]{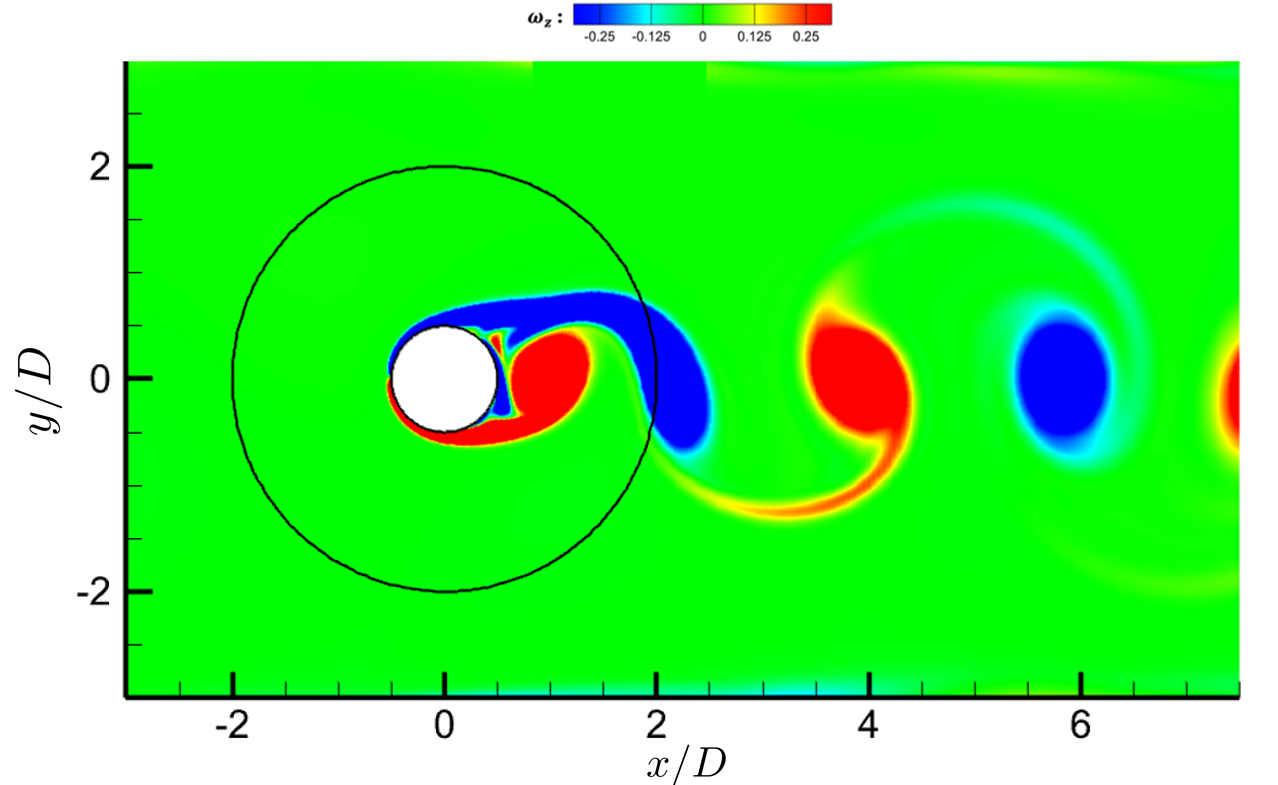}\includegraphics[width=0.54\textwidth]{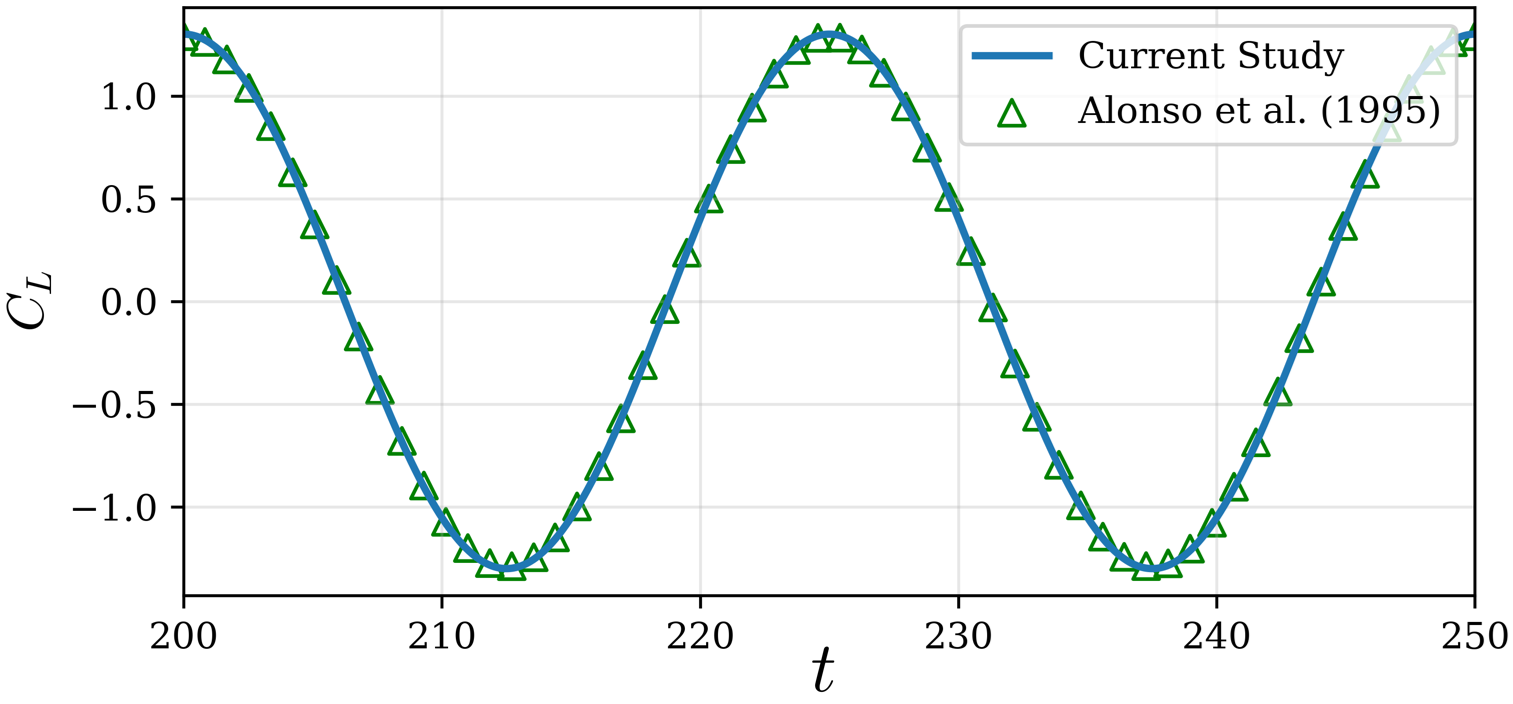}
\par\end{centering}
\begin{centering}
\qquad{}(a)\qquad{}\qquad{}\qquad{}\qquad{}\qquad{}\qquad{}\qquad{}\qquad{}\qquad{}\qquad{}\qquad{}(b)
\par\end{centering}
    \centering
\caption{Circular cylinder at $Re_D=500$: (a) instantaneous z-vorticity ($\omega_z$) contour field, and (b) time history of the lift coefficient, $C_L$.}
    \label{fig:cyl_initial}
\end{figure}

A circular cylinder at Reynolds number $Re_D=500$ is simulated on a static overset-grid configuration for validation. As shown in Fig.~\ref{fig:cyl_initial}(b), the computed $C_L$ time history is in good agreement with the reference results of \cite{Alonso1995MultigridUN}.

\bibliography{Master}
\bibliographystyle{elsarticle-num}

\end{document}